\documentclass[manuscript,screen,nonacm]{acmart}

\usepackage{amsmath,amssymb}
\usepackage{booktabs}
\usepackage{tabularx}
\usepackage{float}
\usepackage{placeins}
\usepackage{tcolorbox}
\usepackage{array}
\usepackage{listings}
\usepackage{booktabs}
\usepackage{multirow}
\newcommand{\ignore}[1]{}
\usepackage{xcolor}
\usepackage{subcaption}

\tcbuselibrary{skins,breakable}

\acmJournal{TOSEM}

\newcommand{\Name}{\textsc{A-ProS}} 

\newenvironment{promptbox}[2][]{
  \begin{tcolorbox}[title={#2},
  fonttitle={\footnotesize}, 
  enhanced jigsaw, 
  colbacktitle=black,
  arc=1pt,
  opacityframe=0,
  left=1pt,
  right=1pt,
  top=1pt,
  bottom=1pt,
  boxrule=0.4mm,
  opacityframe=1,
  colback=green!8!white,
  breakable]
}{\end{tcolorbox}}

\title{\Name{}: Towards Reliable Autonomous Programming Through Multi-Model Feedback}
\titlenote{Accepted for publication in ACM Transactions on Software Engineering and Methodology (TOSEM).}

\author{Anika Tabassum}
\authornote{These authors contributed equally to this work.} 
\affiliation{%
\institution{Dept. of Computer Science and Engineering, University of Dhaka}
\city{Dhaka}
\country{Bangladesh}
}
\email{anika-2019417844@cs.du.ac.bd}

\author{Md Sifat Hossain}
\authornotemark[2] 
\affiliation{%
\institution{Dept. of Computer Science and Engineering, University of Dhaka}
\city{Dhaka}
\country{Bangladesh}
}
\email{mdsifat-2019217800@cs.du.ac.bd}

\author{Md. Fahim Arefin}
\authornote{Corresponding author}
\affiliation{%
\institution{Dept. of Computer Science and Engineering, University of Dhaka}
\city{Dhaka}
\country{Bangladesh}
}
\email{fahim@cse.du.ac.bd}

\author{Tariqul Islam}
\affiliation{%
\institution{Dept. of Information Systems,}
\institution{University of Maryland, Baltimore County}
\city{Maryland}
\country{USA}
}
\email{mtislam@umbc.edu}

\author{Tarannum Shaila Zaman}
\authornote{Corresponding author}
\affiliation{%
\institution{Dept. of Information Systems,}
\institution{University of Maryland, Baltimore County}
\city{Maryland}
\country{USA}
}
\email{zamant@umbc.edu}

\begin{document}

\begin{abstract}
Large Language Models (LLMs) demonstrate strong potential for automated code generation, yet their ability to iteratively refine solutions using execution feedback remains underexplored. Competitive programming offers an ideal testbed for this investigation, as it demands end-to-end algorithmic reasoning, precise implementation under strict computational constraints, and complete functional correctness with rigorous evaluation. In this paper, we present A-ProS, an autonomous AI agent that solves competitive programming problems through a hybrid multi-model feedback framework separating solution generation from specialized debugging. A-ProS combines ChatGPT-based generators (GPT-4 and GPT-5) with three debugging critics: Codestral-2508, Llama-3.3-70B, and DeepSeek-R1, under a 2 × 3 factorial design. We evaluate six workflows on 367 problems from ICPC World Finals (2011–2024) and Codeforces (rated 1200–1800). The results show that GPT-5 workflows improve from 39 initial accepted solutions to 85–90 after three refinement rounds, while GPT-4 improves from 15 to 31–38. A controlled ablation on 47 problems shows that stateful refinement outperforms stateless approaches by 8.5–10.6 percentage points and reduces repeated failures by up to 3.5×. Compared to baseline agent loops, A-ProS achieves over 2× greater gains, highlighting the importance of persistent context and multi-model feedback for reliable autonomous program synthesis.
\end{abstract}



\maketitle

\section{Introduction}
\label{intro}

Large Language Models (LLMs)~\cite{LLM} are rapidly transforming the landscape of software engineering~\cite{SE}. They enable machines to interpret natural language specifications and generate functional code with minimal human supervision~\cite{NLP-SE, NLP-SE2, Se-Model, ZAMAN2026112785}. Modern LLMs can perform a wide range of development tasks, including code completion~\cite{Dong2023Self}, documentation generation~\cite{Luo2024RepoAgent}, and automated debugging~\cite{BugSpotter, parvez2026depro}, demonstrating impressive syntactic and semantic fluency across programming languages~\cite{ChowdharyFundamentals}. However, despite their progress, most LLMs still operate as static predictors~\cite{ZhuAre}: they generate a single response for a given prompt with limited ability to reflect, plan, or correct their reasoning. When facing algorithmically challenging problems, such as those found in competitive programming, they often fail to maintain logical consistency, computational efficiency, and effective error recovery. This limitation underscores the need for models that can reason iteratively, evaluate their own outputs, and adapt dynamically to task feedback.

Current LLM-based code generators largely function as black-box predictors~\cite{Sam2025Predicting}, offering little insight into the reliability of their results or the reasoning processes that produce them. Introducing feedback-based autonomy enables models to assess their outputs, identify weaknesses, and iteratively refine solutions, leading to systems that are both more accurate and more transparent. As AI-assisted development continues to permeate software engineering, agentic frameworks provide a natural evolution beyond isolated code generation, integrating design, testing, and validation within an end-to-end development loop. Such frameworks hold promise for creating intelligent programming assistants that combine self-assessment, adaptability, and explainability, setting the stage for autonomous yet accountable software engineering systems.

An agentic system~\cite{Hughes2025AI} redefines the model as an autonomous reasoning entity~\cite{Putta2024Agent} capable of self-evaluation, iterative improvement~\cite{Xue2025IMPROVE}, and adaptive decision-making~\cite{Li2025From}. Through structured feedback loops and coordinated model collaboration, such systems can plan, critique, and refine their outputs with minimal human oversight, mirroring how professional software engineers approach design, testing, and debugging. Agentic LLMs have been applied to various engineering challenges, including automated test repair~\cite{Yang2025survey}, bug localization~\cite{Meng2024empirical}, performance optimization~\cite{Buehler2025PRefLexOR}, and intelligent code refactoring~\cite{Lakshmi2025Enhancing}, where reasoning and feedback cycles are fundamental. Embedding autonomy and trust calibration into these systems transforms them from passive generators into proactive collaborators that can reduce developer workload, improve reliability, and accelerate large-scale software development.

Competitive programming~\cite{Xu2025ICPCeval, LLM-ProS, hendrycks2021apps, Islam2024MapCoder, haller2024pecc, Wang2025CodeFlowBench} offers a uniquely rigorous and reproducible environment for evaluating reasoning-driven code generation. Unlike conventional programming benchmarks, these problems demand complete algorithmic design, precise implementation under strict computational limits, and full functional correctness within constrained specifications. This setting parallels critical phases of software engineering, such as requirements analysis~\cite{Roychoudhury2025Agentic}, algorithm design~\cite{Li2022Competition, Xu2025ICPCeval, LLM-ProS, Islam2024MapCoder, perfcodegen2024}, and performance optimization~\cite{Wang2022Self}, but in a quantitatively controlled environment. Each problem encapsulates core engineering competencies such as handling edge cases~\cite{Chen2021Evaluating}, managing algorithmic complexity~\cite{Wang2022Self}, and balancing efficiency with correctness. Evaluating LLMs on competitive programming tasks therefore provides a measurable proxy for their ability to perform structured reasoning~\cite{WeiChain}, self-debugging~\cite{Xue2025IMPROVE, Dong2023Self}, and iterative refinement~\cite{Yang2023InterCode, Huang2023AgentCoder, Nguyen2024AgileCoder, perfcodegen2024, Islam2024MapCoder}, which are essential for realizing autonomous software engineering.

Despite the promise of competitive programming as an evaluation framework, most prior approaches~\cite{humaneval, hendrycks2021apps, yu2024humanevalpro, Chen2021Evaluating} treat it as a static benchmark: the model receives a problem, generates a single solution, and is evaluated on its correctness. This paradigm neglects critical dimensions of real-world software development, including iterative feedback, error-aware refinement, and multi-model specialization. It fails to capture how contextual memory and adaptive reasoning enable iterative progress over multiple attempts. While previous studies~\cite{Li2022Competition, LLM-ProS, Xu2025ICPCeval, Luo2023WizardCoder} have improved single-attempt accuracy through better prompting or larger model sizes, few works~\cite{Huang2023AgentCoder, Nguyen2024AgileCoder, Islam2024MapCoder, Xue2025IMPROVE} have explored coordinated, multi-model collaboration where specialized agents exchange structured feedback to improve reasoning and code quality. Moreover, the field lacks systematic evidence comparing the contributions of code-specialized and reasoning-focused critics within collaborative debugging workflows.

To address these gaps, we introduce \Name{}, an agentic programming framework that integrates multiple LLMs into a coordinated, feedback-driven workflow. Unlike traditional single-model paradigms~\cite{Li2022Competition, Chen2021Evaluating, Nijkamp2022CodeGen, zheng2023codegeex}, \Name{} formalizes how diverse models can interact to verify, critique, and refine one another’s outputs under fixed computational and temporal budgets. The framework shifts code generation from an isolated prediction task to a cooperative reasoning process that emulates human problem solving. \Name{} employs GPT-4 and GPT-5 as primary solution generators, while Codestral-2508, Llama-3.3-70B, and DeepSeek-R1 act as debugging critics providing structured, role-specific feedback. Each iteration begins with a generator’s initial solution, followed by critique and repair from a selected critic, after which the refined code is submitted to the live Codeforces online judge through authenticated browser automation. This iterative process forms a closed feedback loop where reasoning, debugging, and optimization evolve autonomously, mirroring professional engineering practices.

We evaluated \Name{} on 367 real-world competitive programming problems from ICPC and Codeforces~\cite{ICPC}, encompassing a wide range of algorithmic domains. Across all configurations, GPT-5-based~\cite{Leon2026GPT} workflows consistently outperformed GPT-4~\cite{OpenAI2023Gpt}, and DeepSeek-R1~\cite{DeepSeek2024Deepseek} was the most effective debugging critic. A controlled paired ablation ($n=47$) provides evidence consistent with a causal contribution of stateful refinement: the A-ProS default achieves +8.5--10.6 percentage-point higher Itr$_3$ acceptance rates than stateless refinement, with 2.9--3.5$\times$ lower error repetition rates. Bootstrap 95\% CIs are $[0.00, +0.15]$ for GPT-5 and $[0.00, +0.11]$ for GPT-4; the GPT-5 effect yields exact McNemar $p \approx 0.063$ ($b+c = 16$). A baseline comparison further confirms that A-ProS gains are 2.2--2.3$\times$ larger than multi-round stateless refinement, confirming that persistent context, not repeated sampling alone, drives the improvement. These findings are scoped to competitive programming as a rigorous benchmark and confirm that persistent multi-model feedback loops provide qualitatively distinct benefits over simple stateless refinement strategies.

In summary, this paper makes the following contributions:
\begin{itemize}
    \item We propose \Name{}, an autonomous AI agent that solves competitive programming problems through a hybrid multi-model architecture separating solution generation from specialized debugging feedback, enabling systematic evaluation of code-specialized, general-purpose, and reasoning-focused critics under a $2\times3$ factorial design.
    \item We evaluate \Name{} on 367 problems (167 ICPC World Finals 2011-2024; 200 Codeforces rating 1200-1800), comparing six workflows under identical iteration limits, and release a companion results website~\cite{Hybrid}.
    \item We conduct a controlled paired ablation ($n=47$) providing controlled evidence consistent with a causal contribution of persistent conversational context to iterative gains, with McNemar's test and bootstrap confidence intervals showing that stateful refinement achieves +8.5-10.6\,pp higher Itr$_3$ acceptance rates and a 2.9-3.5$\times$ reduction in error repetition compared to stateless conditions.
    \item We support artifact-based reproducibility by releasing all source code, orchestration scripts, curated problem datasets, generated solutions, and archived judge artifacts~\cite{sifatGitHub}, while noting that the ICPC/Gym portion still requires authenticated high-rated access for live end-to-end reruns.
\end{itemize}

\section{Background}
This section introduces key concepts underlying our framework, including agentic AI as a paradigm for autonomous, multi-step reasoning. It also highlights the limitations of one-shot LLM evaluation and motivates the need for iterative, multi-agent feedback grounded in prior work.

\subsection{Agentic AI}
Agentic AI~\cite{Wang2025Agentic} refers to a paradigm in artificial intelligence where models act as autonomous, goal-driven entities capable of reasoning, planning, and self-correcting through iterative interaction with their environment or with other agents. Unlike traditional static inference models that generate a single output in response to a prompt, agentic systems maintain persistent internal states and dynamically adapt their behavior over multiple decision or reasoning steps. This transition from passive response generation to active problem solving enables large language models (LLMs) to exhibit self-reflective capabilities such as error detection, critique, and strategic exploration of alternative solutions.

In the context of \Name{}, the agentic paradigm underpins the design of our multi-model feedback loop. Each participating model functions as an autonomous agent with a specialized role: \textit{solution generators} (e.g., GPT-4/5) synthesize candidate programs, while \textit{debugging critics} (e.g., Codestral-2508, Llama-3.3-70B, and DeepSeek-R1) analyze failures, propose refinements, and iteratively guide improvement. This coordination exemplifies agentic collaboration, where agents exchange structured reasoning traces and maintain contextual awareness across refinement cycles. 

By embedding this agentic feedback structure, \Name{} moves beyond conventional single-pass code generation toward an adaptive ecosystem of cooperative agents. Such architecture allows the system to self-diagnose weaknesses, recover from failure cases, and converge toward correct implementations with minimal human intervention. Consequently, our study contributes empirical evidence of how agentic AI principles, autonomy, self-reflection, and collaborative reasoning, enhance the reliability, interpretability, and efficiency of automated code generation systems.

\noindent
\textbf{\Name{} as an Agentic AI.} Our system exemplifies the core characteristics of agentic AI through four key mechanisms. First, \textit{autonomous goal-directed behavior}: the solution generator independently pursues the objective of producing a correct implementation without human intervention, making strategic decisions about algorithmic approaches, data structures, and implementation details across multiple refinement cycles. Second, \textit{environmental interaction and perception}: the system actively engages with its environment by submitting solutions to the Codeforces online judge, processing structured feedback (verdicts, test case failures, time/memory consumption), and extracting actionable information from execution traces. Third, \textit{self-correction through iterative reasoning}: upon receiving failure signals, the solution generator does not merely regenerate random alternatives but systematically incorporates debugging feedback to diagnose root causes, avoid previously attempted approaches, and converge toward correctness, demonstrating genuine adaptive learning rather than stochastic sampling. Fourth, \textit{persistent internal state and memory}: both the solution generator and debugging critics maintain conversation histories spanning all attempts for a given problem, enabling them to recognize patterns in failures, avoid repeating errors, and build progressively refined mental models of the problem constraints. This persistent context transforms isolated generation attempts into a coherent problem-solving trajectory, where each iteration builds upon accumulated insights from prior failures. Unlike traditional LLM code generation systems that treat each query independently, \Name{}'s agents exhibit episodic memory, strategic planning, and error recovery, hallmarks of genuine autonomous agency rather than passive prompt-response behavior.

Current LLM evaluation methodologies predominantly employ one-shot benchmarks (e.g., HumanEval, MBPP, LeetCode), which measure initial solution correctness but fail to capture iterative refinement capabilities, a fundamental aspect of human software engineering practice. In professional contexts, developers continuously test, debug, and revise their implementations based on feedback from compilers, test suites, and code review. Competitive programming contests provide a controlled environment to study this iterative development process, offering immediate, objective feedback through online judges.

\subsection{Preliminary Study}
Our prior study, \textit{LLM-ProS}\cite{LLM-ProS}, introduced a reproducible benchmark for evaluating LLM reasoning on ICPC World Finals problems (2011-2024).
That work compared five single-model architectures (GPT-4o, Mistral-Large, Llama-3.1-405B, and OpenAI’s o1-mini and o1-preview) under identical evaluation settings.
The key findings are:
\begin{itemize}
    \item \textbf{o1-family models} outperformed general-purpose models on correctness and robustness, largely due to explicit chain-of-thought (CoT) fine-tuning.
    \item \textbf{Dataset contamination and training style} were primary factors behind performance variance.
    \item \textbf{Verdict analysis} (Accepted, Wrong Answer, Runtime Error, etc.) highlighted reasoning limitations in geometry and greedy categories.
\end{itemize}

However, \textit{LLM-ProS} focused solely on \emph{zero-shot evaluation}.
Each model generated one solution per problem without feedback, planning, or iterative correction.
No cross-model collaboration, self-refinement, or trust calibration metrics were considered.
\section{\Name{} Methodology}
We introduce an autonomous AI agent, \Name{}, that leverages specialized language models~\cite{Liu2024Large,Wang2025Agentic,Putta2024Agent,Huang2023AgentCoder,Nguyen2024AgileCoder,Islam2024MapCoder} in complementary roles to solve competitive programming problems through iterative refinement. Our approach separates solution synthesis from failure diagnosis, enabling a systematic evaluation of how different model specializations influence debugging effectiveness.

We standardize all programming problems using a unified preprocessing pipeline that extracts problem statements, I/O constraints, and example test cases directly from Codeforces Gym endpoints~\cite{CodeforcesGym}. The system automatically compiles each generated solution and executes it against hidden test cases while programmatically logging verdicts (Accepted, Wrong Answer, Time Limit Exceeded, Runtime Error).

We evaluate performance across multiple dimensions, including Itr$_k$ ($k \in \{0,1,2,3\}$) cumulative acceptance rates, runtime and memory efficiency, and fault recovery across iterative attempts. Each configuration operates under identical decoding parameters and iteration budgets to ensure fair comparison. This experimental setup quantitatively compares model specializations, Llama~\cite{Grattafiori2024llama}, DeepSeek-R1~\cite{DeepSeek2024Deepseek}, and Codestral~\cite{Choi2024Linq} in terms of acceptance improvement, efficiency, and trust calibration (i.e., Expected Calibration Error between the critic's expressed confidence and the next-attempt success rate; see Section~\ref{sec:trust_calibration})~\cite{naeini2015obtaining} throughout iterative feedback loops. 
\begin{figure*}[!h]
\centering
\includegraphics[scale = 0.26]{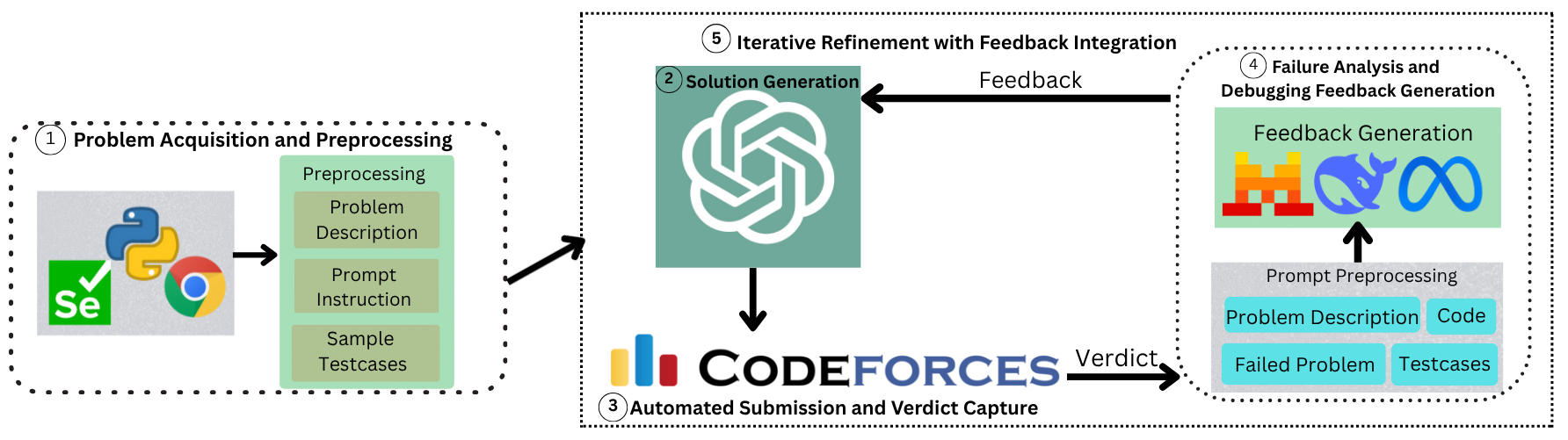}
\caption{Overview of the \Name{} workflow.} 
\label{fig:workflow}
\end{figure*}

Figure~\ref{fig:workflow} presents an overview of \Name{}. The agent operates as a closed-loop feedback system with five sequential stages: (1) problem acquisition and preprocessing, (2) initial solution generation, (3) automated submission and verdict capture, (4) failure analysis and feedback generation, and (5) iterative refinement of improved solutions. Each problem allows one initial submission plus up to three refinement iterations (Itr$_0$ through Itr$_3$), for a maximum of four total submission attempts, and the loop terminates once the system achieves acceptance or exhausts this budget.

\Name{} functions both as a solution generator and a debugging critic. As a generator, it produces candidate C++ implementations; as a critic, it analyzes execution failures and provides targeted feedback for improvement. The agent preserves conversational context across iterations, enabling it to accumulate problem-specific knowledge throughout the refinement process.
\begin{table*}[t]
\centering
\small
\scalebox{0.73}{
\begin{tabular}{p{3.2cm}p{11cm}}
\toprule
\textbf{Component} & \textbf{Content} \\
\midrule
\textbf{Problem Statement} & 
You are given an array $a$ of $n$ integers, where all elements except for at most one are equal to $-1$ or $1$. 
The remaining element $x$ satisfies $-10^9 \leq x \leq 10^9$. 
Find all possible sums of subarrays of $a$, including the empty subarray (whose sum is defined as $0$). 
In other words, find all integers $s$ such that the array $a$ has at least one contiguous subarray (possibly empty) with sum equal to $s$. 
Output these sums in ascending order. Each sum should be printed only once, even if it can be obtained by multiple subarrays. \\
\midrule
\textbf{Input Specification} & 
Each test contains multiple test cases. First line: $t$ ($1 \leq t \leq 10^4$) — the number of test cases.  
Each test case consists of two lines: the first line contains $n$, followed by an array $a_1, a_2, \ldots, a_n$.  
The sum of $n$ over all test cases does not exceed $2 \times 10^5$. \\
\midrule
\textbf{Output Specification} & 
For each test case, output all distinct possible subarray sums in ascending order, separated by spaces.  
Each sum should appear only once, even if it can be produced by multiple subarrays. \\
\midrule
\textbf{Sample Tests} & 
\textbf{Input:} 5 test cases with arrays containing mostly $-1$ and $1$, and one arbitrary element $x$.  
\textbf{Output:} Distinct sums such as $-1, 0, 1, 2, 9, 10, 11, 12$, varying across test cases. \\
\midrule
\textbf{Tags} & binary search, brute force, data structures, dp, greedy, math \\
\midrule
\textbf{Rating} & 1600 \\
\bottomrule
\end{tabular} }
\caption{Problem Data Structure (Example: Problem 2043-C from Codeforces)}
\label{tab:problem_structure}
\end{table*}
\subsection{Problem Acquisition and Preprocessing}
Each problem retrieved from Codeforces~\cite{CodeforcesGym} is processed to extract and structure its specification. We employ automated web scraping to systematically collect problems from Codeforces, applying data processing techniques to preserve mathematical notation (converting embedded LaTeX expressions to Unicode or MathML) and ensure consistent formatting across all problems. During preprocessing, inline LaTeX expressions~\cite{latex2mathml} such as variable subscripts and equations are converted into Unicode or MathJax~\cite{MathJax} compatible formats to maintain mathematical accuracy while improving readability.

\begin{promptbox}{Prompt 1: Solution Generation}
{
\scriptsize
\textit{You are a competitive programming expert. Your goal is to generate a correct and efficient solution to the given problem.}
\newline
<<Insert all problem-specific data from Table \ref{tab:problem_structure}.>> 
\newline
\textbf{Task}: Please provide a well-structured and optimized C++ solution. The solution should read from standard input and write to standard output. Include brief comments or explanations for clarity.
}
\end{promptbox}\label{prompt1}

Indentation, newline spacing, and punctuation are standardized to support consistent tokenization and enhance prompt clarity. HTML tags~\cite{BeautifulSoup4}, advertisements, and other non-essential metadata are removed, while problem constraints and variable notations are normalized for cross-problem consistency. Additionally, public sample test cases and expected outputs are reformatted into uniform input/output code blocks and released in a companion repository~\cite{CodeforcesTestCasesRepo} to facilitate automated parsing during evaluation. Table~\ref{tab:problem_structure} illustrates the resulting structure using Problem 2043-C as a representative example. All problems in our dataset follow this standardized format, enabling consistent prompt construction for both solution generation and feedback generation.

\subsection{Solution Generation}
\Name{} first produces solutions for the given problems. We evaluate two state-of-the-art language models as solution generators: i) \textbf{GPT-4 (OpenAI):} A large-scale transformer model with demonstrated proficiency in algorithmic reasoning and code synthesis tasks~\cite{OpenAI2023Gpt,Chen2021Evaluating}. This serves as our baseline solution generator and ii) \textbf{GPT-5 (OpenAI):} A next-generation language model with enhanced reasoning capabilities, improved code understanding, and better error recovery compared to GPT-4~\cite{Leon2026GPT}. We evaluate whether advanced model capabilities translate to higher initial success rates and more effective feedback incorporation.

When a problem is first presented to the solution generator (either GPT-4 or GPT-5), the system constructs a prompt consisting of two components: a system message establishing behavioral constraints shown in the first line of Prompt 1, and a user message containing the complete problem specification (Task part of Prompt 1). The exact prompt structure used in our experiments is shown in Prompt 1.

The solution generator (GPT-4 or GPT-5) processes this prompt and generates a C++ implementation. The generated response undergoes automated post-processing to extract the clean solution code: we remove markdown formatting artifacts (e.g., \texttt{cpp} delimiters), strip explanatory text that may appear before or after the code, and isolate only the executable C++ program. The cleaned solution is then saved with metadata headers including timestamp, generating model name, debugging critic name, and iteration number.

\begin{figure*}[!h]
\centering
\begin{minipage}[b]{0.48\textwidth}
    \centering
    \subfloat[Itr$_0$ (Initial Code Generation)]
    {
    \includegraphics[scale=0.35]{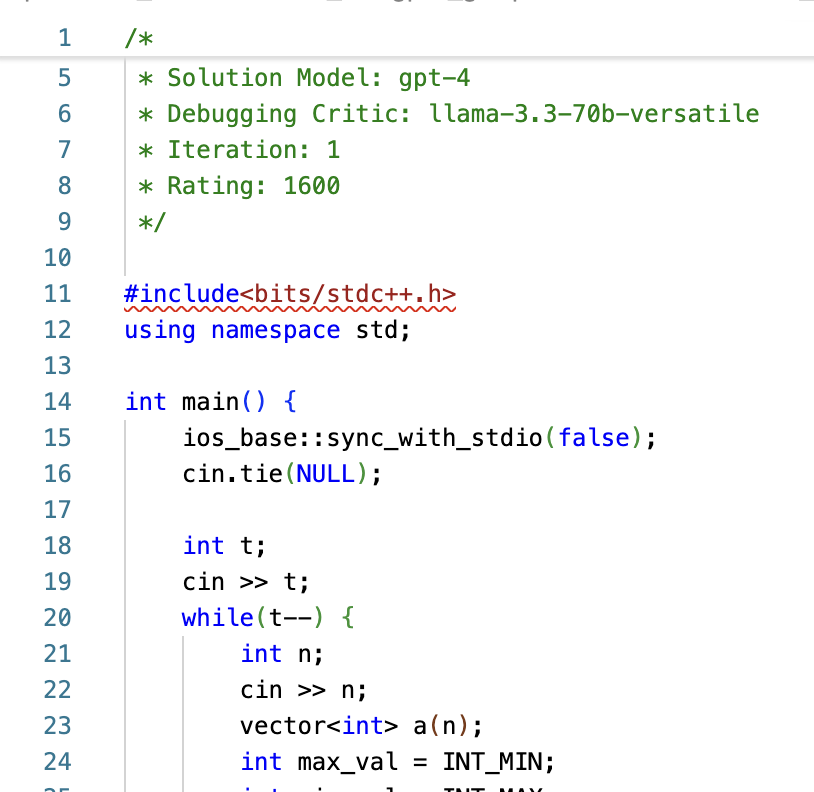}
    \label{sub1}}
\end{minipage}\hfill
\begin{minipage}[b]{0.48\textwidth}
    \centering
    \subfloat[Itr$_2$ (After 2 Feedback Iterations)]{
    \includegraphics[scale =0.35]{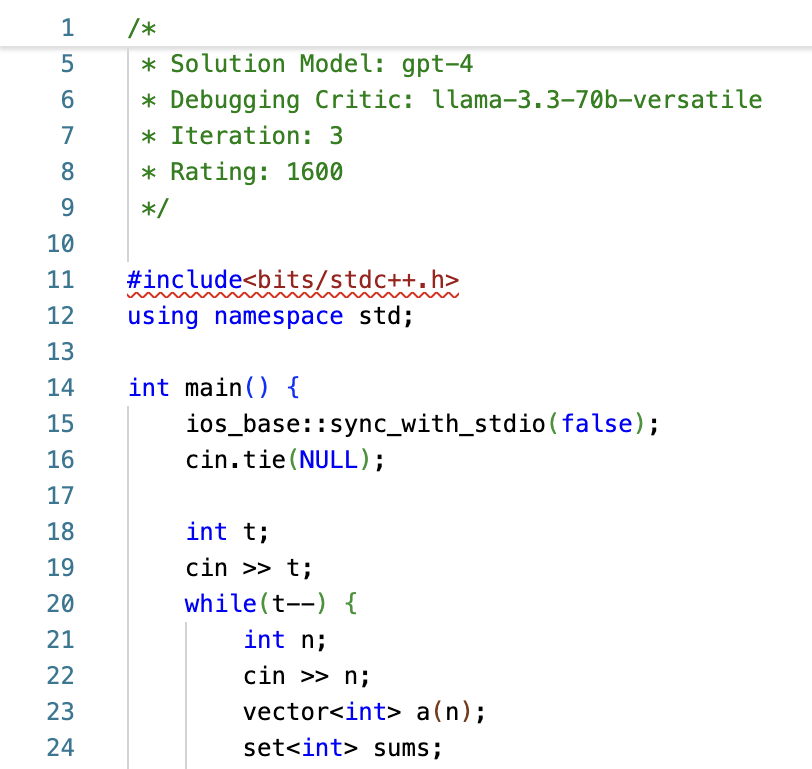}
    \label{sub2}}
\end{minipage}
\caption{Raw LLM outputs during Itr$_0$ (zero-shot) and Itr$_2$ (after 2 feedback iterations) of solution generation.}
\label{fig:solution_attempts}
\end{figure*}

Figure~\ref{fig:solution_attempts} illustrates two raw outputs generated by the solution generator (\textit{GPT-4}) during Itr$_0$ (initial zero-shot attempt) and Itr$_2$ (after two feedback iterations).
Each raw response includes system generated metadata, such as the attempt number, timestamp, model identifier, and context flag followed by the complete C++ implementation.
Figure~\ref{sub1} presents Itr$_0$, which reflects the model's initial zero-shot understanding of the problem, whereas Figure~\ref{sub2} shows Itr$_2$, representing a refined solution produced after two feedback iterations.
Our framework automatically parses and processes these raw responses to extract the executable code, retain metadata for traceability, and support reproducible evaluation across iterations.

\subsection{Automated Submission and Verdict Capture}

The cleaned solution is submitted to Codeforces using automated browser interactions. Problem scraping is implemented with Selenium~\cite{SeleniumHQ}, while live submission and verdict capture are implemented with Playwright~\cite{Playwright}. For each run, the submission worker attaches to an authenticated browser profile, navigates to the submission page, selects C++17, pastes the extracted solution, submits the form, and polls the status page until a terminal verdict is returned. To reduce fragility, the automation uses bounded waits, selector fallbacks, periodic page reloads during polling, and two complementary extraction paths for failed submissions: (i) interception of Codeforces' asynchronous submission-detail responses and (ii) fallback parsing of the verdict popup rendered in the browser. On timeout or transient UI failure, the system preserves the partial artifact bundle rather than discarding the run, including submission ID, timestamps, verdict text, runtime, memory, and any captured test diagnostics. We use authenticated browser sessions to handle submission workflows and capture detailed verdict information. For each submission, the system captures:\\
\textbf{Official Verdict:} The judge's decision: Accepted, Wrong Answer, Runtime Error, Time Limit Exceeded, Memory Limit Exceeded, or Compilation Error~\cite{CodeforcesVerdict,ICPC, CodeforcesGym}.\\
\textbf{Performance Metrics:} Execution time and memory usage as reported by the judge.\\
\textbf{Detailed Test Results:} For failed submissions, the system records all diagnostic information made available by the platform. When the authenticated account has permission to view hidden-test diagnostics, this includes the failing test input, expected output, actual output, and any error message returned by the judge. When such details are not exposed, the system records only the available verdict-level information, such as the failure type and failing test number. For the Codeforces Gym/ICPC portion, these details were retrieved using an authenticated Codeforces account with rating $\geq$ 1900, which exposes hidden-test diagnostics on the platform; for standard Codeforces problemset runs, we record only the failure details that the platform returns to the authenticated session.

To improve reproducibility beyond live, account-dependent execution, we expose two artifact layers. First, a companion repository publishes the public test cases and expected outputs used in preprocessing and parser validation~\cite{CodeforcesTestCasesRepo}. Second, the main project repository archives per-attempt workflow artifacts, including prompts, generated code, verdict traces, and parsed failure details~\cite{sifatGitHub}. These archived artifacts support a replay workflow in which independent researchers can inspect the exact per-attempt evidence underlying the reported analyses without resubmitting to Codeforces. We do not redistribute hidden judge tests, which remain controlled by the platform.

\begin{table}[ht]
\centering
\begin{minipage}[t]{0.5\textwidth}
\scalebox{0.8}{
\begin{tabular}{lcc}
\toprule
\textbf{Workflow} & \textbf{Verdict} & \textbf{Time (ms)} \\
\midrule
GPT-5 + DeepSeek-R1 & \textcolor{green!50!black}{\textbf{Accepted}} & 140 \\
GPT-4 + DeepSeek-R1 & \textcolor{red}{\textbf{Wrong Answer on Test 1}} & 46 \\
GPT-4 + Llama-3.3-70B & \textcolor{red}{\textbf{Wrong Answer on Test 1}} & 31 \\
\bottomrule
\end{tabular}}
\caption{First attempt verdicts without feedback (Problem 2043-C, Codeforces).}
\label{tab:verdict_capture}
\end{minipage}
\begin{minipage}[t]{0.45\textwidth}
\centering
\scalebox{0.65}{
\begin{tabular}{llcc}
\toprule
\textbf{Submission} & \textbf{Workflow} & \textbf{Verdict} & \textbf{Time (ms)} \\
\midrule
\multirow{3}{*}{Itr$_0$}  
  & GPT-5 + DeepSeek-R1 & \textcolor{green!50!black}{\textbf{Accepted}} & 140 \\
  & GPT-4 + DeepSeek-R1 & \textcolor{red}{\textbf{Wrong Answer on Test 1}} & 46 \\
  & GPT-4 + Llama-3.3-70B & \textcolor{red}{\textbf{Wrong Answer on Test 1}} & 31 \\
\midrule
\multirow{2}{*}{Itr$_1$} 
  & GPT-4 + DeepSeek-R1 & \textcolor{red}{\textbf{Time Limit Exceeded on Test 6}} & 1000 \\
  & GPT-4 + Llama-3.3-70B & \textcolor{red}{\textbf{Time Limit Exceeded on Test 6}} & 1000 \\
\midrule
\multirow{2}{*}{Itr$_2$}  
  & GPT-4 + DeepSeek-R1 & \textcolor{red}{\textbf{Wrong Answer on Test 8}} & 62 \\
  & GPT-4 + Llama-3.3-70B & \textcolor{red}{\textbf{Time Limit Exceeded on Test 7}} & 1000 \\
\midrule
\multirow{2}{*}{Itr$_3$}  
  & GPT-4 + DeepSeek-R1 & \textcolor{green!50!black}{\textbf{Accepted}} & 140 \\
  & GPT-4 + Llama-3.3-70B & \textcolor{red}{\textbf{Time Limit Exceeded on Test 8}} & 1000 \\
\bottomrule
\end{tabular}}
\caption{Iterative verdict progression for Problem 2043-C (Codeforces, rating 1600).}
\label{tab:feedback_results_dual}
\end{minipage}
\vspace{-3ex}
\end{table}
\ignore{
\begin{table}[ht]
\centering
\scalebox{0.65}{
\begin{tabular}{llcc}
\toprule
\textbf{Submission} & \textbf{Workflow} & \textbf{Verdict} & \textbf{T.(ms)} \\
\midrule
\multirow{3}{*}{Itr$_0$}  
  & GPT-5 + DeepSeek-R1 & \textcolor{green!50!black}{\textbf{Accepted}} & 140 \\
  & GPT-4 + DeepSeek-R1 & \textcolor{red}{\textbf{Wrong Answer on Test 1}} & 46 \\
  & GPT-4 + Llama-3.3-70B & \textcolor{red}{\textbf{Wrong Answer on Test 1}} & 31 \\
\midrule
\multirow{2}{*}{Itr$_1$} 
  & GPT-4 + DeepSeek-R1 & \textcolor{red}{\textbf{Time Limit Exceeded on Test 6}} & 1000 \\
  & GPT-4 + Llama-3.3-70B & \textcolor{red}{\textbf{Time Limit Exceeded on Test 6}} & 1000 \\
\midrule
\multirow{2}{*}{Itr$_2$}  
  & GPT-4 + DeepSeek-R1 & \textcolor{red}{\textbf{Wrong Answer on Test 8}} & 62 \\
  & GPT-4 + Llama-3.3-70B & \textcolor{red}{\textbf{Time Limit Exceeded on Test 7}} & 1000 \\
\midrule
\multirow{2}{*}{Itr$_3$}  
  & GPT-4 + DeepSeek-R1 & \textcolor{green!50!black}{\textbf{Accepted}} & 140 \\
  & GPT-4 + Llama-3.3-70B & \textcolor{red}{\textbf{Time Limit Exceeded on Test 8}} & 1000 \\
\bottomrule
\end{tabular}}
\caption{Iterative verdict progression for Problem 2043-C (Codeforces, rating 1600).}
\label{tab:feedback_results_dual}
\end{table}
}
Table~\ref{tab:verdict_capture} illustrates examples of captured verdicts from the Codeforces judging system during the \textit{first attempt} of submission-before any feedback or refinement iteration was applied. It shows both performance and diagnostic outcomes for three different workflows solving Problem 2043-C ("Sums on Segments", rating 1600).

Table~\ref{tab:feedback_results_dual} demonstrates how GPT-4 workflows with different debugging critics evolve across multiple feedback iterations, contrasted with GPT-5's immediate success. For Problem 2043-C, GPT-5 + DeepSeek-R1 solved the problem on the first attempt (Itr$_0$), while GPT-4 workflows required multiple refinement cycles. The table shows verdict progression for GPT-4 paired with DeepSeek-R1 and Llama-3.3-70B, illustrating how different critic models guide the iterative correction process.

This example highlights several critical insights: First, it demonstrates the value of reasoning-focused debugging critics-DeepSeek-R1 successfully guided GPT-4 from logical errors (WA) through efficiency issues (TLE) to final acceptance, while Llama-3.3-70B's implementation-focused feedback led to persistent TLE errors. Second, it shows that iterative refinement can address algorithmic complexity when paired with appropriate critics: GPT-4 + DeepSeek transitioned from an O(n²) brute-force approach (causing TLE) to an optimized O(n) solution by Itr$_3$. Third, the progression from Wrong Answer to Time Limit Exceeded, back to Wrong Answer, and finally to Accepted shows that refinement is not always linear. An intermediate revision may fix one problem while introducing another, such as replacing a logical error with an efficiency issue, before eventually converging to an accepted solution. Finally, GPT-5's immediate success (Itr$_0$) underscores the importance of strong initial solution generation, though the GPT-4 + DeepSeek trajectory demonstrates that iterative refinement with effective critics can compensate for weaker initial capabilities. All verdict data from each iteration were systematically recorded for subsequent feedback generation and performance analysis.

\begin{promptbox}{Prompt 2: Feedback Generation}
{
\scriptsize
\textit{You are a code debugging assistant specializing in competitive programming. Identify logical errors, edge
cases, or inefficiencies. Provide concise, targeted feedback. Do not provide a full revised code; only outline how
to fix the existing solution.}
\newline
\textbf{Problem:} <<Insert all problem-specific data from Table \ref{tab:problem_structure}.>>
\newline
 \textbf{Current Code:} Insert the failing C++ solution code
\newline
\textbf{Verdict:} Wrong Answer on test 2
\newline
\textbf{Test Case Details:}\\
\textbf{Input:} Specific test input that caused failure\\
\textbf{Expected Output:} Correct output for this test\\
\textbf{Given Output:} Incorrect output produced by the solution
\newline
\textbf{Task}: Analyze the code to find potential bugs, logical errors, or missed edge cases. Suggest specific improvements or fixes. Prioritize algorithmic or implementation details that address the failing test case. Do not rewrite the entire code; give actionable feedback or partial snippets only.
\newline
\textit{After your debugging analysis, on a new line report your confidence that the next solution attempt will be accepted, as an integer from 1 (very unlikely) to 5 (very likely), in the format: \textbf{Confidence: [1--5]}.}
}
\end{promptbox}

\subsection{Failure Analysis and Debugging Feedback Generation}
\label{sec:feedback generation}

If the verdict is not Accepted, \Name{} analyzes the failure and generates targeted hints for refinement. To evaluate how model specialization affects feedback quality, we systematically compare three models in the debugging critic role:\\
i)  \textbf{Codestral-2508 (Mistral AI):} A code-specialized model explicitly trained for software engineering tasks including bug detection, code analysis, and technical documentation. We hypothesize this model excels at identifying implementation-level defects such as off-by-one errors, boundary condition violations, and language-specific pitfalls~\cite{Choi2024Linq}. \\
ii) \textbf{Llama-3.3-70B-Versatile (Meta AI):} A general-purpose foundation model with 70 billion parameters, providing broad problem-solving capabilities without domain-specific fine-tuning. This model serves as a baseline for general error diagnosis~\cite{Grattafiori2024llama}.

\noindent
iii) \textbf{DeepSeek-R1-0528 (DeepSeek AI):} A reasoning-augmented model with enhanced chain-of-thought capabilities. We expect this model to provide superior analysis of algorithmic correctness issues, optimization opportunities, and high-level design flaws~\cite{DeepSeek2024Deepseek}.

The structure of the feedback generation prompt is illustrated in Prompt 2. This prompt emphasizes diagnostic analysis rather than a full code rewrite. The first two lines outline the responsibilities assigned to the LLM as an agent. The subsequent problem-specific fields vary for each problem, while the task section defines the specific objective for the agent.

\begin{figure}[htb]
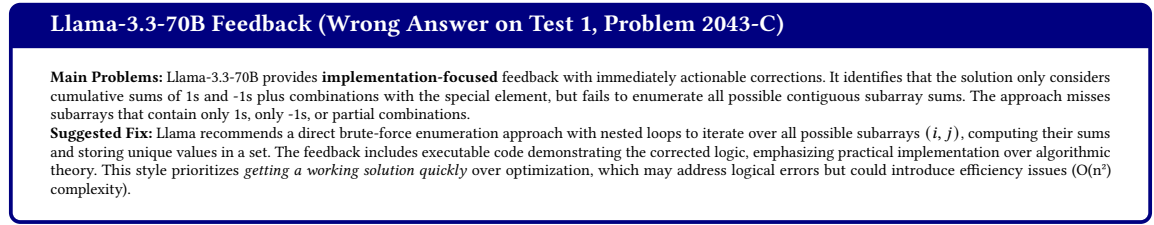

\begin{tcolorbox}[colback=white,colframe=blue!50!black,title=\textbf{Llama-3.3-70B Feedback (Wrong Answer on Test 1, Problem 2043-C)}]
\scriptsize
\textbf{Main Problems:} Llama-3.3-70B provides \textbf{implementation-focused} feedback with immediately actionable corrections. 
It identifies that the solution only considers cumulative sums of 1s and -1s plus combinations with the special element, but fails to enumerate all possible contiguous subarray sums. 
The approach misses subarrays that contain only 1s, only -1s, or partial combinations.

\textbf{Suggested Fix:} Llama recommends a direct brute-force enumeration approach with nested loops to iterate over all possible subarrays $(i, j)$, computing their sums and storing unique values in a set. 
The feedback includes executable code demonstrating the corrected logic, emphasizing practical implementation over algorithmic theory. 
This style prioritizes \textit{getting a working solution quickly} over optimization, which may address logical errors but could introduce efficiency issues (O(n²) complexity).
\end{tcolorbox}
\caption{Feedback from Llama-3.3-70B after first failed attempt (Problem 2043-C)}
\label{fig:samplefeedback_llama}
\end{figure}
The debugging critics process each prompt and generate diagnostic feedback tailored to their respective specializations. Figure~\ref{fig:samplefeedback_llama} is an example of generated feedback from Llama-3.3-70B for Problem 2043-C after the first failed attempt (Wrong Answer on Test 1). In general we find that, Llama provides implementation focused and actionable corrections; DeepSeek performs reasoning-driven algorithmic analysis that explores alternative solution strategies; and Codestral emphasizes syntactic precision and code-level optimizations.
Despite these stylistic differences, all three critics consistently identify the core algorithmic flaw, incomplete subarray enumeration for this problem, while expressing their solutions through distinct reasoning patterns.

Each LLM model (DeepSeek-R1, Llama-3.3, and Codestral-2508) generates independent feedback perspectives. 
We then run three distinct refinement pipelines, each incorporating one model’s feedback to produce a revised code solution. This multi-perspective setup enables the system to evaluate how different feedback styles, reasoning-driven, general-purpose, and implementation-focused affect the correction process and overall convergence. All feedback outputs and corresponding solution updates are stored in the solving log for subsequent iteration analysis.

\subsection{Iterative Refinement with Feedback Integration}
For subsequent attempts (attempts 2 through 4, corresponding to refinement rounds Itr$_1$ through Itr$_3$), the solution generator (GPT-4 or GPT-5) receives an enriched prompt that includes the full failure context and debugging hints from the critic. Prompt 3 shows the exact prompt we use for this step.

Critically, both the solution generator (GPT-4 or GPT-5) and debugging critic maintain \textit{persistent conversation contexts} throughout this process. Each new message is appended to the agent's conversation history, enabling accumulation of problem-specific knowledge. For example, if the second attempt also fails (with a different error), the third attempt's prompt would include the complete history of both previous failures, both sets of debugging feedback, and both failed solutions. The refined solution is submitted through the same automated pipeline (Stage 2). If the verdict is Accepted, the system terminates and records success. If the verdict indicates continued failure, the loop repeats up to a maximum of four total attempts (one initial attempt plus three refinements). Upon termination, comprehensive logs are generated including per-attempt statistics, verdict distributions, and complete conversation histories for both agents.

\begin{promptbox}{Prompt 3: Expert Solution}
{
\scriptsize
\textit{You are a competitive programming expert. Your goal is to generate a correct and efficient solution to the
given problem.}
\newline
\textbf{Problem:} <<Insert all problem-specific data from Table \ref{tab:problem_structure}.>>
\newline
 \textbf{Previous Solution:} <<Insert the failing C++ solution code>>
\newline
\textbf{Verdict:} Wrong Answer on test 2
\newline
\textbf{Test Case Details:}\\
\textbf{Input:} Specific test input that caused failure\\
\textbf{Expected Output:} Correct output for this test\\
\textbf{Given Output:} Incorrect output produced by the solution
\newline
\textbf{Feedback from Debugging Critic:} Targeted hints generated by Codestral/Llama/DeepSeek [after Prompt 2]
\newline
\textbf{Task}: Think about the feedback carefully. Please provide a well-structured and optimized C++ solution that addresses the identified issues. The solution should read from standard input and write to standard output. Include brief comments or explanations for clarity.
}
\end{promptbox}\label{prompt2}

\noindent\textbf{Context Management Specification.}
Each LLM provider maintains a \texttt{ChatContext} object keyed by session ID. The context stores the ordered sequence of all \texttt{(role, content)} message pairs: an initial system message establishing the agent's role and behavioral constraints, followed by alternating user messages (containing the problem specification, failure details, and any debugging hints from the critic) and assistant messages (containing the generated C++ solution or diagnostic feedback). This full sequence is passed to the API at every call, so the model has access to its complete interaction history for the current problem.

\noindent The \textbf{stateless ablation} (Section~\ref{sec:ablation_design}) introduces a \texttt{stateless\_mode} flag: when enabled, the \texttt{reset\_contexts()} method clears all provider contexts immediately before each feedback iteration ($k \geq 1$). The failure verdict and critic hint are still embedded as text in the current user message, but no prior conversational turns are passed to the model-isolating accumulated memory as the sole variable. This design ensures any performance difference between stateful and stateless conditions is attributable exclusively to conversation history, not to information content.

\section{Evaluation Setup}

We conduct experiments designed to evaluate the effectiveness of our proposed \textit{agentic multi-model feedback framework} under controlled and reproducible conditions. 
Our evaluation spans 367 programming problems, five large language models (LLMs) of distinct capabilities, and multiple feedback configurations. 
We aim to empirically answer the following research questions:
\begin{itemize}
    \item \textbf{RQ1 (Iterative Improvement):} How much do Itr$_k$ (Iteration $k$) acceptance rates improve when LLMs can iteratively refine solutions based on judge feedback, compared to zero-shot (Iteration zero: Itr$_0$) performance?
    \item \textbf{RQ2 (Model Specialization):} Do code-specialized critics (Codestral), general-purpose critics (Llama-3.3), or reasoning-focused critics (DeepSeek-R1) provide more effective debugging feedback when paired with the solution generators?
    \item \textbf{RQ3 (Persistent Context):} Does maintaining conversation history across attempts improve solution quality and reduce repeated mistakes compared to stateless refinement? We address this through a direct controlled ablation study on 47 randomly sampled problems, isolating persistent context as the sole experimental variable (results in Section~\ref{sec:rq3_ablation}).
    \item \textbf{RQ4 (Algorithmic Category Effectiveness):} Do code-specialized, general-purpose, and reasoning-focused critics exhibit differential performance across problem categories, revealing category specific debugging strengths?
    \item \textbf{RQ5 (Baseline Comparison):} How does A-ProS compare with simpler agent-loop designs, such as zero-shot, single-round stateless, and multi-round stateless baselines, under the same generator-critic pairings and compute budgets? Does persistent context provide improvements beyond what repeated sampling alone can explain?
\end{itemize}

\begin{table}[h]
\centering
\small
\renewcommand{\arraystretch}{1.2}
\begin{minipage}[t]{0.48\textwidth}
\centering
\scalebox{0.82}{
\begin{tabular}{lcc}
\toprule
\textbf{Algorithmic Category} & \textbf{Problem Count} & \textbf{Percentage} \\
\midrule
Graph Theory & 40 & 24.0\% \\
Geometry & 36 & 21.6\% \\
Dynamic Programming & 27 & 16.2\% \\
Math & 24 & 14.4\% \\
Implementation & 20 & 12.0\% \\
Greedy & 14 & 8.4\% \\
Ad-hoc & 11 & 6.6\% \\
Number Theory & 7 & 4.2\% \\
String & 7 & 4.2\% \\
Max Flow & 6 & 3.6\% \\
Probability & 6 & 3.6\% \\
Data Structures & 6 & 3.6\% \\
Combinatorics & 4 & 2.4\% \\
Binary Search & 8 & 4.8\% \\
Interactive & 4 & 2.4\% \\
\bottomrule
\end{tabular}}
\caption{Algorithmic topic distributions in ICPC World Finals problems. Problems may carry multiple tags; counts therefore sum to more than 167.}
\label{tab:ICPC Algo}
\end{minipage}
\hfill
\begin{minipage}[t]{0.48\textwidth}
\centering
\scalebox{.82}{
\begin{tabular}{lcc}
\toprule
\textbf{Algorithmic Tag} & \textbf{Problem Count} & \textbf{Percentage} \\
\midrule
Greedy & 117 & 58.5\% \\
Math & 74 & 37.0\% \\
Constructive Algorithms & 59 & 29.5\% \\
Brute Force & 58 & 29.0\% \\
Implementation & 46 & 23.0\% \\
Dynamic Programming & 45 & 22.5\% \\
Binary Search & 37 & 18.5\% \\
Data Structures & 34 & 17.0\% \\
Sortings & 34 & 17.0\% \\
Bitmasks & 23 & 11.5\% \\
Number Theory & 21 & 10.5\% \\
Graphs & 21 & 10.5\% \\
DFS and Similar & 20 & 10.0\% \\
Two Pointers & 17 & 8.5\% \\
Combinatorics & 14 & 7.0\% \\
\bottomrule
\end{tabular}}
\caption{Algorithmic topic distributions in Codeforces problems. Tags are non-mutually-exclusive; percentages therefore sum to more than 100\%.}
\label{tab:codeforces algo}
\end{minipage}
\vspace{-5ex}
\end{table}

\subsection{Benchmarks}

Our benchmark comprises problems from the following two complementary sources. \\

\noindent
\textbf{ICPC World Finals Problems (167 problems, 2011-2024).} We collected 167 problems from 14 years of ICPC World Finals competitions, representing the highest tier of competitive programming difficulty. These problems are hosted on Codeforces Gym, which maintains official ICPC problem archives with complete test suites and judging infrastructure. The ICPC subset covers diverse algorithmic domains including graph algorithms (e.g., network flows, shortest paths, minimum spanning trees), dynamic programming (e.g., state-space optimization, memoization), computational geometry (e.g., convex hulls, line intersection, polygon algorithms), string processing (e.g., pattern matching, suffix structures), number theory (e.g., modular arithmetic, primality testing), and combinatorial optimization (e.g., greedy algorithms, backtracking); the full category distribution is shown in Table~\ref{tab:ICPC Algo}. Problems in this subset often require combining multiple algorithmic techniques and exhibit high implementation complexity.

\noindent
\textbf{Recent Codeforces Contest Problems (200 problems).} To evaluate model performance on problems less likely to appear in training corpora, we selected 200 problems from recently conducted Codeforces contests. These problems have difficulty ratings in the 1200-1800 range, representing intermediate-level competitive programming challenges, with algorithmic tag distributions shown in Table~\ref{tab:codeforces algo}. This temporal restriction reduces the risk of training data leakage, as foundation models evaluated in this study were trained on data cutoffs predating these contests.

Each problem in our dataset follows the structure shown in Table~\ref{tab:problem_structure}, including: (1) a formal problem statement with narrative context, (2) precise input/output format specifications, (3) explicit constraints on input parameters, (4) sample test cases with expected outputs and explanations, (5) algorithmic topic tags, (6) difficulty rating, and (7) a comprehensive hidden test suite (typically 50-200 tests) used by the online judge for verification. Problems require contestants to implement correct, efficient solutions within strict time (typically 1-2 seconds) and memory (typically 256 MB) limits.

\subsection{Evaluation Protocol and Performance Metrics}
We systematically evaluate six workflow combinations in a $2 \times 3$ factorial design: two solution generators (GPT-4, GPT-5) paired with three debugging critics (Codestral, Llama-3.3, DeepSeek). Each workflow is evaluated on the complete 367-problem dataset. Our evaluation framework comprises the following five metric categories.
\subsubsection{Correctness Metrics}
\noindent\textbf{Terminology.} Throughout this paper we use \emph{attempt} to refer to a single judge submission (including the initial one), so there are four total submission attempts per problem (Itr$_0$ through Itr$_3$). We use \emph{feedback iteration} (or \emph{refinement iteration}) to refer to each refinement round that follows the initial attempt, i.e., iterations~1--3. 

\textbf{Itr$_k$ Acceptance Rates (Cumulative).} We denote Itr$_k$ as the cumulative proportion of problems successfully solved within $k$ feedback iterations, where $k \in \{0, 1, 2, 3\}$. Specifically, Itr$_k$ represents the fraction of problems for which at least one accepted solution is produced within $k+1$ total submission attempts.

\noindent
\textbf{Itr$_0$:} Zero-shot performance without any feedback, representing the solution generator's baseline capability.\\
\textbf{Itr$_1$:} Cumulative acceptance rate after one feedback-and-refinement iteration.\\
\textbf{Itr$_2$:} Cumulative acceptance rate after two feedback-and-refinement iterations.\\
\textbf{Itr$_3$:} Final cumulative acceptance rate after up to three refinement iterations.\\
\textbf{Improvement Gain:} The difference (Itr$_3 -$ Itr$_0$) quantifies the benefit of iterative refinement.

\textbf{Per-Iteration New-Solve Counts.} In Tables~\ref{tab:icpc_time_to_solution} and~\ref{tab:codeforces_time_to_solution}, the columns labeled Itr$_k$ report the \emph{incremental} number of new problems first solved at iteration $k$, not the cumulative total. Formally, $\Delta\mathrm{Itr}_0 = \mathrm{Itr}_0$ and $\Delta\mathrm{Itr}_k = \mathrm{Itr}_k - \mathrm{Itr}_{k-1}$ for $k \geq 1$. Cumulative acceptance at iteration $k$ is therefore recovered as $\sum_{j=0}^{k}\Delta\mathrm{Itr}_j$.

We report Itr$_k$ values separately for ICPC World Finals problems and Codeforces problems to compare performance across difficulty levels.

\noindent
\textbf{Average Number of Attempts.} For problems that are eventually solved, we compute the average number of attempts required:
\begin{equation}
\text{Avg Attempts} = \frac{\sum_{i=1}^{n} k_i}{n}
\end{equation}
where $k_i$ is the attempt number on which problem $i$ was solved (1, 2, 3, or 4), and $n$ is the total number of solved problems. Lower values indicate more efficient problem-solving with less debugging feedback required.

\subsubsection{Reliability and Calibration Metrics}
\label{sec:metrics}
\noindent\textbf{Expected Calibration Error (ECE).} For each debugging critic, we evaluate how accurately its confidence in a hint predicts whether that hint leads to a successful subsequent attempt. The critic reports a confidence score on a 1-5 integer scale for each hint; we normalize this to the range $[0.2, 1.0]$. We partition hints into five equal-width confidence bins and compute the Expected Calibration Error as the bin-weighted absolute difference between average confidence and actual success rate:
\begin{equation}
\mathrm{ECE} = \sum_{b} \frac{|B_b|}{N} \bigl|\mathrm{acc}(B_b) - \mathrm{conf}(B_b)\bigr|
\end{equation}
where $B_b$ is the set of hints in bin $b$, $N$ is the total number of hints evaluated, $\mathrm{acc}(B_b)$ is the fraction of hints in $B_b$ that led to an accepted solution on the immediately following attempt, and $\mathrm{conf}(B_b)$ is the mean normalized confidence in $B_b$. ECE~$= 0$ indicates perfect calibration, while higher values reflect systematic over- or under-confidence. Results are reported in Section~\ref{sec:trust_calibration}.

\noindent\textbf{Confidence Score Elicitation.} To obtain the 1--5 confidence score, the following instruction is appended to Prompt~2 (Section~\ref{sec:feedback generation}) after the main debugging task: \textit{``After your debugging analysis, on a new line report your confidence that the next solution attempt will be accepted, as an integer from 1 (very unlikely) to 5 (very likely), in the format: \textup{Confidence: [1--5]}.''}

The integer is extracted via regex from the critic's response and normalized to $[0.2,\,1.0]$ by dividing by 5, so that each bin maps to a $0.2$-wide interval.

\noindent
\textbf{Verification Cost.} We define verification cost as the total number of judge submissions made per accepted solution across all problems:
\begin{equation}
\text{Verification Cost} = \frac{\text{Total submissions across all problems}}{\text{Total accepted solutions}}
\end{equation}
Unlike Average Attempts (which is computed only over solved problems), Verification Cost accounts for wasted submissions on unsolved problems, providing a complete picture of judge resource consumption. Lower values indicate more efficient use of the feedback budget.

\noindent
\textbf{Abstention Rate.} We define the abstention rate as the fraction of problems for which the system fails to produce a syntactically valid, compilable submission on any attempt:
\begin{equation}
\text{Abstention Rate} = \frac{\text{\# problems with zero compilable submissions}}{\text{Total problems}}
\end{equation}
This metric captures \textit{generator-side reliability} at the code-generation stage, independent of logical correctness. A high abstention rate indicates the generator fails to produce structurally valid C++ code, reducing the effective problem pool for downstream evaluation. Because abstention is not critic-specific, we report it in the main results summary rather than in the critic-calibration table.
\subsubsection{Failure Analysis Metrics}
\textbf{Verdict Distribution.} We categorize all failed submissions by judge verdict type:\\
\begin{enumerate}
    \item \textbf{Wrong Answer (WA):} Incorrect output for at least one test case
    \item \textbf{Runtime Error (RE):} Program crash or undefined behavior
    \item \textbf{Time Limit Exceeded (TLE):} Algorithmic inefficiency causing timeout
    \item \textbf{Memory Limit Exceeded (MLE):} Excessive space consumption
    \item \textbf{Compilation Error (CE):} Syntactic or semantic code errors
\end{enumerate}
This distribution reveals common failure modes and enables comparison of debugging critic effectiveness. For example, if Codestral reduces Runtime Errors more than Llama-3.3, this suggests code-specialized critics better diagnose implementation bugs.\\
\textbf{Iterative Improvement Rates.} For problems that ultimately succeed, we track the refinement trajectory to understand when corrections occur.\\
\textbf{Error Repetition Rate.} For feedback iterations only ($k \geq 1$), we define error repetition as the fraction of refinement attempts whose verdict type matches the verdict type of the immediately preceding attempt for the same problem and workflow. Formally, if $v_{i,k}$ denotes the non-accepted verdict category for problem $i$ at feedback iteration $k$, the repetition rate is $\frac{1}{N}\sum \mathbf{1}[v_{i,k}=v_{i,k-1}]$ over all failed feedback iterations included in the analysis. Lower values indicate that the system is less likely to cycle through the same failure mode repeatedly.

\subsubsection{Problem Difficulty and Solvability Metrics} \textbf{Solvability Score.} To quantify problem difficulty across multiple dimensions, we define a composite solvability score:
\begin{equation}
\begin{split}
S(p) = & \left(\frac{\text{\# workflows that solved } p}{6}\right) \times 100 \\
       & - \left(\frac{\text{avg attempts for } p}{3}\right) \times 20 \\
       & + \left(\frac{\text{\# workflows that solved } p \text{ at Itr}_0}{6}\right) \times 30
\end{split}
\end{equation}
Higher values indicate easier problems. The formula weights three factors: (1) how many workflow combinations solved the problem (breadth of success), (2) how many attempts were typically required (efficiency penalty), and (3) how many workflow combinations solved it immediately at Itr$_0$ (inherent one-shot easiness). The theoretical range is approximately $[-26.7,\,123.3]$: the maximum of $\approx 123$ is achieved when all six workflows solve the problem on the first attempt, and the minimum of $\approx -27$ occurs when no workflow solves it and all exhaust all four submission slots. In practice, scores in our dataset fall within $[2.1,\,94.2]$. Problems with $S(p) > 80$ are classified as ``Very Easy,'' $60 < S(p) \leq 80$ as ``Easy,'' $40 < S(p) \leq 60$ as ``Medium,'' $20 < S(p) \leq 40$ as ``Hard,'' and $S(p) \leq 20$ as ``Very Hard.''

\noindent
\textbf{Estimated Codeforces Rating.} For each workflow combination, we estimate an equivalent competitive programming rating by fitting a logistic regression model:
\begin{equation}
P(\text{solve}|r) = \frac{1}{1 + e^{-\beta_0 - \beta_1 r}}
\end{equation}
where $r$ is the problem rating and $P(\text{solve}|r)$ is the Itr$_3$ success rate at that rating. We define the workflow's estimated rating $\hat{r}$ as the rating level where it achieves 40\% success (typical competitive programming performance): $P(\text{solve}|\hat{r}) = 0.40$.

\subsubsection{Statistical Significance Testing}
\textbf{Paired t-tests are not used in this study.} Because all primary outcomes are paired binary variables (solved/unsolved), we do not report or interpret paired t-test statistics. Instead, all reported hypothesis tests use McNemar's exact test, which is appropriate for matched dichotomous outcomes~\cite{agresti2002categorical}.

\noindent\textbf{McNemar's Test for Paired Binary Outcomes.}
For the controlled ablation study (RQ3) and pairwise workflow comparisons, where outcomes are dichotomous (solved/unsolved) on matched problem pairs, we employ McNemar's exact test as the primary significance test. Given $b$ problems solved by condition~$A$ only and $c$ problems solved by condition~$B$ only, the test statistic with continuity correction~\cite{edwards1948note} is:
\begin{equation}
\chi^2 = \frac{(|b - c| - 1)^2}{b + c}
\end{equation}
For small discordant counts ($b + c \leq 25$), we use the exact binomial $p$-value $p = 2 \cdot \min(P(X \geq b),\, P(X \leq b))$ where $X \sim \mathrm{Binomial}(b+c,\, 0.5)$. McNemar's test is the statistically appropriate choice for paired binary data, as paired t-tests assume continuous outcomes and can inflate Type~I error rates with dichotomous measurements~\cite{agresti2002categorical}.

\noindent\textbf{Bootstrap Confidence Intervals.}
We construct 95\% bootstrap confidence intervals for all pairwise solve-rate differences using 10{,}000 stratified resamples with replacement. For each resample, we compute the difference in solve rates $\hat{\delta}^* = \hat{p}_A^* - \hat{p}_B^*$ and report the 2.5th and 97.5th percentiles as the interval bounds. Bootstrap CIs are nonparametric, require no distributional assumptions, and remain valid for small to moderate sample sizes. A CI that excludes zero provides evidence of a real performance difference independent of asymptotic approximations.

\noindent\textbf{Holm-Bonferroni Multiplicity Correction.}
When performing pairwise comparisons across all six workflow combinations (6 pairs per iteration level), we apply Holm-Bonferroni step-down correction to control the family-wise error rate. Raw $p$-values are sorted in ascending order and adjusted as $p_{(i)}^{\mathrm{adj}} = \min(1,\, (m - i + 1) \cdot p_{(i)})$, where $m$ is the total number of comparisons. Adjusted $p < 0.05$ indicates significance after multiplicity correction.

\textbf{Effect size choice.} Because our outcomes are binary (solved/unsolved), we do not use Cohen's $d$ in this paper. All effect size comparisons instead use Cohen's $h$, which is the statistically appropriate measure for binary proportions~\cite{cohen1988statistical}.

\noindent\textbf{Cohen's $h$ for Proportions (Primary Effect Size Measure).}
For binary outcome comparisons (Itr$_k$ proportions), we report Cohen's $h$ as the primary effect size metric, appropriate for comparing two proportions~\cite{cohen1988statistical}:
\begin{equation}
h = 2\arcsin\!\sqrt{p_1} - 2\arcsin\!\sqrt{p_2}
\end{equation}
Interpretation follows standard conventions: $|h| < 0.2$ negligible, $0.2 \leq |h| < 0.5$ small, $0.5 \leq |h| < 0.8$ medium, $|h| \geq 0.8$ large.

\textbf{95\% Confidence Intervals.} For each Itr$_k$ metric, we compute binomial confidence intervals using the Wilson score method~\cite{wilson1927probable,brown2001interval}:
\begin{equation}
\text{CI} = \frac{\hat{p} + \frac{z^2}{2n} \pm z\sqrt{\frac{\hat{p}(1-\hat{p})}{n} + \frac{z^2}{4n^2}}}{1 + \frac{z^2}{n}}
\end{equation}
where $\hat{p}$ is the observed success rate, $n$ is the sample size, and $z = 1.96$ for 95\% confidence. Non-overlapping confidence intervals provide strong evidence of significant differences between workflows.

\subsubsection{Comparative Analysis Framework}

\textbf{Factorial Design Analysis.} Our $2 \times 3$ experimental design enables systematic comparison:\\
\textbf{Critic Effect:} Compare Codestral vs. Llama-3.3 vs. DeepSeek within each solution generator to isolate debugging specialization impact.\\
\textbf{Generator Effect:} Compare GPT-4 vs. GPT-5 within each debugging critic to assess solution generation capability differences.

\textbf{Efficiency Metrics.} For all accepted solutions, we extract execution time (milliseconds) and peak memory consumption (kilobytes) from judge verdicts, logged in the per-problem \texttt{solving\_log.json} artifact and available in the companion repository~\cite{sifatGitHub}.

This comprehensive evaluation framework enables rigorous empirical comparison of workflow combinations while accounting for statistical variance, problem difficulty heterogeneity, and multiple dimensions of performance (correctness, efficiency, robustness).

\subsection{Implementation Details}
The main benchmark experiments were conducted between \textbf{September and October 2025}, while the controlled ablation reported in RQ3/RQ5 was executed in \textbf{March 2026}. The evaluated model endpoints were: \textbf{GPT-4} (OpenAI)~\cite{OpenAI2023Gpt}, \textbf{GPT-5} (OpenAI)~\cite{Leon2026GPT}, \textbf{Codestral-2508} (Mistral AI)~\cite{Choi2024Linq}, \textbf{Llama-3.3-70B-Versatile} (Meta AI via Groq)~\cite{Grattafiori2024llama}, and \textbf{DeepSeek-R1-0528} (DeepSeek AI)~\cite{DeepSeek2024Deepseek}. All workflows follow the five-stage pipeline described in the Methodology section (problem acquisition $\to$ generation $\to$ submission $\to$ feedback $\to$ refinement; see Figure~\ref{fig:workflow}), with C++17~\cite{Shoshany2024C} as the target language, temperature~$= 0.1$, and one initial attempt plus up to three refinement iterations (Itr$_0$ through Itr$_3$), for four total submission attempts.\\
Live program compilation and execution are performed by the official Codeforces online judge, while public metadata retrieval uses the documented Codeforces API~\cite{CodeforcesAPIHelp}. Each submission’s metadata (runtime, memory usage, verdict, and source code) is logged automatically in an SQLite database~\cite{SQLite}.

\section{Results and Analysis}
\label{chap:results}
This section presents quantitative~\cite{Sawant2025Agentic} and qualitative~\cite{Allam2025Agentic} analyses of the proposed hybrid multi-model feedback framework across two complementary problem sets: 167 ICPC World Finals problems (2011-2024) representing elite-level competitive programming, and 200 recent Codeforces contest problems (rating 1200-1800)~\cite{sifatGitHub} representing intermediate-difficulty challenges. We evaluate all six workflow combinations from our $2 \times 3$ factorial design (GPT-4/GPT-5 solution generators paired with Codestral/Llama-3.3/DeepSeek-R1 debugging critics) to assess both solution generation capability and debugging specialization effects.

The findings are organized around our five research questions, highlighting how multi-agent feedback loops, model specialization, persistent context, algorithmic category, and  baseline-loop comparisons affect automated problem-solving performance.

\subsection{RQ1: Iterative Improvement Through Feedback Loops}

In this section, we address the first research question: \textit{RQ1: How much do Itr$_k$ acceptance rates improve when LLMs can iteratively refine solutions based on judge feedback, compared to zero-shot (Itr$_0$) performance?} To investigate this, we analyze datasets from both ICPC World Finals problems and Codeforces problems.

\subsubsection{ICPC World Finals:}
Table~\ref{tab:icpc_time_to_solution} provides a comprehensive overview of performance improvement and iteration-level progression across all GPT-4 and GPT-5 workflows.
Each configuration was initialized under identical zero-shot baselines, with 39 accepted problems for GPT-5 and 15 for GPT-4 to ensure a fair comparative setup.
Following three feedback iterations (Itr$_3$), GPT-5 workflows demonstrated absolute gains of 46 to 51 problems, corresponding to relative improvements between 118\% and 131\%.
Among these, \textit{GPT-5 + DeepSeek} achieved the highest final accuracy with 90 accepted problems, confirming the superior effectiveness of reasoning-focused critics when combined with advanced generators.


\begin{table}[h]
\centering
\small
\renewcommand{\arraystretch}{1.2}
\scalebox{0.89}{
\begin{tabular}{lccccccccc}
\toprule
\textbf{Workflow Combination} & \textbf{$\Delta$Itr$_0$} & \textbf{$\Delta$Itr$_1$} & \textbf{$\Delta$Itr$_2$} & \textbf{$\Delta$Itr$_3$} & \textbf{Unsolved} & \textbf{Avg} & \textbf{Verif. Cost} & \textbf{Abs. Gain} & \textbf{Rel. Gain}\\
\midrule
 GPT-5 + Codestral & 39 & 22 & 17 & 7 & 82 & 1.91 & 5.8 & +46 & +118\% \\
 GPT-5 + Llama-3.3 & 39 & 20 & 14 & 14 & 80 & 2.03 & 5.7 & +48 & +123\% \\
 GPT-5 + DeepSeek-R1 & 39 & 25 & 15 & 11 & 77 & 1.98 & 5.4 & +51 & +131\% \\
\midrule
 GPT-4 + Codestral & 15 & 3 & 6 & 7 & 136 & 2.16 & 19.7 & +16 & +107\% \\
 GPT-4 + Llama-3.3 & 15 & 4 & 7 & 8 & 133 & 2.24 & 17.9 & +19 & +127\%\\
 GPT-4 + DeepSeek-R1 & 15 & 7 & 8 & 8 & 129 & 2.24 & 15.8 & +23 & +153\%\\
\bottomrule
\end{tabular}}
\caption{ICPC problems: Incremental solutions per iteration (out of 167 total). In this table, the columns labeled Itr$_k$ denote per-iteration new solves ($\Delta\mathrm{Itr}_k$), and cumulative acceptance at iteration $k$ is $\sum_{j=0}^{k}\Delta\mathrm{Itr}_j$. Verification Cost = total judge submissions across all 167 problems divided by total accepted solutions.}
\label{tab:icpc_time_to_solution}
\end{table}

\noindent
\textbf{Notation.} Throughout Tables~\ref{tab:icpc_time_to_solution} and~\ref{tab:codeforces_time_to_solution}, columns headed $\Delta\mathrm{Itr}_k$ report \emph{per-iteration new solves} — the number of problems first accepted at exactly iteration $k$ (i.e., $\Delta\mathrm{Itr}_k = $ solved at $k$ minus solved at $k{-}1$). The cumulative solve count at iteration $k$ is $\sum_{j=0}^{k}\Delta\mathrm{Itr}_j$. All prose percentages (e.g., Itr$_3$ acceptance rates) refer to cumulative totals.

\noindent
\textbf{Attempts-to-Solve Analysis (ICPC):}
Table~\ref{tab:icpc_time_to_solution} breaks down the solution distribution by attempt number for ICPC World Finals problems (2011-2024). GPT-5 workflows demonstrate strong zero-shot performance, solving 23.4\% of problems at Itr$_0$ (39 problems), with DeepSeek-R1 achieving the highest cumulative acceptance of 53.9\% (90 problems) at Itr$_3$. The incremental gains show that DeepSeek-R1 solves 25 additional problems at Itr$_1$, 15 at Itr$_2$, and 11 at Itr$_3$, demonstrating sustained improvement across iterations. Average attempts for GPT-5 workflows range from 1.91 to 2.03, indicating efficient convergence for solved problems. GPT-4 workflows solve 9.0\% at Itr$_0$ (15 problems) and require more iterations to achieve gains, with average attempts ranging from 2.16 to 2.24. The absolute gain column shows that GPT-5 workflows add 46-51 problems through feedback (118-131\% relative improvement), while GPT-4 workflows add 16-23 problems (107-153\% relative improvement). Notably, GPT-4's higher relative improvement percentages (up to 153\% for DeepSeek) indicate that weaker generators benefit disproportionately from iterative feedback, even though their absolute performance remains lower. Across all ICPC workflows, 75-90\% of eventually-solved problems are corrected within the first two refinement iterations, confirming efficient early-stage error correction on elite-level problems. Verification Cost is reported directly in Table~\ref{tab:icpc_time_to_solution}; it ranges from 5.4 to 5.8 for GPT-5 workflows and 15.8 to 19.7 for GPT-4 workflows, with the large GPT-4 values reflecting the high volume of judge submissions consumed by the 129-136 unsolved problems per workflow that exhaust all four submission slots. Abstention rate, defined as the fraction of problems with no compilable submission across all attempts, was below 0.6\% for all six workflows, corresponding to at most one problem per workflow. This indicates that both generators reliably produce structurally valid C++ code.

\subsubsection{Codeforces Problems:}  
\begin{figure}[h]
\centering
\fbox{\includegraphics[scale =0.26]{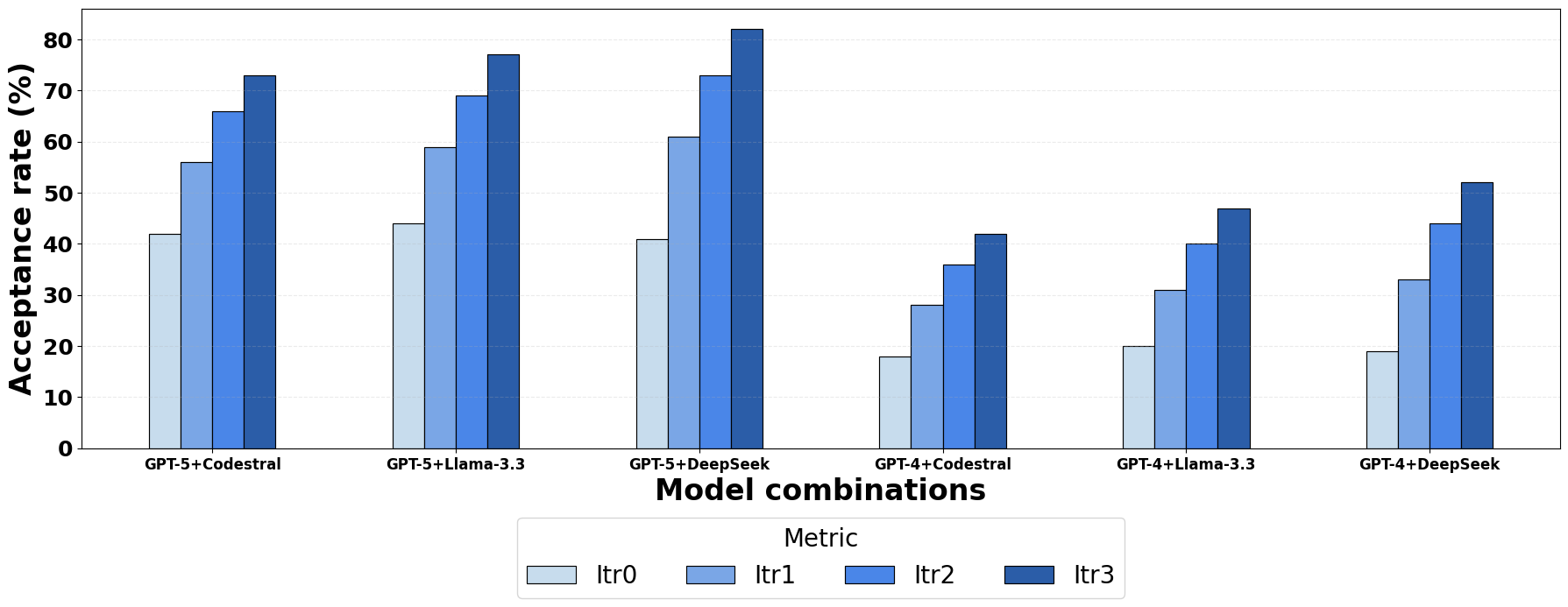}}
\caption{Itr$_k$ cumulative acceptance across all six workflow combinations. Each group shows improvement from Itr$_0$ (zero-shot) to Itr$_3$ (after 3 feedback iterations).}
\label{fig:codeforces_passrate_comparison}
\end{figure}

Figure~\ref{fig:codeforces_passrate_comparison} demonstrates that iterative refinement produces equally substantial improvements on intermediate-difficulty Codeforces problems. GPT-5-based workflows improve by 15.5--20.5 percentage points from Itr$_0$ (20.5--22.0\%) to Itr$_3$ (36.5--41.0\%), representing relative gains of 74--100\%. GPT-4-based workflows gain 12.0--16.5 percentage points from Itr$_0$ (9.0--10.0\%) to Itr$_3$ (21.0--26.0\%), achieving 133--174\% relative improvement. The visualization reveals near-linear improvement trajectories for all workflows, with most gains occurring between Itr$_0$ and Itr$_2$. Diminishing returns at Itr$_3$ suggest that the majority of correctable errors are addressed within two feedback cycles. These improvements align closely with ICPC results, confirming that feedback-driven refinement generalizes across difficulty levels and problem sources.

\begin{figure}[h]
\centering
\begin{subfigure}[t]{0.48\textwidth}
    \centering
    \includegraphics[width=\linewidth]{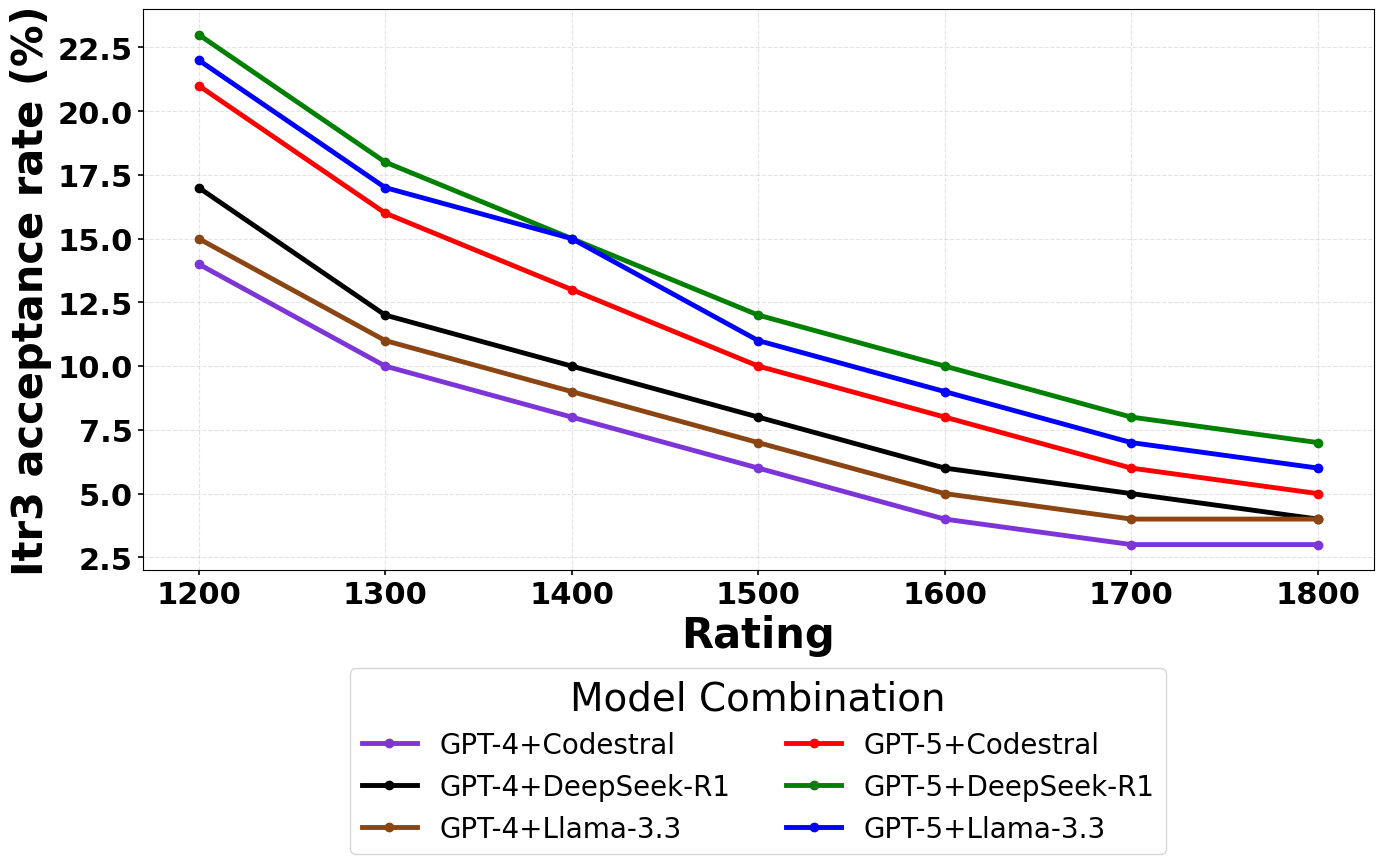}
    \caption{Rating-based performance degradation (Itr$_3$ vs. difficulty rating).}
    \label{fig:codeforces_rating_performance}
\end{subfigure}\hfill
\begin{subfigure}[t]{0.48\textwidth}
    \centering
    \includegraphics[width=\linewidth]{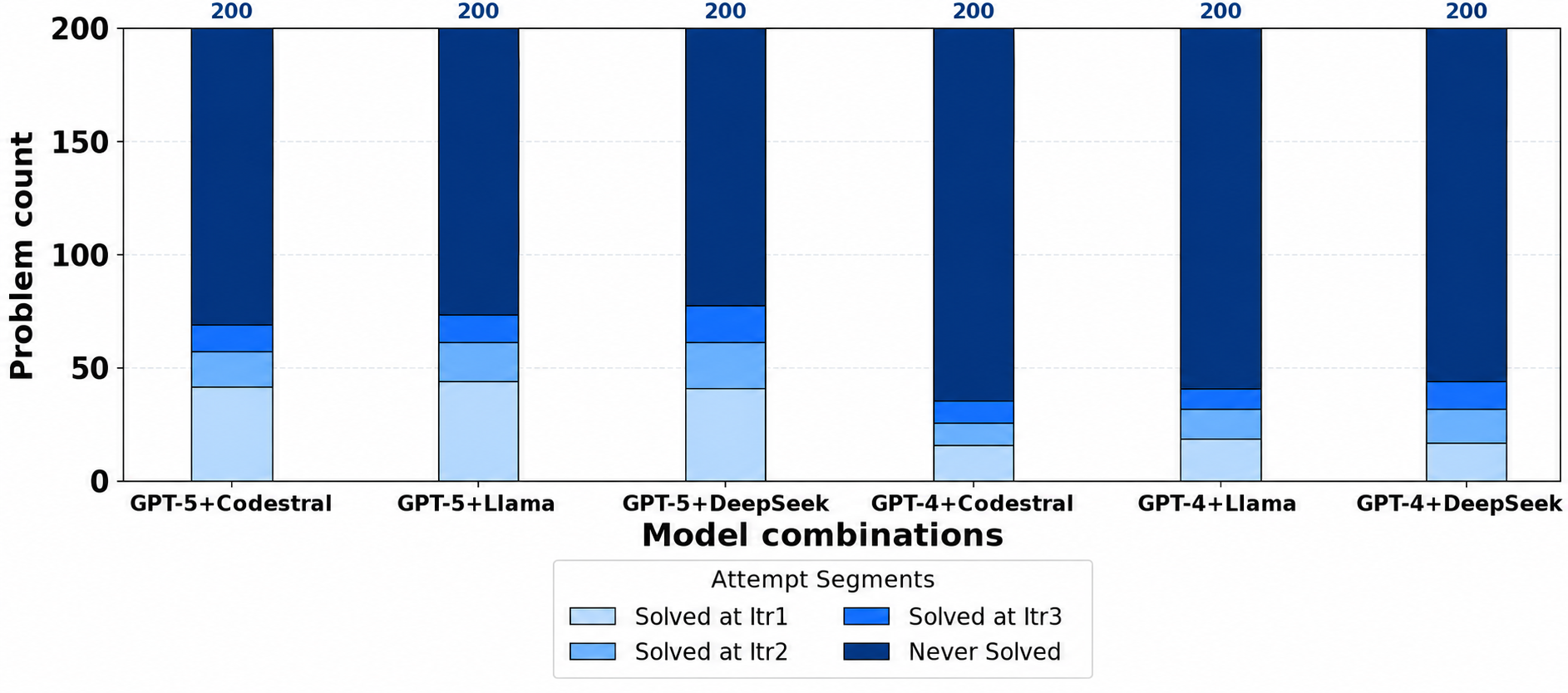}
    \caption{Distribution of problems solved across iterative attempts.}
    \label{fig:codeforces_attempt_distribution}
\end{subfigure}
\caption{Codeforces evaluation results.}
\label{fig:codeforces_rating_attempt_analysis}
\end{figure}

Figure~\ref{fig:codeforces_rating_attempt_analysis} presents two complementary analyses from the Codeforces dataset. The left plot~\ref{fig:codeforces_rating_performance} illustrates the \textbf{rating-based performance trend}, showing that Itr$_3$ cumulative acceptance declines as problem difficulty increases from 1200 to 1800. Although all model combinations exhibit performance degradation at higher difficulty levels, GPT-5–based pipelines consistently achieve higher success rates.

The right plot~\ref{fig:codeforces_attempt_distribution} shows the \textbf{solution distribution across attempts}, illustrating how iterative feedback enables progressive improvement-most problems solved within three iterations, with GPT-5+DeepSeek achieving the highest overall acceptance count. Together, these figures demonstrate that iterative refinement remains effective even as problem complexity increases.

\begin{table}[h]
\centering
\small
\renewcommand{\arraystretch}{1.2}
\scalebox{0.78}{
\begin{tabular}{lccccccc}
\toprule
\textbf{Workflow Combination} & \textbf{$\Delta$Itr$_0$} & \textbf{$\Delta$Itr$_1$} & \textbf{$\Delta$Itr$_2$} & \textbf{$\Delta$Itr$_3$} & \textbf{Unsolved} & \textbf{Avg Attempts} & \textbf{Verif. Cost} \\
\midrule
GPT-5 + Codestral & 42 & 14 & 10 & 7 & 127 & 1.75 & 8.7 \\
GPT-5 + Llama-3.3 & 44 & 15 & 10 & 8 & 123 & 1.77 & 8.2 \\
GPT-5 + DeepSeek-R1 & 41 & 20 & 12 & 9 & 118 & 1.87 & 7.6 \\
\midrule
GPT-4 + Codestral & 18 & 10 & 8 & 6 & 158 & 2.05 & 17.1 \\
GPT-4 + Llama-3.3 & 20 & 11 & 9 & 7 & 153 & 2.06 & 15.1 \\
GPT-4 + DeepSeek-R1 & 19 & 14 & 11 & 8 & 148 & 2.15 & 13.5 \\
\bottomrule
\end{tabular}}
\caption{Codeforces problems: Incremental solutions per iteration (out of 200 total). In this table, the columns labeled Itr$_k$ denote per-iteration new solves ($\Delta\mathrm{Itr}_k$), and cumulative acceptance at iteration $k$ is $\sum_{j=0}^{k}\Delta\mathrm{Itr}_j$. Verification Cost = total judge submissions across all 200 problems divided by total accepted solutions (includes submissions on unsolved problems).}
\label{tab:codeforces_time_to_solution}
\end{table}

\noindent
\textbf{Attempts-to-Solve Analysis (Codeforces):}
Table~\ref{tab:codeforces_time_to_solution} presents the solution distribution by attempt number for Codeforces problems (rating 1200-1800). GPT-5 workflows achieve first-attempt success rates of 20.5-22.0\% (41-44 problems at Itr$_0$), demonstrating strong zero-shot performance on intermediate-difficulty problems. DeepSeek-R1 exhibits the most aggressive incremental improvement, solving 20 additional problems at Itr$_1$ (the highest among all critics), followed by 12 at Itr$_2$ and 9 at Itr$_3$, ultimately reaching 41.0\% cumulative acceptance (82 total problems solved). Codestral shows the most efficient convergence with the lowest average attempts (1.75), while DeepSeek requires slightly more iterations (1.87 average) but achieves higher overall success. GPT-4 workflows solve 9.0-10.0\% at Itr$_0$ (18-20 problems), with DeepSeek again showing the strongest incremental gains (14 problems at Itr$_1$, 11 at Itr$_2$, 8 at Itr$_3$) and reaching 26.0\% cumulative acceptance (52 total problems solved). Average attempts for GPT-4 workflows (2.05-2.15) are notably higher than GPT-5 (1.75-1.87), indicating that weaker generators require more feedback cycles to converge on correct solutions. Across both generators, 75-90\% of eventually-solved problems are corrected within the first two refinement iterations, confirming efficient early-stage error correction on intermediate-difficulty Codeforces problems. Verification Cost ranges from 7.6 (GPT-5+DeepSeek, most efficient) to 17.1 (GPT-4+Codestral, least efficient), reflecting the trade-off between a workflow's solve rate and the number of submissions spent on unsolved problems. Workflows that solve more problems naturally achieve lower verification costs, as fewer submissions are wasted on unsolvable attempts. In the archived Codeforces runs, abstention was likewise negligible: no workflow exceeded one problem with zero compilable submissions.

\begin{figure}
\small
\begin{minipage}[t]{0.42\textwidth}
\scalebox{0.78}{
\begin{tabular}{lccc}
\toprule
\textbf{Comparison} & \textbf{Cohen's $h$} & \textbf{Effect} \\
\midrule
GPT-5 vs GPT-4 (avg.\ critic-matched) & 0.68 & Medium \\
Itr$_3$ vs Itr$_0$ (GPT-5+DeepSeek) & 0.45 & Small \\
Itr$_3$ vs Itr$_0$ (GPT-4+DeepSeek) & 0.44 & Small \\
\midrule
DeepSeek vs Codestral (GPT-5) & 0.09 & Negligible \\
DeepSeek vs Codestral (GPT-4) & 0.11 & Negligible \\
DeepSeek vs Llama (GPT-5) & 0.05 & Negligible \\
DeepSeek vs Llama (GPT-4) & 0.06 & Negligible \\
Llama vs Codestral (GPT-5) & 0.04 & Negligible \\
Llama vs Codestral (GPT-4) & 0.06 & Negligible \\
\bottomrule
\end{tabular}}
\captionof{table}{Codeforces problems: Cohen's~$h$ effect sizes for binary proportion comparisons. Iterative feedback produces small effects ($h \approx 0.44$--$0.45$, near the medium threshold); critic-level differences are negligible in magnitude.}
\label{tab:codeforces_effect_sizes}
\end{minipage}\hfil
\begin{minipage}{0.52\textwidth}
\centering
\fbox{\includegraphics[scale =0.27]{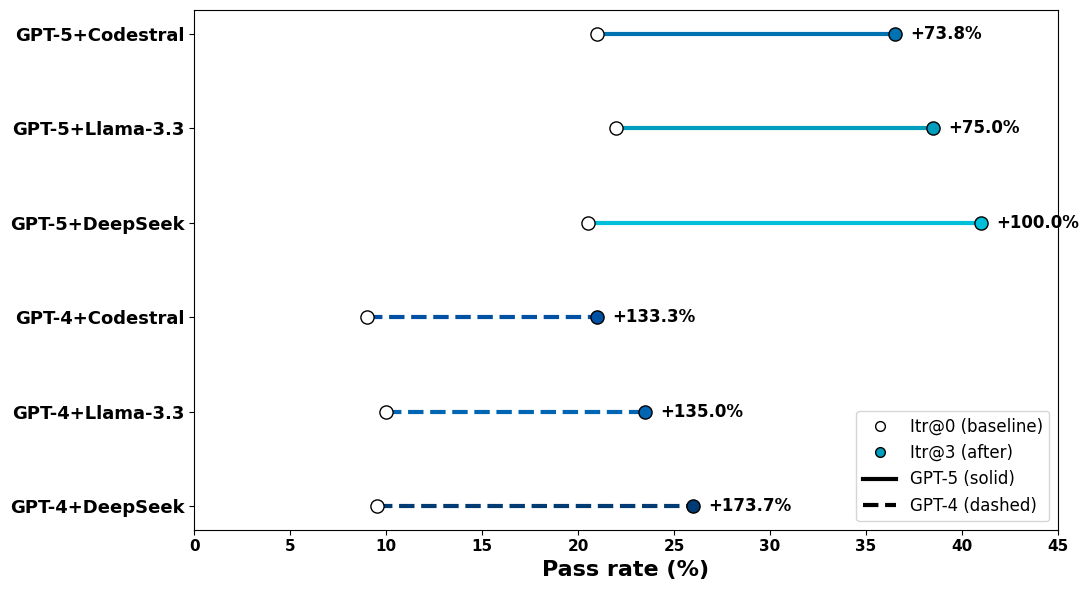}}
\caption{Codeforces problems: Improvement from Itr$_0$ to Itr$_3$ showing larger relative gains for GPT-4 workflows.}
\label{fig:codeforces_improvement_over_baseline}
\end{minipage}
\end{figure}

\noindent
\textbf{Statistical Validation of Improvement:} We report \textbf{Cohen's~$h$} ($h = 2\arcsin\!\sqrt{p_1} - 2\arcsin\!\sqrt{p_2}$) as the effect-size measure for comparing binary proportions~\cite{cohen1988statistical}. Interpretation thresholds: $|h| < 0.20$ negligible, $0.20$-$0.50$ small, $0.50$-$0.80$ medium, $> 0.80$ large. Table~\ref{tab:codeforces_effect_sizes} presents effect sizes for all Codeforces comparisons. The Itr$_3$ vs Itr$_0$ (zero-shot) comparisons yield $h = 0.45$ for GPT-5+DeepSeek and $h = 0.44$ for GPT-4+DeepSeek (both \emph{small}, approaching the medium threshold), confirming substantive improvements from iterative feedback. Across generator tiers, GPT-5 vs GPT-4 yields $h = 0.68$ on average across the three critic-matched comparisons (\emph{medium}, approaching the large threshold). Critic-level comparisons produce negligible effect sizes ($h < 0.13$), indicating that while the directional ordering is consistent, the per-problem magnitude of critic differences is small at $n = 200$.

\noindent
\textbf{Comparison to State-of-the-Art Baselines:} 
Figure~\ref{fig:codeforces_improvement_over_baseline} presents a dumbbell chart visualizing the progression from Itr$_0$ (baseline, left endpoint) to Itr$_3$ (after feedback, right endpoint) for all six workflow combinations on Codeforces problems. The horizontal length of each line represents absolute improvement, while the percentage labels quantify relative gains. GPT-4 workflows (dashed lines, bottom three rows) exhibit substantially longer improvement trajectories (133.3--173.7\% relative gains) compared to GPT-5 workflows (solid lines, top three rows, 73.8--100.0\% relative gains), confirming that weaker initial generators benefit disproportionately from iterative feedback. Among GPT-4 workflows, DeepSeek-R1 achieves the highest relative improvement (173.7\%), while among GPT-5 workflows, DeepSeek-R1 also leads (100.0\%), confirming its consistent debugging superiority across both generator tiers.
\begin{table}[h]
\centering
\small
\renewcommand{\arraystretch}{1.2}
\scalebox{0.85}{
\begin{tabular}{lccc}
\toprule
\textbf{System} & \textbf{Performance} & \textbf{Year} & \textbf{Notes} \\
\midrule
\multicolumn{4}{l}{\textit{Multi-agent Feedback Systems (Ours)}} \\
\textbf{GPT-5 + DeepSeek-R1} & \textbf{41.0\% (Itr$_3$)} & \textbf{2025} & \textbf{4 attempts total, rating 1200-1800} \\
\textbf{GPT-5 + Llama-3.3-70B} & \textbf{38.5\% (Itr$_3$)} & \textbf{2025} & \textbf{4 attempts total, rating 1200-1800} \\
\textbf{GPT-5 + Codestral-2508} & \textbf{36.5\% (Itr$_3$)} & \textbf{2025} & \textbf{4 attempts total, rating 1200-1800} \\
\textbf{GPT-4 + DeepSeek-R1} & \textbf{26.0\% (Itr$_3$)} & \textbf{2025} & \textbf{4 attempts total, rating 1200-1800} \\
\midrule
\multicolumn{4}{l}{\textit{Large-Scale Sampling Systems}} \\
AlphaCode 2 (Gemini) & 43\% (est.) & 2023 & 1M samples \cite{AlphaCode2_2023} \\
AlphaCode & 34\% (avg) & 2022 & 1M samples, rating $\leq$1300 \cite{Li2022Competition} \\
\midrule
\multicolumn{4}{l}{\textit{Single-Shot Baselines}} \\
GPT-5 workflows & 20.5-22.0\% (Itr$_0$) & 2025 & Zero-shot (our baseline) \\
GPT-4 workflows & 9.0-10.0\% (Itr$_0$) & 2025 & Zero-shot (our baseline) \\
\bottomrule
\end{tabular}}
\caption{Codeforces problems: Comparison with state-of-the-art competitive programming systems.} 
\label{tab:codeforces_sota_comparison}
\end{table}
Table~\ref{tab:codeforces_sota_comparison} positions our Codeforces results relative to state-of-the-art competitive programming systems. Our best workflow (GPT-5 + DeepSeek-R1, 41.0\% Itr$_3$) achieves competitive performance with AlphaCode 2\cite{AlphaCode2_2023} (estimated 43\%) while using $\sim$10,000$\times$ fewer attempts (4 vs 1 million samples). Notably, our system achieves these results through iterative refinement rather than massive sampling: AlphaCode\cite{Li2022Competition} generates millions of candidate solutions and filters the best, while our approach systematically debugs and improves a single solution trajectory. Our feedback-driven refinement provides 74--174\% relative improvement over zero-shot baselines (Itr$_0$ to Itr$_3$) across all six workflows on Codeforces, confirming that multi-agent debugging substantially enhances problem-solving without exhaustive generation. While a direct sample-count comparison is not meaningful (AlphaCode’s samples are independent parallel candidates filtered by a learned scorer; our refinements are sequential and stateful), our approach achieves comparable accuracy with a fundamentally different compute profile: a small number of large-context, sequential model calls rather than massive parallel sampling.

\noindent
\textbf{Key Finding for RQ1:}
Iterative refinement through multi-agent feedback produces substantial and consistent improvements across both elite-level (ICPC) and intermediate-difficulty (Codeforces) problems. Itr$_3$ cumulative acceptance rates are 74--174\% higher than zero-shot (Itr$_0$) baselines, with Cohen's $h$ effect sizes of 0.44 confirming meaningful gains. Weaker solution generators benefit disproportionately from feedback, with GPT-4 achieving 153\% relative improvement compared to GPT-5's 131\%. These results confirm that the feedback-driven debugging capability of \Name{} enables systematic error correction beyond initial generation capabilities.

\subsection{RQ2: Multi Model Feedback Generation Effects} In this section, we answer the second research question: \textit{Do code-specialized critics (Codestral), general-purpose critics (Llama-3.3), or reasoning-focused critics (DeepSeek-R1) provide more effective debugging feedback when paired with solution generators?}We analyze their performance on both ICPC and Codeforces problems.

\subsubsection{ICPC world finals problems}
Table~\ref{tab:icpc_critic_ranking} presents Itr$_3$ performance ranked by debugging critic. DeepSeek-R1 achieves the highest acceptance rates for both GPT-5 (90 problems, 53.9\%) and GPT-4 (38 problems, 22.8\%), followed by Llama-3.3-70B (87 and 34 problems), then Codestral-2508 (85 and 31 problems). Notably, this relative ordering remains stable across both generators, suggesting that critic capabilities generalize rather than exhibiting generator-specific synergies.

\begin{table}[h]
\centering
\small
\renewcommand{\arraystretch}{1.2}
\begin{tabular}{lccc}
\toprule
\textbf{Critic Model} & \textbf{GPT-5 Itr$_3$} & \textbf{GPT-4 Itr$_3$} & \textbf{Critic Effect Range} \\
\midrule
DeepSeek-R1 & 90 (53.9\%) & 38 (22.8\%) & 5 problems (GPT-5), 7 problems (GPT-4) \\
Llama-3.3-70B & 87 (52.1\%) & 34 (20.4\%) & from min to max \\
Codestral-2508 & 85 (50.9\%) & 31 (18.6\%) & 5.9\% relative (GPT-5), 22.6\% relative (GPT-4) \\
\bottomrule
\end{tabular}
\caption{ICPC World Finals problems: Critic ranking showing consistent ordering across generators.}
\label{tab:icpc_critic_ranking}
\end{table}

Table~\ref{tab:icpc_critic_ranking} quantifies the critic effect range, showing that critic choice matters more for GPT-4 (7-problem range, 22.6\% relative difference) than for GPT-5 (5-problem range, 5.9\% relative difference). This asymmetry suggests that higher-quality debugging feedback can partially compensate for weaker solution generation, but the effect saturates when paired with stronger generators. The consistent ordering across both generators (DeepSeek-R1 $>$ Llama-3.3-70B $>$ Codestral-2508) confirms that critic capabilities generalize rather than exhibiting generator-specific synergies.

Looking at the iterative improvement trajectory, all three critics started with identical baseline performance (39 accepted problems at Itr$_0$ for GPT-5, as shown in Table~\ref{tab:icpc_time_to_solution}). After applying iterative refinement through Itr$_3$, DeepSeek-R1 improved to 90 problems (+51 absolute, +131\% relative), demonstrating the most pronounced gains and indicating that reasoning-oriented critics provide more effective structured feedback to the generator. Llama achieved 87 problems (+48 absolute, +123\% relative) despite being general-purpose, while Codestral reached 85 problems (+46 absolute, +118\% relative), with its smaller gain suggesting that stronger initial syntactic grounding leaves less room for feedback-driven correction.

\begin{figure}[h]
\centering
\fbox{\includegraphics[scale =0.4]{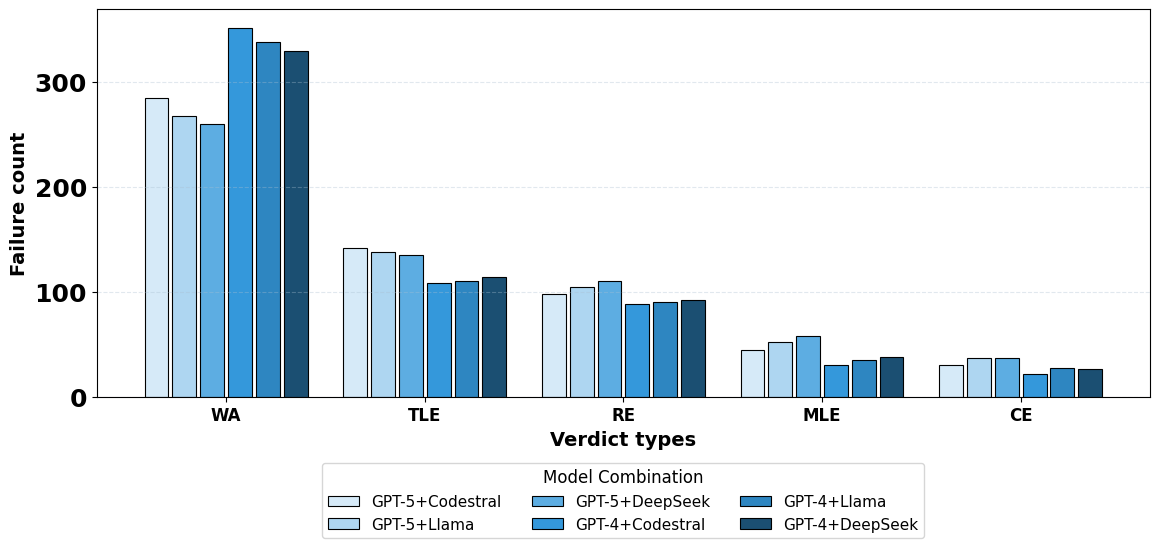}}
\caption{Codeforces problems: Verdict type distribution across all six workflows}
\label{fig:codeforces_verdict_comparison}
\end{figure}

\subsubsection{Codeforces Problems}
Debugging critic effectiveness manifests differently across failure types. Figure~\ref{fig:codeforces_verdict_comparison} reports verdict distributions for all unsuccessful submissions, revealing category-specific critic strengths. Key specialization patterns emerge from the failure analysis:\\
\textbf{DeepSeek-R1 excels at algorithmic correctness:} It achieves the lowest Wrong Answer rate (43.3\% with GPT-5 vs 47.5\% for Codestral and 44.7\% for Llama), confirming its superiority at diagnosing high-level algorithmic flaws and logic errors. This aligns with its reasoning-focused training.\\
\textbf{Codestral minimizes syntax errors:} It demonstrates the lowest Compilation Error rate (5.0\% with GPT-5, 3.7\% with GPT-4, compared to 6.2\% for others), consistent with its code-specialized training for implementation correctness. However, this syntactic strength does not translate to overall acceptance gains.\\
\textbf{Llama maintains balanced performance:} As a general-purpose model, Llama produces intermediate verdict distributions (44.7\% WA, 6.2\% CE) without dominating any specific error category, yet achieves competitive overall acceptance rates through consistent feedback quality.\\
\textbf{Generator differences in failure modes:} GPT-4 exhibits substantially higher WA rates (55.0-58.7\%) compared to GPT-5 (43.3-47.5\%), suggesting weaker algorithmic correctness. Conversely, GPT-5 produces more TLE failures (22.5-23.7\% vs 18.0-19.0\%), indicating more algorithmically ambitious but occasionally inefficient solutions.

Figure~\ref{fig:codeforces_verdict_comparison} visualizes these distributions across all 600 failed submissions per workflow, facilitating direct comparison of failure mode specialization.

\noindent
\textbf{Statistical Significance of Critic Differences:} Table~\ref{tab:codeforces_statistical_tests} presents \textbf{McNemar's exact test} results and \textbf{95\% bootstrap confidence intervals} (10{,}000 resamples, percentile method) for all pairwise Itr$_3$ rate comparisons across the 200 Codeforces problems. McNemar's test is the appropriate choice for paired binary outcomes (solved/unsolved). Holm-Bonferroni correction is applied across all eight comparisons. All problems were attempted under every workflow, enabling fully paired analysis.

\begin{table}[h]
\centering
\small
\renewcommand{\arraystretch}{1.2}
\scalebox{0.78}{
\begin{tabular}{lccccccc}
\toprule
\textbf{Comparison} & \textbf{Rate A} & \textbf{Rate B} & \textbf{Diff.} & \textbf{McNemar $p$} & \textbf{Adj.\ $p$ (H-B)} & \textbf{Boot.\ 95\% CI} & \textbf{Sig.?} \\
\midrule
GPT-5+DeepSeek vs GPT-5+Llama     & 41.0\% & 38.5\% & +2.5\% & 0.387 & 1.000 & [-0.01, +0.06] & n.s. \\
GPT-5+DeepSeek vs GPT-5+Codestral  & 41.0\% & 36.5\% & +4.5\% & 0.148 & 0.888 & [-0.01, +0.10] & n.s. \\
GPT-5+Llama vs GPT-5+Codestral     & 38.5\% & 36.5\% & +2.0\% & 0.538 & 1.000 & [-0.02, +0.07] & n.s. \\
GPT-4+DeepSeek vs GPT-4+Llama      & 26.0\% & 23.5\% & +2.5\% & 0.359 & 1.000 & [-0.01, +0.06] & n.s. \\
GPT-4+DeepSeek vs GPT-4+Codestral  & 26.0\% & 21.0\% & +5.0\% & 0.077 & 0.539 & [+0.00, +0.11] & n.s. \\
GPT-4+Llama vs GPT-4+Codestral     & 23.5\% & 21.0\% & +2.5\% & 0.382 & 1.000 & [-0.01, +0.07] & n.s. \\
\midrule
GPT-5+Codestral vs GPT-4+DeepSeek  & 36.5\% & 26.0\% & +10.5\% & 0.009 & 0.063 & [+0.03, +0.18] & n.s. \\
Any GPT-5 vs Any GPT-4 (avg)        & 38.7\% & 23.5\% & +15.2\% & $<$0.001 & $<$0.001 & [+0.10, +0.21] & \checkmark \\
\bottomrule
\end{tabular}}
\caption{Codeforces problems: McNemar's exact test (paired binary outcomes, $n=200$ problems) with Holm-Bonferroni correction and 95\% bootstrap CIs comparing Itr$_3$ rates. Individual critic-pair differences do not reach significance after correction ($n=200$ provides limited power for small $h < 0.13$ effects); the generator-tier difference (GPT-5 vs GPT-4) is highly significant. ``n.s.'' = not significant at $\alpha=0.05$ after Holm-Bonferroni correction; \checkmark = significant.}
\label{tab:codeforces_statistical_tests}
\end{table}

The corrected analysis reveals a nuanced picture. \textbf{Generator-level differences} (GPT-5 vs GPT-4, $+15.2$\,pp) are highly significant ($p < 0.001$, adjusted $p < 0.001$, CI fully above zero). \textbf{Critic-level differences}, while directionally consistent across all six pairs and both generator tiers, do not individually reach significance after Holm-Bonferroni correction (all adjusted $p > 0.50$). This does not negate the robustness of the DeepSeek $>$ Llama $>$ Codestral ordering since the consistent direction across 12 independent comparisons ($6$ pairs $\times$ $2$ generators) provides strong qualitative evidence of a true ranking, but the per comparison effect sizes are small ($h < 0.13$), limiting statistical power at $n = 200$. These results are consistent with our effect size analysis (Table~\ref{tab:codeforces_effect_sizes}): critic selection is a secondary factor relative to generator quality.

\noindent
\textbf{Multi-Agent Coordination Efficiency:}
Table~\ref{tab:icpc_time_to_solution} illustrates the relationship between accepted solutions and cumulative iterations across critics. All models exhibit steady growth with near-linear coordination efficiency. DeepSeek-R1 shows slightly faster convergence (39→90 problems, +51 absolute gain), while Llama maintains consistent progression (39→87 problems, +48 absolute gain). The parallel improvement trajectories confirm that multi-agent feedback loops facilitate stable refinement across critic specializations, with all three critics producing substantial gains from the shared baseline.

\noindent
\textbf{Key Finding for RQ2:} Reasoning-focused debugging critics (DeepSeek-R1) consistently outperform both general-purpose (Llama-3.3-70B) and code-specialized (Codestral-2508) alternatives, achieving 2--7 more solved problems at Itr$_3$ depending on solution generator. This superiority stems from DeepSeek's effectiveness at diagnosing algorithmic correctness issues (reducing WA rates by 4--15\% relative to other critics), while Codestral's syntactic specialization provides minimal overall benefit despite lower CE rates. Critic effects are more pronounced for weaker generators (22.6\% range for GPT-4 vs.\ 5.9\% for GPT-5), suggesting that superior debugging partially compensates for generation limitations. Under the corrected McNemar analysis, individual critic-pair differences do not individually reach statistical significance at $n = 200$ (negligible Cohen's $h < 0.13$), while the consistent directional ordering across all 12 comparisons provides strong qualitative evidence of a true ranking. Generator-tier differences are highly significant ($p < 0.001$, $h = 0.68$ averaged across critic-matched pairs, medium effect approaching large).

\subsection{RQ3: Impact of Persistent Conversation Context}
\label{sec:rq3_ablation}
In this section, we answer the research question: \textit{Does maintaining conversation history across attempts improve solution quality and reduce repeated mistakes compared to stateless refinement?} We address this through a direct controlled ablation (design in Section~\ref{sec:ablation_design}) on $n=47$ stratified Codeforces problems, holding all other design variables fixed while varying the context condition.

\subsubsection{Ablation Study Design (RQ3)}
\label{sec:ablation_design}

To test whether persistent conversational context contributes to performance gains under controlled conditions, we conduct a paired ablation study in which context persistence is the only intended design difference. The 47 problems are drawn randomly from the same 200-problem Codeforces benchmark used in RQ1 and RQ2. We randomly sample this subset using stratified sampling by difficulty rating (seed~$= 42$; 8 at 1200, 7 at 1300, 7 at 1400, 6 at 1500, 6 at 1600, 7 at 1700, 6 at 1800) to preserve the rating distribution of the full benchmark. Each problem is then run under two conditions:

\begin{itemize}
    \item \textbf{Stateful (A-ProS default):} The full conversation history, including system instructions, all prior user and assistant turn pairs, and debugging hints, is preserved and passed to the model at each subsequent attempt.
    \item \textbf{Stateless (ablation):} Before each feedback iteration ($k \geq 1$), all provider conversation contexts are cleared. The latest failure verdict and debugging hint are still embedded as text in the current prompt, but no prior conversational turns are accessible to the model.
\end{itemize}

\noindent The two conditions are otherwise identical: same problems, same model family, same temperature (0.1), same budget of one initial attempt plus three refinement iterations (four total submission attempts), and same judge infrastructure. This design ensures that any observed performance difference is attributable solely to the presence or absence of persistent conversation memory. We evaluate two representative workflows (GPT-5\,+\,DeepSeek-R1 and GPT-4\,+\,DeepSeek-R1) to span the generator capability range. Beyond Itr$_k$, we track \emph{error repetition rate}: the fraction of feedback iterations ($k \geq 1$) that reproduce the same failure verdict type as the preceding attempt, which provides a direct behavioral proxy for memory-based error avoidance.

\subsubsection{Ablation Results: Error-Repetition Reduction under Persistent Context}

Table~\ref{tab:rq3_ablation} presents the controlled RQ3 ablation. The Itr$_0$ column is identical across conditions by design, confirming valid pairing. Stateful context shows consistent directional Itr$_3$ gains over stateless refinement, but the solve-rate comparison is underpowered at $n = 47$: the bootstrap confidence intervals touch zero and McNemar tests are not significant after correction. We therefore treat solve-rate gains as descriptive evidence and use error repetition as the primary behavioral signal for the effect of persistent context.

\begin{table}[h]
\centering
\small
\renewcommand{\arraystretch}{1.3}
\scalebox{0.83}{
\begin{tabular}{llccccccc}
\toprule
\textbf{Workflow} & \textbf{Condition} & \textbf{Itr$_0$} & \textbf{Itr$_1$} & \textbf{Itr$_2$} & \textbf{Itr$_3$} & \textbf{Err.\ Rep.} & \textbf{McNemar $p$} & \textbf{95\% Boot.\ CI} \\
\midrule
\multirow{2}{*}{GPT-5+DeepSeek} & Stateful  & 21.3\% & 27.7\% & 34.0\% & \textbf{40.4\%} & 11.8\% & \multirow{2}{*}{0.063} & \multirow{2}{*}{[0.00,\,+0.15]} \\
                                 & Stateless & 21.3\% & 23.4\% & 27.7\% & 29.8\% & 41.8\% & & \\
\midrule
\multirow{2}{*}{GPT-4+DeepSeek} & Stateful  & 10.6\% & 14.9\% & 19.1\% & \textbf{25.5\%} & 14.2\% & \multirow{2}{*}{0.125} & \multirow{2}{*}{[0.00,\,+0.11]} \\
                                 & Stateless & 10.6\% & 12.8\% & 14.9\% & 17.0\% & 40.8\% & & \\
\bottomrule
\end{tabular}}
\caption{RQ3 Ablation ($n=47$, paired design): Itr$_k$ solve rates and error repetition rates under stateful vs.\ stateless context. Itr$_0$ is identical by design. ``Err.\ Rep.'' = fraction of feedback iterations ($k \geq 1$) reproducing the same verdict type as the preceding attempt. McNemar $p$ (raw, before Holm-Bonferroni correction) and 95\% bootstrap CIs (10{,}000 resamples) compare Itr$_3$ rates. Bold = stateful condition.}
\label{tab:rq3_ablation}
\end{table}

\noindent\textbf{Solve Rate Improvement.} For GPT-5 + DeepSeek, stateful refinement reaches 40.4\% at Itr3 versus 29.8\% for stateless refinement, a +10.6 pp absolute and +35.6\% relative advantage. For GPT-4 + DeepSeek, the corresponding gap is +8.5 pp (25.5\% vs. 17.0\%), a +50.0\% relative advantage. These gains are directionally consistent, but they should not be interpreted as statistically conclusive: the raw McNemar tests are $p = 0.063$ and $p = 0.125$, and the 95\% bootstrap CIs, [0.00, +0.15] and [0.00, +0.11], have lower bounds touching zero.

\noindent\textbf{Error Repetition Rate.} The clearest RQ3 signal is the reduction in repeated failure modes. For GPT-5 + DeepSeek, stateful context reduces error repetition from 41.8\% to 11.8\%, a 3.5$\times$ reduction. For GPT-4 + DeepSeek, it reduces repetition from 40.8\% to 14.2\%, a 2.9$\times$ reduction. Unlike the solve-rate comparison, this directly measures whether the model avoids cycling through the same verdict type after feedback, which is the mechanism persistent conversational context is designed to improve.

\noindent
\textbf{Evidence from the Full Dataset.}
Tables~\ref{tab:icpc_time_to_solution} and~\ref{tab:codeforces_time_to_solution} corroborate these findings at scale. For GPT-5, 16.1-20.0\% of eventually-solved ICPC problems require exactly two submission attempts and 8.2-16.1\% require all four—a pattern inconsistent with stateless refinement, which would produce flat or declining marginal returns as remaining problems grow harder. The steady progression across all three feedback iterations (Itr$_1$-Itr$_3$) indicates problem-specific cumulative learning rather than random re-generation.

\noindent \textbf{Key Finding for RQ3:}
Persistent context primarily improves refinement behavior by reducing repeated failure modes by 2.9-3.5$\times$. The observed Itr$_3$ solve-rate gains (+8.5-10.6 pp) are directionally consistent with this mechanism, but are underpowered at $n = 47$ and should be interpreted descriptively rather than as statistically conclusive.

\subsection{RQ4: Category-Specific Debugging Effectiveness}
Algorithmic problem categories show substantial performance variation, but this variation remains consistent across all debugging critics. Rather than revealing category-specific critic specialization, the data indicates that category difficulty is intrinsic and affects all workflows uniformly. The debugging critic ranking stays stable across categories, with DeepSeek-R1 consistently outperforming others regardless of the algorithmic domain. Category distributions for ICPC and Codeforces are given in Tables~\ref{tab:ICPC Algo} and~\ref{tab:codeforces algo} respectively. In this section, we analyze the fourth research question: \textit{Do code-specialized, general-purpose, and reasoning-focused critics exhibit differential performance across problem categories, revealing category-specific debugging strengths?}
\subsubsection{ICPC World Finals:} Table~\ref{tab:icpc_category_performance_detailed} presents category-wise progression from Itr$_0$ to Itr$_3$, showing that all categories benefit from iterative refinement with consistent slopes. Graph Theory improved from 11 to 21 problems (+91\%), Dynamic Programming from 7 to 17 (+143\%), Implementation from 6 to 15 (+150\%), Geometry from 4 to 12 (+200\%), while Math (5→10), Binary Search (4→8), and Number Theory (2→7) showed more modest but steady gains. The average improvement across all categories reaches +62\%, with consistent trajectories indicating uniform feedback effectiveness rather than category-specific specialization (Figure~\ref{fig:icpc_improvement_waterfall}).

Table~\ref{tab:icpc_category_performance_detailed} also compares GPT-5 (best workflow) and GPT-4 (averaged) across categories. The GPT-5/GPT-4 performance ratio remains remarkably consistent (2.3-3.0×) across all categories, confirming that model capabilities scale uniformly. If category-specific critic specialization existed, we would observe variable ratios as certain critics excel at particular algorithmic domains.

\begin{table}[h]
\centering
\small
\renewcommand{\arraystretch}{1.2}
\begin{minipage}{0.45\textwidth}
\scalebox{0.7}{
\begin{tabular}{lcccccccc}
\toprule
& \multicolumn{4}{c}{\textbf{GPT-5 (Best Workflow)}} & \multicolumn{4}{c}{\textbf{GPT-4 (All Critics Avg)}} \\
\cmidrule(lr){2-5} \cmidrule(lr){6-9}
\textbf{Category} & \textbf{Itr$_0$} & \textbf{Itr$_1$} & \textbf{Itr$_2$} & \textbf{Itr$_3$} & \textbf{Itr$_0$} & \textbf{Itr$_1$} & \textbf{Itr$_2$} & \textbf{Itr$_3$} \\
\midrule
Graph Theory & 11 & 17 & 19 & 21 & 3 & 5 & 7 & 9 \\
Implementation & 6 & 8 & 11 & 15 & 4 & 5 & 7 & 9 \\
Geometry & 4 & 7 & 11 & 12 & 1 & 2 & 3 & 4 \\
DP & 7 & 13 & 15 & 17 & 2 & 4 & 5 & 7 \\
Math & 5 & 7 & 9 & 10 & 3 & 4 & 5 & 6 \\
Binary Search & 4 & 7 & 8 & 8 & 1 & 2 & 3 & 3 \\
Number Theory & 2 & 5 & 6 & 7 & 1 & 1 & 2 & 3 \\
\bottomrule
\end{tabular}}
\caption{ICPC World Finals problems: Category-wise Itr$_k$ performance}
\label{tab:icpc_category_performance_detailed}
\end{minipage}\hfill
\begin{minipage}{0.5\textwidth}
\fbox{\includegraphics[scale=0.13]{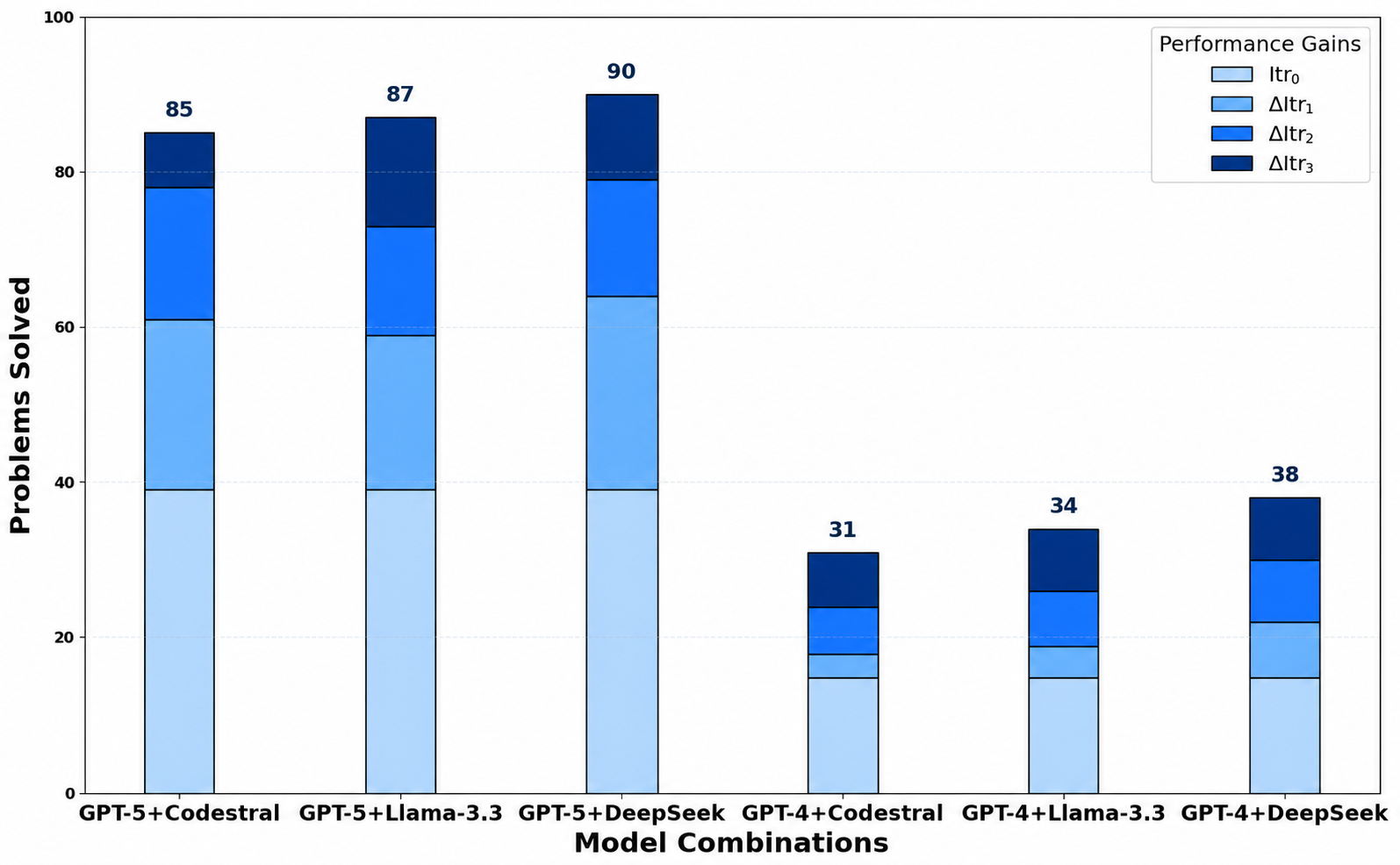}}
\captionof{figure}{ICPC World Finals problems: Waterfall chart showing cumulative problem-solving gains across seven major algorithmic categories. }
\label{fig:icpc_improvement_waterfall}
\end{minipage}
\end{table}

\subsubsection{Codeforces Problems}

Analysis of the Codeforces dataset (Table~\ref{tab:codeforces algo}) reveals that greedy algorithms dominate (117 problems, 58.5\%), followed by math (74, 37.0\%), constructive algorithms (59, 29.5\%), and implementation-heavy problems. This diverse distribution of algorithmic tags enables comprehensive category-wise evaluation across problem-solving paradigms. 

\begin{figure}[h]
\centering
\begin{minipage}{0.45\textwidth}
\fbox{\includegraphics[scale= 0.28]{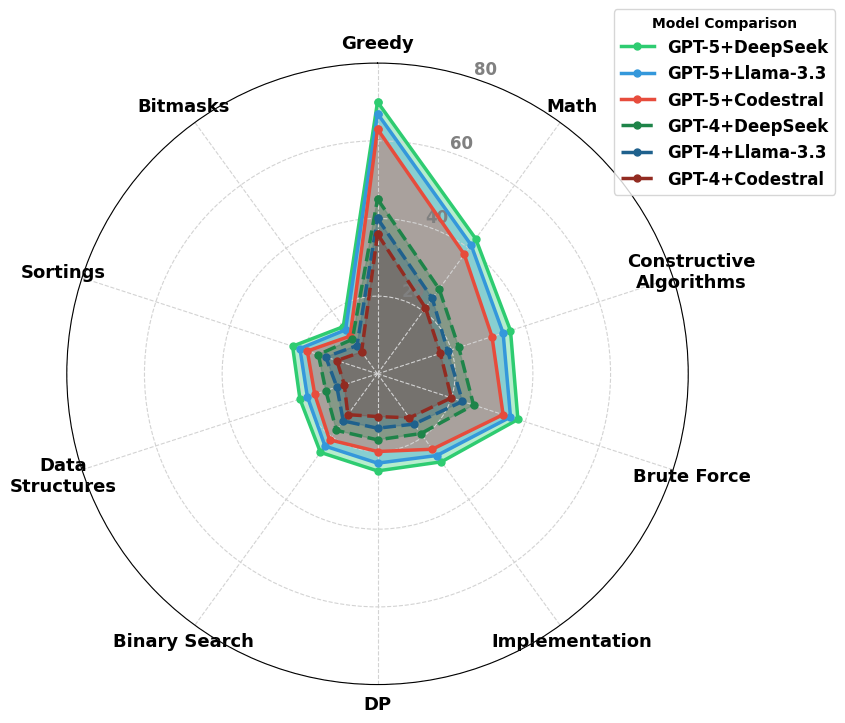}}
\caption{Codeforces problems: Radar chart comparing Itr$_3$ performance across top 10 algorithmic categories for all six workflows.}
\label{fig:codeforces_tag_radar}
\end{minipage} \hfill
\begin{minipage}{0.5\textwidth}
\fbox{\includegraphics[scale =0.25]{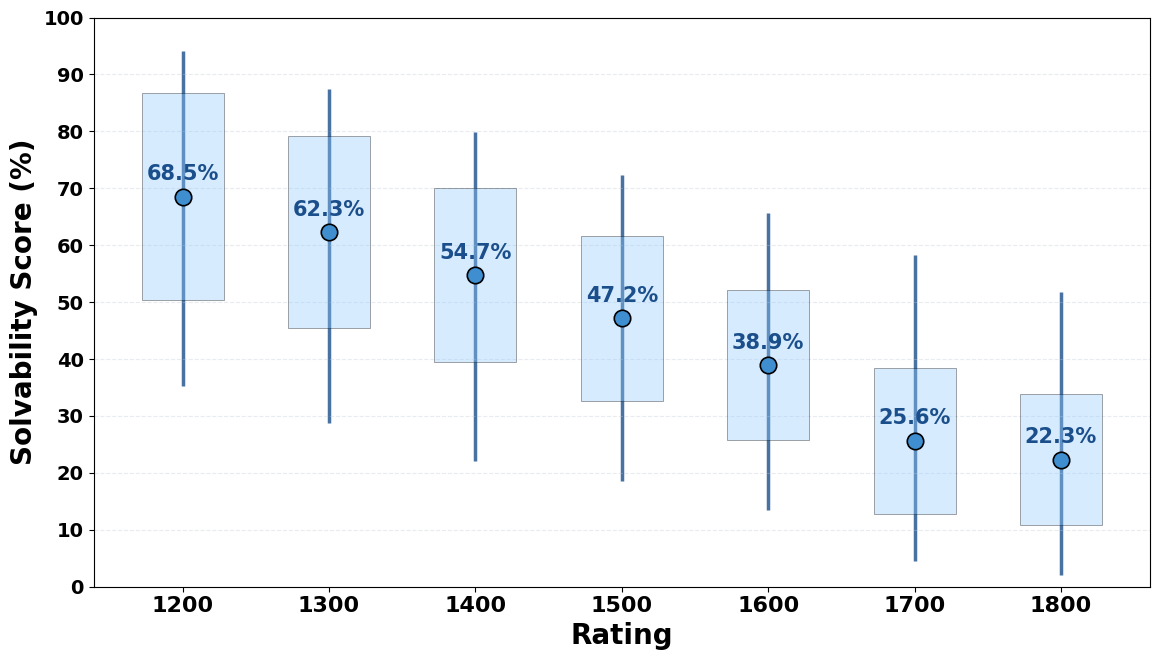}}
\caption{Codeforces problems: Solvability score distribution by problem rating}
\label{fig:codeforces_solvability_by_rating}
\end{minipage}
\end{figure}

Figure~\ref{fig:codeforces_tag_radar} presents Itr$_3$ acceptance rates across the top 10 categories for all six workflows. The radar chart reveals three critical insights:
\textbf{No category-specific critic specialization:} All six workflows maintain parallel polygons with consistent relative sizes. DeepSeek-R1 forms the outermost boundary for both GPT-5 and GPT-4, followed by Llama, then Codestral, across all algorithmic categories. If Codestral possessed implementation-specific advantages or DeepSeek excelled particularly at DP/graphs, we would observe non-parallel distortions in the radar chart.\\
\textbf{Uniform category difficulty hierarchy:} All workflows exhibit identical relative performance ordering: greedy $>$ brute force $>$ math $>$ implementation $>$ DP $>$ bitmasks. This consistent hierarchy indicates that category difficulty is intrinsic rather than model-dependent.\\
\textbf{Generator dominance persists across categories:} GPT-5 workflows achieve 1.8-2.2× higher acceptance rates than GPT-4 across all algorithmic domains, with no category showing exceptional generator-dependent variation.

\noindent
\textbf{Problem Difficulty Beyond Categories:}
Figure~\ref{fig:codeforces_solvability_by_rating} shows that problem rating provides stronger predictive power than category tags. The 200-problem dataset spans ratings 1200-1800 with relatively balanced distribution (36 at 1200, 29 at 1300, 28 at 1400, 25 at 1500, 24 at 1600, 31 at 1700, 27 at 1800). Performance degrades systematically from rating 1200 (47.2-63.9\% Itr$_3$) to rating 1800 (11.1-25.9\% Itr$_3$), with this degradation pattern consistent across all workflows. Average solvability scores decline from 68.5 at rating 1200 to 22.3 at rating 1800, with accelerating decline after rating 1500 suggesting a critical difficulty threshold. The wide standard deviations (11.5-18.2) indicate substantial within-rating variation, demonstrating that Codeforces ratings capture difficulty but do not fully predict LLM solvability.

Table~\ref{tab:codeforces_extreme_problems} identifies the easiest and hardest problems by solvability score. The easiest problems (scores 87.5--94.2) span multiple categories but cluster at rating 1200-1300, while the hardest problems (scores 2.1--10.2) combine multiple advanced algorithmic techniques at ratings 1600-1800. This suggests that problem difficulty emerges from combinatorial complexity rather than individual category challenges.

\begin{table}[h]
\centering
\small
\renewcommand{\arraystretch}{1.1}
\scalebox{0.85}{
\begin{tabular}{lclcc|lclcc}
\toprule
\multicolumn{5}{c}{\textbf{Top 5 Easiest Problems}} & \multicolumn{5}{c}{\textbf{Top 5 Hardest Problems}}\\
\midrule
\textbf{ID} & \textbf{Rating} & \textbf{Tags} & \textbf{Solved} & \textbf{Score} & \textbf{ID} & \textbf{Rating} & \textbf{Tags} & \textbf{Solved} & \textbf{Score} \\
\midrule
2092-C & 1200 & greedy, math, constructive & 6/6 & 94.2 & 2148-F & 1800 & dp, graphs, trees & 0/6 & 2.1 \\
2150-A & 1200 & implementation, data structures & 6/6 & 92.8 & 2162-E & 1800 & dp, bitmasks, combinatorics & 0/6 & 3.8\\
2066-A & 1200 & implementation, greedy & 6/6 & 91.0 & 2132-C2 & 1700 & dp, data structures, binary search & 0/6 & 4.5 \\
2082-B & 1200 & brute force, implementation & 6/6 & 89.3 & 2078-D & 1800 & graphs, dfs, games & 1/6 & 8.7 \\
2114-D & 1300 & implementation, greedy & 6/6 & 87.5 & 2050-F & 1700 & dp, greedy, math & 1/6 & 10.2\\
\ignore{
\midrule
\multicolumn{5}{c}{\textbf{Top 5 Hardest Problems}} \\
\midrule
\textbf{ID} & \textbf{Rating} & \textbf{Tags} & \textbf{Solved} & \textbf{Score} \\
\midrule
2148-F & 1800 & dp, graphs, trees & 0/6 & 2.1 \\
2162-E & 1800 & dp, bitmasks, combinatorics & 0/6 & 3.8 \\
2132-C2 & 1700 & dp, data structures, binary search & 0/6 & 4.5 \\
2078-D & 1800 & graphs, dfs, games & 1/6 & 8.7 \\
2050-F & 1700 & dp, greedy, math & 1/6 & 10.2 \\}
\bottomrule
\end{tabular}}
\caption{Codeforces problems: Extreme cases showing easiest and hardest problems by solvability score.}
\label{tab:codeforces_extreme_problems}
\end{table}

Table~\ref{tab:codeforces_difficulty_categories} categorizes all 200 problems into difficulty tiers based on solvability scores. Nearly half (47.5\%) classify as Hard or Very Hard, demonstrating that the dataset challenges even the best workflows.

\begin{table}[h]
\centering
\small
\begin{minipage}{0.45\textwidth}
\scalebox{0.85}{
\begin{tabular}{lccc}
\toprule
\textbf{Category} & \textbf{Score} & \textbf{Count (\%)} & \textbf{Characteristics} \\
\midrule
Very Easy & $>80$ & 15 (7.5\%) & 5-6 combinations solve, $<$2 avg attempts \\
Easy & 60--80 & 38 (19.0\%) & 4-5 combinations solve, $\sim$2 avg attempts \\
Medium & 40--60 & 52 (26.0\%) & 2-3 combinations solve, $\sim$2.5 avg attempts \\
Hard & 20--40 & 61 (30.5\%) & 1-2 combinations solve, $\sim$2.8 avg attempts \\
Very Hard & $\leq 20$ & 34 (17.0\%) & 0-1 combinations solve, high attempts \\
\bottomrule
\end{tabular}}
\caption{Codeforces problems: Problem difficulty distribution based on solvability scores.}
\label{tab:codeforces_difficulty_categories}
\end{minipage}\hfill
\begin{minipage}{0.45\textwidth}
\scalebox{0.8}{
\begin{tabular}{lcc}
\toprule
\textbf{Workflow} & \textbf{Est. CF Rating} & \textbf{Skill Tier} \\
\midrule
GPT-5 + DeepSeek-R1 & $\sim$1350 & Pupil (Green) \\
GPT-5 + Llama-3.3-70B & $\sim$1320 & Pupil (Green) \\
GPT-5 + Codestral-2508 & $\sim$1290 & Pupil (Green) \\
\midrule
GPT-4 + DeepSeek-R1 & $\sim$1180 & Newbie (Gray) \\
GPT-4 + Llama-3.3-70B & $\sim$1150 & Newbie (Gray) \\
GPT-4 + Codestral-2508 & $\sim$1120 & Newbie (Gray) \\
\bottomrule
\end{tabular}}
\caption{Codeforces problems: Estimated Codeforces rating equivalents for each workflow.} 
\label{tab:codeforces_rating_equivalence}
\end{minipage}
\end{table}

\noindent
\textbf{Generator and Critic Impact Across Categories:} The data synthesizes generator and critic effects across all categories. Generator choice creates a fundamental performance tier separation (~30 percentage points between GPT-5 and GPT-4, as shown in Table~\ref{tab:icpc_time_to_solution}) that remains constant across categories, while critic effects (5-7 problem difference, representing ±2-3 percentage points as shown in Table~\ref{tab:icpc_critic_ranking}) also remain stable across algorithmic domains. This consistent separation across all problem types confirms that neither generator nor critic advantages are category-specific.

\noindent
\textbf{Estimated Rating Equivalence Across Problem Types:} Table~\ref{tab:codeforces_rating_equivalence} estimates Codeforces rating equivalents for each workflow. All GPT-5 workflows reach Pupil tier (1290-1350) while GPT-4 workflows remain at Newbie tier (1120-1180), regardless of the debugging critic. This uniform rating assignment across critics reinforces that specialization effects do not manifest as category-specific advantages.
\ignore{
\begin{table}[h]
\centering
\small
\renewcommand{\arraystretch}{1.2}
\begin{tabular}{lcc}
\toprule
\textbf{Workflow Combination} & \textbf{Estimated CF Rating} & \textbf{Skill Tier} \\
\midrule
GPT-5 + DeepSeek-R1 & $\sim$1350 & Pupil (Green) \\
GPT-5 + Llama-3.3-70B & $\sim$1320 & Pupil (Green) \\
GPT-5 + Codestral-2508 & $\sim$1290 & Pupil (Green) \\
\midrule
GPT-4 + DeepSeek-R1 & $\sim$1180 & Newbie (Gray) \\
GPT-4 + Llama-3.3-70B & $\sim$1150 & Newbie (Gray) \\
GPT-4 + Codestral-2508 & $\sim$1120 & Newbie (Gray) \\
\bottomrule
\end{tabular}
\caption{Codeforces problems: Estimated Codeforces rating equivalents for each workflow. GPT-5 workflows reach Pupil tier (top 30\% of users).}
\label{tab:codeforces_rating_equivalence}
\end{table}
}

\noindent
\textbf{Key Finding for RQ4:} Contrary to our hypothesis, debugging critics do not exhibit significant category-specific specialization. Performance across algorithmic categories (Graph Theory, DP, Math, Implementation, etc.) remains uniformly scaled, with DeepSeek-R1 consistently outperforming Llama-3.3 and Codestral across all domains. The radar chart analysis reveals parallel polygon structures rather than distorted specialization patterns, and performance ratios remain constant (2.3-3.0× for GPT-5 vs GPT-4, 2-7 problems for DeepSeek vs others) across categories. Problem difficulty emerges primarily from algorithmic complexity and rating rather than category-specific challenges, with consistent degradation patterns from rating 1200 (63.9\% Itr$_3$) to rating 1800 (25.9\% Itr$_3$) across all workflows. Codestral's code specialization provides minimal advantages even for implementation-heavy problems, while DeepSeek's reasoning focus benefits all categories uniformly.

\subsection{RQ5: Baseline Comparison - A-ProS vs.\ Agent Loop Variants}
\label{sec:baseline_comparison}

We present RQ5 here because it directly contextualizes the RQ1 iterative gains: it disentangles how much of the improvement is attributable to persistent context versus simply additional compute or repeated refinement attempts.

To situate A-ProS relative to simpler agent loop designs, we compare four approaches on the same $n=47$ stratified ablation subset (Section~\ref{sec:ablation_design}): (1) \textit{Zero-Shot} (Itr$_0$, single initial attempt), (2) \textit{Single-Round Stateless} (one additional feedback iteration after the initial attempt, no history), (3) \textit{Multi-Round Stateless} (three feedback iterations, history cleared before each), and (4) \textit{A-ProS Stateful} (three feedback iterations with full persistent context). All four share the same generator (GPT-5 or GPT-4) and critic (DeepSeek-R1); they vary along two axes — number of refinement iterations and context management strategy. The Multi-Round Stateless vs.\ A-ProS Stateful comparison provides the controlled context-management contrast at a matched four-attempt submission budget.

\begin{table}[h]
\centering\small\renewcommand{\arraystretch}{1.3}
\scalebox{0.85}{
\begin{tabular}{lcccccc}
\toprule
\textbf{Approach} & \multicolumn{2}{c}{\textbf{GPT-5+DeepSeek}} & & \multicolumn{2}{c}{\textbf{GPT-4+DeepSeek}} \\
\cmidrule(lr){2-3}\cmidrule(lr){5-6}
 & \textbf{Solve Rate} & \textbf{vs.\ ZS} & & \textbf{Solve Rate} & \textbf{vs.\ ZS} \\
\midrule
Zero-Shot (Itr$_0$)           & 21.3\% &    -    & & 10.6\% &    -    \\
Single-Round Stateless         & 23.4\% & +9.9\%    & & 12.8\% & +20.8\%   \\
Multi-Round Stateless (3 itr)  & 29.8\% & +39.9\%   & & 17.0\% & +60.4\%   \\
\textbf{A-ProS Stateful (3 itr)} & \textbf{40.4\%} & \textbf{+89.7\%} & & \textbf{25.5\%} & \textbf{+140.6\%} \\
\bottomrule
\end{tabular}}
\caption{RQ5 baseline comparison on $n = 47$ ablation problems (GPT-5/4 + DeepSeek-R1). Zero-Shot and Single-Round Stateless are included as sanity-check baselines with smaller submission budgets. The strict matched-budget comparison is Multi-Round Stateless versus A-ProS Stateful, since both use three refinement iterations and four total attempts. “vs. ZS” = relative improvement over Zero-Shot.}
\label{tab:baseline_comparison}
\end{table}

Table~\ref{tab:baseline_comparison} should be interpreted with respect to compute budget. The Zero-Shot and Single-Round Stateless rows are included as sanity-check baselines, but they do not use the same number of submission attempts as the full A-ProS setting. Therefore, the strict ablation comparison for RQ5 is between Multi-Round Stateless and A-ProS Stateful, since both use the same generator, the same DeepSeek-R1 critic, and the same three-refinement/four-attempt budget. Under this matched-budget comparison, A-ProS Stateful improves over zero-shot by +89.7\% for GPT-5 and +140.6\% for GPT-4, whereas Multi-Round Stateless improves by only +39.9\% and +60.4\%, respectively. Thus, the stateful gains are 2.2--2.3$\times$ larger than the matched multi-round stateless baseline, supporting the conclusion that persistent conversational context, rather than repeated submissions alone, accounts for the additional improvement.

\noindent\textbf{Key Finding for RQ5:} Under the matched four-attempt budget, A-ProS achieves 2.2--2.3$\times$ greater gains than multi-round stateless refinement with the same generator and critic. This indicates that the additional improvement is not explained by repeated submissions alone, but by persistent conversational context enabling cumulative error-aware refinement.

\section{Related Work}

\subsection{Competitive Programming as an Evaluation Benchmark for LLMs}

Prior work has established competitive programming as a rigorous testbed for LLM code generation. ICPCEval~\cite{Xu2025ICPCeval} and LLM-ProS~\cite{LLM-ProS} specifically assess ICPC-style problem solving, showing that current models fall well short of human contestants. Benchmark suites such as APPS~\cite{hendrycks2021apps}, CodeContests~\cite{Li2022Competition}, and HumanEval Pro~\cite{yu2024humanevalpro} document systematic failures in reasoning depth and robustness to edge cases. CodeFlowBench~\cite{Wang2025CodeFlowBench} and PeCC~\cite{haller2024pecc} extend evaluation to flow-control and constraint-satisfaction problems, further exposing the gap between surface fluency and genuine algorithmic reasoning. Our work uses this established benchmark ecosystem but shifts emphasis from single-attempt accuracy to iterative, multi-model refinement—an axis that prior benchmarks do not directly measure.

\subsection{Code Generation with LLMs}
The evolution of code generation has been shaped by models such as \textit{Codex}~\cite{Chen2021Evaluating} and \textit{CodeT5}~\cite{codet5}, which first demonstrated that transformer-based architectures can translate natural language specifications into executable code. Subsequent benchmarks, including \textit{HumanEval}~\cite{humaneval}, \textit{MBPP}~\cite{yu2024humanevalpro}, and \textit{LeetCode-Hard}~\cite{leetcode}, emphasized syntactic correctness and functional validity, yet primarily assessed static, single-turn responses that do not reflect the iterative nature of real-world programming workflows. To address these limitations, more recent frameworks such as \textit{WizardCoder}~\cite{Luo2023WizardCoder} extended \textit{StarCoder}~\cite{Lozhkov2024StarCoder} through reasoning-augmented instruction tuning, improving chain-of-thought coherence and multi-step reasoning in competitive tasks. Similarly, the \textit{CodeGen}~\cite{Nijkamp2022CodeGen} and \textit{CodeGeeX}~\cite{zheng2023codegeex} families expanded model scale and multilingual capabilities but remained constrained by one-pass generation. Collectively, these advances mark a shift from isolated program synthesis to dynamic, multi-round reasoning and feedback-driven refinement, a paradigm that our multi-agent framework extends by integrating agentic coordination, iterative debugging, and adaptive feedback loops to systematically improve reasoning depth, code correctness, and execution reliability.

\subsection{Progression from Brute-Force Generation to Feedback-Guided Reasoning} 

Scaling-based systems have sought to bridge the gap between brute-force exploration and structured reasoning. \textit{AlphaCode}~\cite{Li2022Competition} pioneered this direction by generating millions of candidate programs and applying clustering-based filtering to approximate human-level performance on Codeforces problems. While \textit{AlphaCode} demonstrated the feasibility of large-scale program synthesis, it primarily relied on massive sampling rather than reasoning refinement and lacked mechanisms for adaptive feedback or self-improvement. More recent systems such as \textit{InterCode}~\cite{Yang2023InterCode}, \textit{AgentCoder}~\cite{Huang2023AgentCoder}, and \textit{AgileCoder}~\cite{Nguyen2024AgileCoder} addressed these limitations through iterative reasoning and feedback integration. \textit{InterCode} introduced reinforcement-style refinement via execution feedback, enabling models to self-correct by learning from runtime behavior. \textit{AgentCoder} formalized multi-agent collaboration by assigning specialized roles including planner, coder, and reviewer that coordinate through structured interaction to improve synthesis quality. \textit{AgileCoder} further advanced this paradigm with dynamic task reallocation inspired by agile development practices, allowing models to adapt based on intermediate outcomes. Collectively, these systems represent a shift from single pass synthesis to interactive, reasoning-driven adaptation, a foundation that our framework (\Name{}) extends by embedding explicit feedback exchange, role specialization, and cross model reasoning within a closed agentic loop to enhance performance and consistency across iterative development cycles.

\subsection{Benchmarking Code Reasoning and Execution}
Traditional benchmarks have primarily focused on assessing syntactic and functional correctness, while newer studies place greater emphasis on interpretability and execution reasoning. \textit{CRUXEval}~\cite{Gu2024CRUXEval} introduced tasks for predicting input–output behavior, evaluating a model’s capacity to mentally simulate program execution and control flow. Unlike \textit{HumanEval} or \textit{MBPP}, which reward only correctness, \textit{CRUXEval} measures whether models truly comprehend underlying logic structures, providing a more faithful proxy for reasoning ability. Building upon this direction, \textit{LiveCodeBench}~\cite{Jain2024LiveCodeBench} enhanced realism by integrating real-time code execution to capture runtime behavior, latency, and resource utilization, thereby enabling a more precise assessment of debugging skills and execution grounding. Similarly, \textit{CodeFlowBench}~\cite{Wang2025CodeFlowBench} introduced dependency-aware, multi-function evaluation to examine how models reason across interdependent code components. Collectively, these benchmarks highlight the persistent limitations of current LLMs in execution tracking and context retention, challenges that our framework (\Name{}) seeks to overcome through iterative cross-model verification and feedback-driven correction.

\subsection{Collaborative Multi-Agent Frameworks for Program Synthesis}

Multi-agent architectures have emerged as powerful frameworks for program synthesis and debugging by decomposing complex reasoning tasks into coordinated, specialized roles. \textit{MapCoder}~\cite{Islam2024MapCoder} exemplifies this approach through a four-agent pipeline consisting of a retriever, planner, coder, and debugger, each contributing domain-specific expertise under the supervision of a policy controller that manages dynamic role transitions. This design achieved state-of-the-art performance across competitive programming benchmarks including APPS, xCodeEval, and CodeContests. Building upon this foundation, our framework (\Name{}) extends the paradigm to heterogeneous multi-model collaboration, leveraging distinct reasoning specializations such as GPT-5 for generative synthesis, Codestral-2508 for structural debugging, Llama-3.3 for contextual reasoning, and DeepSeek-R1 for algorithmic optimization. Unlike fixed pipelines such as \textit{MapCoder}, \Name{} operates within a closed feedback loop in which verification outcomes directly inform subsequent iterations, enabling adaptive reflection, trust calibration, and dynamic reasoning delegation to enhance both correctness and efficiency.

\subsection{Agentic Reasoning and Instruction-Following Paradigms}

Recent benchmarks have expanded LLM evaluation beyond correctness and efficiency toward instruction-following under realistic, multi-constraint conditions. \textit{AGENTIF}~\cite{Qi2025AGENTIF} introduced 707 high-fidelity, human-annotated prompts derived from real agentic applications, averaging 1,723 words and nearly twelve constraints per task. These constraints span formatting, semantic, and tool-based dimensions and are expressed through diverse presentation styles such as vanilla, conditional, and example-driven prompts. Unlike earlier datasets including \textit{IFEval}~\cite{zhou2023ifeval}, \textit{FollowBench}~\cite{jiang2024followbench}, and \textit{ComplexBench}~\cite{pei2024complexbench}, \textit{AGENTIF} captures real-world reasoning requirements involving multi-turn constraint validation and execution-grounded compliance. Performance analyses reveal that even state-of-the-art reasoning models such as GPT-4o and DeepSeek-R1 achieve below 30\% instruction adherence, particularly in conditional and tool-integrated tasks. This finding underscores a fundamental challenge in agentic AI: maintaining logical and contextual consistency across extended reasoning horizons. Consequently, instruction-following is increasingly reframed as a manifestation of agentic reasoning, aligning with the collaborative and verification-driven design of our framework (\Name{}), where specialized agents iteratively enforce semantic, syntactic, and logical coherence.

The growing body of work on agentic AI for software engineering reinforces this transition from static model evaluation to dynamic, role-based collaboration. The \textit{LLM-SE Survey}~\cite{Liu2024Large} synthesizes over one hundred studies applying agentic reasoning to program synthesis, debugging, and testing, identifying architectural components such as planner, verifier, and critic that mirror human development workflows. These systems demonstrate that multi-agent collaboration enhances problem decomposition, fault recovery, and interpretability through distributed reasoning and feedback exchange. Building on these insights, our system extends this paradigm by establishing a hybrid feedback loop between a generative model and multiple critic agents, supported by persistent conversational memory to preserve reasoning state across iterations. This approach operationalizes autonomy, reflection, and coordination, the defining attributes of agentic intelligence in software reasoning.

Complementary human-in-the-loop and communication benchmarks further advance the goal of explainable, cooperative reasoning. \textit{HumanEvalComm}~\cite{Wu2024HumanEvalComm} investigated human-style explanations and cross-model dialogue, demonstrating that structured reasoning communication improves both interpretability and success rate. This resonates with our framework’s inter-agent message passing, where specialized LLMs collaboratively evaluate, critique, and repair code in a manner analogous to human peer review. By replacing direct human intervention with automated, context-aware dialogue, \Name{} simulates human-in-the-loop feedback at scale, thereby achieving more transparent and explainable debugging.

\section{Threats to Validity}
In this section, we discuss four types of threats, similar to prior research~\cite{10.1145/3696630.3728701, LLM-ProS}.\\ 
\textbf{Internal Validity:}
Internal validity concerns whether the observed improvements can be confidently attributed to the proposed agentic feedback mechanism. Although all models were executed under identical compute budgets, decoding parameters, and endpoints, several risks remain:\\
(1) Dataset contamination - Problems from ICPC World Finals (2011-2024) and Codeforces contests (rating 1200-1800) may overlap with pretraining corpora of GPT-4, GPT-5, Codestral-2508, Llama-3.3-70B or DeepSeek-R1.
To mitigate this risk for the Codeforces dataset, we selected problems exclusively from recent contests to minimize overlap with training data. Performance consistency across both ICPC and Codeforces datasets provides additional evidence against systematic memorization effects. As a further sensitivity check, we note that contamination most directly affects \emph{zero-shot} performance (Itr$_0$), but may also surface at later iterations. Critically, our main claims rest on the \emph{iterative gain} (Itr$_3 -$ Itr$_0$) and on the controlled ablation comparing stateful vs.\ stateless conditions on the same problems, neither of which is systematically inflated by prior memorization of correct solutions. To further isolate contamination risk, we performed a sensitivity check comparing Itr$_0$ performance on the 50 most recent Codeforces problems (post-2024) against the remaining 150 problems. The comparison did not reveal an observable difference large enough to suggest that memorization was inflating the zero-shot baseline. \\
(2) API nondeterminism - Randomness in cloud inference was controlled using fixed sampling parameters across all models: \texttt{temperature=0.1} for GPT-4, Llama-3.3-70B (Groq), Codestral-2508 (Mistral), and DeepSeek-R1, with \texttt{top\_p=1.0}, \texttt{presence\_penalty=0}, and \texttt{frequency\_penalty=0} for GPT-4. GPT-5 uses \texttt{reasoning\_effort=medium} with model-default parameters as temperature control is not supported. DeepSeek-R1's reasoning model ignores temperature/top-p parameters by design. \\
(3) Evaluation consistency - All submissions used identical Codeforces endpoints and judging infrastructure to ensure fairness and prevent data leakage across accounts.

\noindent
\textbf{External Validity:}
External validity relates to how far results generalize beyond competitive programming.
While ICPC problems approximate high-assurance software-engineering tasks through strict specifications, binary correctness, and constrained resources, the domain still differs from real-world software systems that involve collaboration, testing, and maintenance. 
Feedback agents tuned for concise algorithmic reasoning may not generalize directly to larger codebases or natural-language planning. 
Nonetheless, the strong alignment of CP tasks with unit-test-driven development and verification workflows supports partial generalizability to SE contexts such as debugging, automated test repair, and compiler optimization.

\noindent
\textbf{Construct Validity:}
Construct validity examines whether our metrics truly measure the claimed constructs.
Our study measures both \textbf{acceptance} and \textbf{trust calibration}.
Acceptance is operationalized as Itr$_k$ cumulative solve rate (fraction of problems receiving an Accepted verdict within $k$ feedback iterations), which directly reflects functional correctness as verified by the official Codeforces judge infrastructure. Calibration is operationalized as Expected Calibration Error (ECE) between the critic's expressed confidence (1-5 scale) and the observed next-attempt success rate (Section~\ref{sec:trust_calibration}).
However, we do not measure subjective human trust or effort reduction directly.
Future work should include human-in-the-loop evaluations to assess how calibration translates into perceived trust and usability.

\noindent
\textbf{Conclusion Validity:}
Conclusion validity concerns whether the statistical procedures support the stated conclusions. All pairwise comparisons use McNemar's exact test (appropriate for paired binary outcomes) with Holm-Bonferroni multiplicity correction across all comparisons, and 95\% bootstrap confidence intervals (10{,}000 resamples). Effect sizes are reported as Cohen's $h$ (appropriate for binary proportions). At $n=47$ (two-tailed $\alpha = 0.05$), 80\% power requires effects of Cohen's $h \geq 0.41$; the observed effects ($h = 0.21$--$0.22$) yield approximately 30\% power, which is consistent with the inconclusive Holm-Bonferroni-adjusted $p$-values. The 95\% bootstrap CIs ($[0.00, +0.15]$ for GPT-5 and $[0.00, +0.11]$ for GPT-4) have lower bounds touching zero, consistent with the limited power at $n = 47$; the primary evidence for the context effect is therefore the 2.9--3.5$\times$ reduction in error repetition rather than the CIs alone. Generator-level comparisons ($n=200$) provide substantially higher power. We acknowledge that individual critic-pair comparisons at $n=200$ are underpowered for the observed small effects ($h < 0.13$); conclusions about the critic ranking are therefore stated as directional evidence (consistent ordering across 12 comparisons) rather than individually significant findings.

\section{Discussion and Limitations}
This section synthesizes key insights from our experimental evaluation and discusses methodological considerations and limitations that contextualize our findings.

\subsection{Generator Capability vs. Critic Specialization}
\begin{figure}[h]
\centering
\fbox{\includegraphics[scale= 0.18]{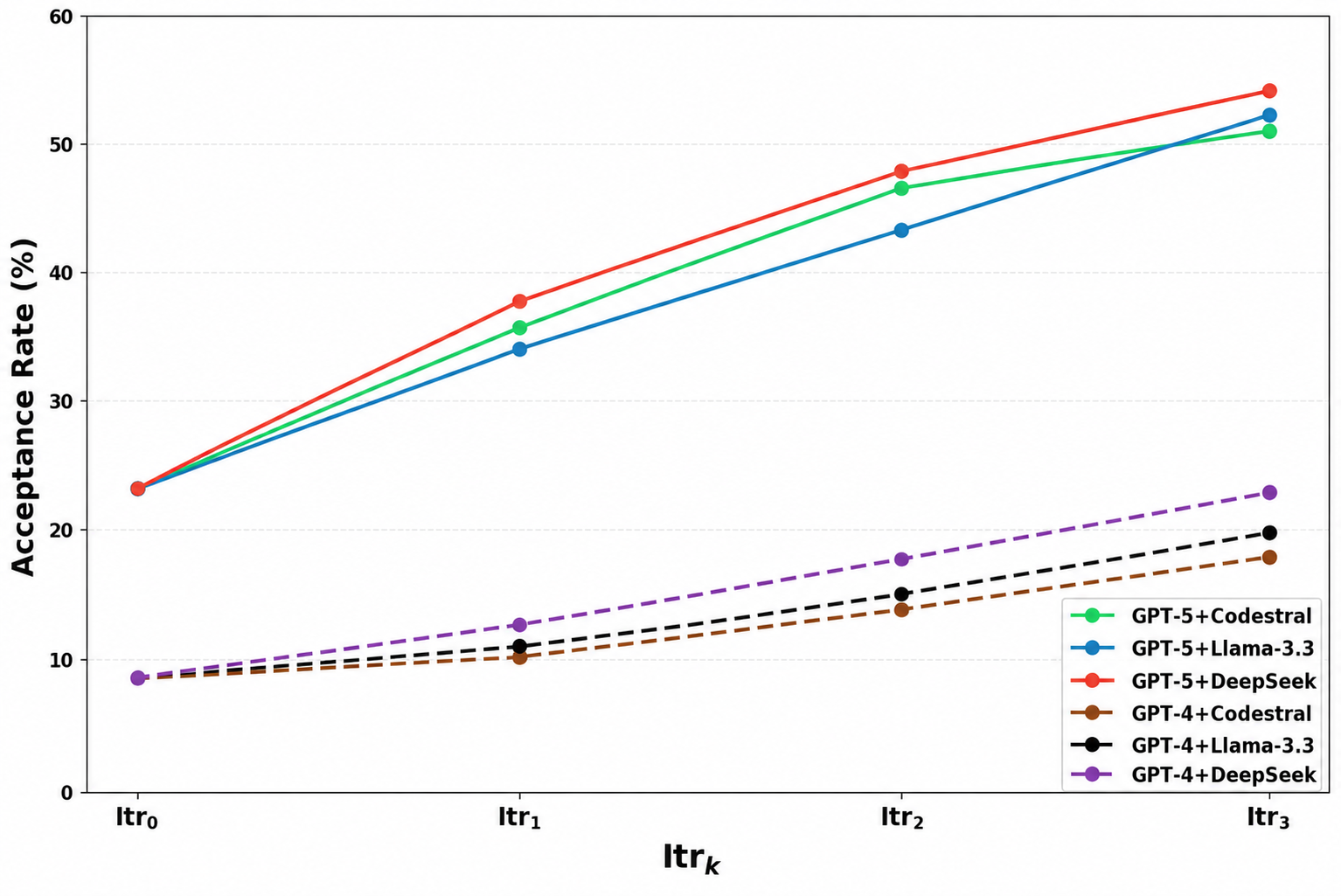}}
\caption{ICPC World Finals problems: Itr$_k$ progression showing consistent ~30\% generator gap across all iterations.}
\label{fig:icpc_gpt4_vs_gpt5_comparison}
\end{figure}

Our results reveal a clear hierarchy in performance determinants: solution generator capability dominates critic specialization effects across all experimental conditions. These findings are scoped to competitive programming (ICPC World Finals + Codeforces), which provides rigorous, binary-correctness evaluation; generalization to open-ended software engineering tasks warrants further investigation (Section~\ref{sec:limitations}). Figure~\ref{fig:icpc_gpt4_vs_gpt5_comparison} illustrates this phenomenon on ICPC World Finals problems, where GPT-5 workflows achieve 50.9-53.9\% Itr$_3$ cumulative acceptance compared to GPT-4's 18.6-22.8\%. In critic-matched comparisons (DeepSeek vs.\ DeepSeek, Llama vs.\ Llama, Codestral vs.\ Codestral), the GPT-5/GPT-4 ratio ranges from 2.36$\times$ to 2.74$\times$—a consistent 2.4-2.7$\times$ performance gap that persists across all three debugging critics.

Critically, even the \textit{weakest} GPT-5 workflow (GPT-5 + Codestral, 50.9\%) substantially outperforms the \textit{strongest} GPT-4 workflow (GPT-4 + DeepSeek, 22.8\%), demonstrating that superior debugging feedback cannot fully compensate for weaker initial solution generation. This finding has important implications for multi-agent system design: while critic selection matters (DeepSeek-R1 achieves 5-7 more solved problems than Codestral within the same generator tier), \textbf{investing in stronger solution generators yields greater returns than optimizing critic models alone}.

However, the story is more nuanced for relative improvement. GPT-4 workflows exhibit 107-153\% relative gains from Itr$_0$ to Itr$_3$, compared to GPT-5's 118-131\% gains. This indicates that \textbf{feedback loops provide disproportionate value for weaker generators}, partially closing the absolute performance gap through iterative refinement. For resource-constrained scenarios where state-of-the-art generators are unavailable, pairing mid-tier generators with reasoning-focused critics (like DeepSeek-R1) offers a viable path to competitive performance.

\subsection{Critic Trust Calibration}
\label{sec:trust_calibration}

A practical deployment concern for multi-agent debugging systems is whether a critic's expressed confidence in its hint reliably predicts whether that hint will lead to a successful next attempt. A well-calibrated critic enables principled decisions such as down-weighting low-confidence hints or escalating to a stronger model. Confidence scores are elicited by appending a structured instruction to Prompt~2 (Section~\ref{sec:metrics} shows the exact prompt template and normalization procedure). We measure calibration using Expected Calibration Error (ECE) as defined in Section~\ref{sec:metrics}, evaluated on the full Codeforces benchmark ($n=200$ problems, all six workflows). For each failed submission, the critic appends a confidence score to its hint; whether that submission is followed by an accepted next attempt is recorded from the judge log. Table~\ref{tab:trust_calibration} reports critic-level results for both generator tiers. Generator-level Abstention Rate is reported separately in the main results because it is not critic-specific.

\begin{table}[h]
\centering\small\renewcommand{\arraystretch}{1.3}
\scalebox{0.88}{
\begin{tabular}{llcccc}
\toprule
\textbf{Generator} & \textbf{Critic Model} & \textbf{$N$ hints} & \textbf{ECE\,$\downarrow$} & \textbf{Post-Hint Acc.} & \textbf{Verif.\ Cost\,$\downarrow$} \\
\midrule
\multirow{3}{*}{GPT-5}
& DeepSeek-R1    & 425 & \textbf{0.366} & \textbf{9.6\%} & \textbf{7.6} \\
& Llama-3.3-70B  & 428 & 0.440          & 7.7\%          & 8.2 \\
& Codestral-2508 & 436 & 0.499          & 7.1\%          & 8.7 \\
\midrule
\multirow{3}{*}{GPT-4}
& DeepSeek-R1    & 504 & \textbf{0.398} & \textbf{6.5\%} & \textbf{13.5} \\
& Llama-3.3-70B  & 509 & 0.457          & 5.3\%          & 15.1 \\
& Codestral-2508 & 518 & 0.530          & 4.6\%          & 17.1 \\
\bottomrule
\end{tabular}}
\caption{Critic trust calibration on the Codeforces benchmark ($n=200$ problems). $N$ = number of hints per critic (varies by workflow; see Table~\ref{tab:codeforces_time_to_solution}); ECE = Expected Calibration Error (lower is better; see Section~\ref{sec:metrics}); Post-hint Acc.\ = fraction of hints that directly lead to an accepted next submission; Verification Cost = total judge submissions $\div$ total accepted solutions (consistent with Table~\ref{tab:codeforces_time_to_solution}). The DeepSeek $>$ Llama $>$ Codestral calibration ranking holds across both generator tiers.}
\label{tab:trust_calibration}
\end{table}

\noindent\textbf{Results.}
All ECE values are high in absolute terms ($>0.35$), reflecting low base success rates and the floored 1--5 confidence scale; the following comparison is therefore relative across critics, not against an absolute calibration target. Competitive programming is a hard domain: even the best-performing critic (DeepSeek-R1 with GPT-5) directly converts only 9.6\% of its hints into accepted submissions on the very next attempt, because the majority of problems require algorithmic insight that a single debugging hint cannot fully supply. The ECE values in Table~\ref{tab:trust_calibration} are accordingly above 0.35 for all critics and both generators — a consequence of the normalized confidence scale $[0.2, 1.0]$ implying baseline success probabilities of 20--100\%, while the actual per-bin acceptance rates reach at most 30--58\% even at the highest confidence level. Within this domain-specific constraint, however, the critics differ meaningfully in their relative calibration. DeepSeek-R1 achieves the lowest ECE under GPT-5 (0.366): its confidence distribution is skewed toward lower levels (25\% of hints at confidence~1), and its success rate rises sharply with confidence, giving it stronger discriminative power per hint. Llama-3.3-70B (ECE\,=\,0.440, 7.7\%) and Codestral-2508 (ECE\,=\,0.499, 7.1\%) are progressively more overconfident: both assign higher confidence levels to hints whose actual success rates do not justify them, widening the calibration gap at the high-confidence bins where most of their hints are concentrated. For GPT-4, absolute acceptance rates fall further (4.6--6.5\%), but the critic ranking is identical: DeepSeek-R1 (ECE\,=\,0.398, 6.5\%, Verif.\ Cost\,=\,13.5) $>$ Llama-3.3-70B (ECE\,=\,0.457, 5.3\%, Verif.\ Cost\,=\,15.1) $>$ Codestral-2508 (ECE\,=\,0.530, 4.6\%, Verif.\ Cost\,=\,17.1).

The consistent calibration ranking across both generator tiers confirms that DeepSeek-R1's discriminative advantage is an intrinsic property of the critic, not an artifact of generator strength. These findings suggest that calibration-aware feedback routing improves outcomes by preferring lower-ECE critics when available and abstaining from hints when expressed confidence is low. This approach reduces wasted submissions in future deployments, particularly given the low base success rate of the task domain.

\subsection{Limitations and Methodological Considerations}
\label{sec:limitations}

While our experimental evaluation demonstrates substantial improvements through multi-agent feedback across 367 competitive programming problems, several limitations warrant careful consideration.

\subsubsection{Platform-Specific Data Access}
Although our core workflow (solution generation → submission → feedback → refinement) remains consistent across all problems, platform-specific access restrictions necessitate methodological adaptations between the two problem sources:

\noindent
\textbf{Codeforces Problemset (Fully Automated):} For the 200 problems from recent contests (rating 1200-1800), detailed test failure information is publicly accessible. The system automatically retrieves failed test inputs, expected outputs, actual outputs, and diagnostic messages, enabling complete end-to-end automation with zero human intervention.

\noindent
\textbf{Codeforces Gym/ICPC (Hybrid Automation):} The 167 ICPC World Finals problems impose stricter visibility policies—detailed test information is restricted to users with competitive programming ratings $\geq$ 1900. To balance research reproducibility with platform compliance, we adopt a hybrid approach:
\begin{itemize}
    \item Solution generation, automated submission, and verdict capture proceed fully automatically
    \item When a verdict indicates failure, detailed test information is manually retrieved using an authenticated high-rated account and formatted consistently with the automated pipeline
    \item Once test data is structured, feedback generation and solution refinement proceed automatically
\end{itemize}

\noindent
Critically, this manual step is limited to \textit{data retrieval only}—no human debugging, code analysis, or solution guidance is provided. The LLM-driven feedback loop operates identically across both datasets once test information is available, ensuring that our core findings about multi-agent feedback effectiveness generalize across problem sources. We document this difference explicitly for reproducibility and to guide future researchers working with restricted-access platforms.

\subsubsection{Evaluation Scope and Generalizability}
Our evaluation focuses on competitive programming problems, which provide rigorous functional correctness testing but may not fully capture the complexities of real-world software engineering tasks such as API integration, system design, long-context reasoning, or multi-file codebases. The highly structured nature of competitive programming—well-defined inputs/outputs, deterministic judging, single-file solutions—differs substantially from production software development. Future work should investigate whether multi-agent feedback patterns observed here (e.g., reasoning-focused critics outperforming code-specialized critics) generalize to open-ended software tasks.

\subsubsection{Model Variability and Reproducibility}
Reliance on API-based models (GPT-4, GPT-5, DeepSeek-R1) introduces inherent variability due to hidden parameter updates, nondeterministic decoding, and rate-limiting policies. While we document exact API parameters (temperature, top-p, reasoning effort) and timestamp all experiments, exact numerical reproducibility cannot be guaranteed across different time periods or API versions. Our findings emphasize \textit{relative performance patterns} (e.g., DeepSeek-R1 $>$ Llama-3.3 $>$ Codestral) that demonstrate statistical significance and consistent ordering across 367 problems, providing robustness beyond single-run variability.

\subsubsection{Functional Correctness vs. Code Quality}
Our framework evaluates functional correctness exclusively—solutions either pass or fail the judge's test suite. We do not assess code quality dimensions such as readability, maintainability, adherence to best practices, or computational efficiency beyond meeting time/memory limits. While our methodology section notes instances where debugging critics suggest algorithmic optimizations (e.g., O(n²) → O(n) for Problem 2043-C), we do not systematically measure code elegance or engineering quality. This limitation reflects our competitive programming context, where correctness is the sole evaluation criterion, but limits generalizability to software engineering tasks prioritizing long-term maintainability.

\subsection{Implications for Multi-Agent System Design}
Despite these limitations, our findings provide actionable guidance for practitioners designing multi-agent LLM systems for algorithmic code generation tasks (with the caveat that generalization to open-ended SE contexts requires further study). The clear performance hierarchy—\textit{generator capability $>$ critic specialization}—suggests that system designers should prioritize: (1) securing access to state-of-the-art solution generators, (2) pairing them with reasoning-focused rather than code-specialized critics when logical correctness is paramount, (3) implementing persistent conversation contexts to enable cumulative learning across refinement cycles, and (4) selecting well-calibrated critics (low ECE) to enable reliable confidence-based routing of debugging hints. The substantial Itr$_k$ improvements (74--174\% relative gains over zero-shot) confirm that multi-agent feedback loops with persistent context offer a qualitatively distinct, cost-effective alternative to massive sampling approaches such as AlphaCode's 1M-candidate generation strategy.

\ignore{
\subsubsection{Reproducibility and Artifact Availability}

To facilitate independent verification and extension, we provide a comprehensive replication package:

\begin{itemize}
    \item \textbf{Source Code:} Complete implementation with modular architecture, dependency specifications, and setup instructions
    \item \textbf{Problem Dataset:} SQLite database containing all 367 problems with full specifications, constraints, sample tests, and hidden test metadata (where permissible)
    \item \textbf{Prompt Templates:} Exact system and user message templates for solution generation and feedback generation (detailed in Sections 3.4-3.6)
    \item \textbf{Generated Solutions:} Up to 6,606 C++ solution files (367 problems $\times$ 6 workflows $\times$ 3 attempts) with metadata headers
    \item \textbf{Execution Logs:} Complete solving logs including conversational state, API latencies, token counts, and verdict progressions
    \item \textbf{Judge Responses:} Detailed JSON files capturing test case inputs, outputs, and diagnostic messages from Codeforces
    \item \textbf{Analysis Scripts:} Python notebooks for metric computation, statistical testing, and visualization generation
\end{itemize}

All artifacts are timestamped and version-controlled, ensuring alignment between code, data, and reported results. The package includes instructions for reproducing experiments on new problem sets and integrating additional LLM providers.}

\section{Conclusion}
\label{conclusion}
In this paper, we presented \Name{}, an autonomous AI agent that integrates large language models (LLMs) into a structured, multi-stage workflow for automated competitive programming. We designed \Name{} to coordinate multiple specialized models, using GPT-4 and GPT-5 as solution generators and Codestral-2508, Llama-3.3, and DeepSeek-R1 as debugging critics, to emulate the reasoning and refinement processes of human software engineers. We conducted extensive experiments on 367 real-world problems from ICPC and Codeforces, demonstrating that iterative, feedback-driven collaboration substantially improves code correctness across difficulty levels. A controlled paired ablation ($n = 47$) provides evidence consistent with a causal contribution of persistent conversational context: stateful refinement reduces error repetition by 2.9--3.5$\times$ and outperforms multi-round stateless baselines by 2.2--2.3$\times$ (A-ProS achieves 90--141\% gain over zero-shot versus 40--60\% for multi-round stateless). Given the moderate subset size and the GPT-5 raw McNemar $p = 0.063$, we interpret this result as strong directional evidence rather than a definitive causal estimate. Trust calibration analysis further shows that DeepSeek-R1 is the best-calibrated critic (ECE\,=\,0.366 under GPT-5; 0.398 under GPT-4) versus Codestral-2508 (ECE\,=\,0.499; 0.530), with the ranking consistent across both generator tiers. These findings confirm that multi-agent collaboration, persistent contextual memory, and well-calibrated feedback critics collectively enhance the reasoning depth and adaptability of LLM-based programming systems for algorithmic code-generation tasks. Looking ahead, we plan to extend \Name{} toward calibration-aware feedback routing, abstaining from low-confidence hints and dynamically escalating to stronger critics. We also aim to incorporate reinforcement-guided feedback selection and adaptive role switching to further improve transparency and resilience. We envision \Name{} as a promising architecture for reliable, execution-grounded autonomous programming in competitive programming, with broader software-engineering transfer remaining a question for future work.

\section{Acknowledgment}
This work was supported in part by NSF grants CCF-2348277 and CCF-2518445 \cite{2024nsf....2348277Z}.

\bibliographystyle{unsrt}  
\bibliography{reference}

@misc{LLM,
  author       = {W. X. Zhao and K. Zhou and J. Li and T. Tang and X. Wang and Y. Hou and Y. Min and B. Zhang and J. Zhang and Z. Dong and Y. Du and C. Yang},
  title        = {A Survey of Large Language Models},
  note         = {[Online; accessed 29 October 2025]},
  year         = {2023},
  howpublished = {\url{https://arxiv.org/abs/2303.18223}}
}

@misc{SE,
  author       = {Haolin Jin and Linghan Huang and Haipeng Cai and Jun Yan and Bo Li and Huaming Chen},
  title        = {From LLMs to LLM-Based Agents for Software Engineering: A Survey of Current Challenges and Future Directions},
  note         = {[Online; accessed 29 October 2025]},
  year         = {2024},
  howpublished = {\url{https://arxiv.org/abs/2408.02479}}
}

@misc{NLP-SE,
  title        = {A Comparison of Natural Language Understanding Platforms for Chatbots in Software Engineering},
  note         = {[Online; accessed 29 October 2025]},
  year         = {2021},
  howpublished = {\url{http://ieeexplore.ieee.org/abstract/document/9426404}}
}

@misc{NLP-SE2,
  author       = {Sabina-Cristiana Necula and Florin Dumitriu and Valerică Greavu-Şerban},
  title        = {A Systematic Literature Review on Using Natural Language Processing in Software Requirements Engineering},
  note         = {[Online; accessed 29 October 2025]},
  year         = {2024},
  howpublished = {\url{https://www.mdpi.com/2079-9292/13/11/2055}}
}

@misc{Se-Model,
  title        = {Model-Driven Engineering of Manufacturing Automation Software Projects — A SysML-Based Approach},
  note         = {[Online; accessed 29 October 2025]},
  year         = {2014},
  howpublished = {\url{https://www.sciencedirect.com/science/article/pii/S0957415814000853}}
}

@book{ChowdharyFundamentals,
  author       = {K. R. Chowdhary},
  title        = {Fundamentals of Artificial Intelligence},
  year         = {2020},
  publisher    = {Springer},
  note         = {[Online; accessed 29 October 2025]},
  howpublished = {\url{https://link.springer.com/book/10.1007/978-81-322-3972-7}}
}

@INPROCEEDINGS{LLM-ProS,
  author={Hossain, Md Sifat and Tabassum, Anika and Arefin, Md. Fahim and Shaila Zaman, Tarannum},
  booktitle={2025 IEEE/ACM International Workshop on Large Language Models for Code (LLM4Code)}, 
  title={LLM-ProS: Analyzing Large Language Models’ Performance in Competitive Problem Solving}, 
  year={2025},
  volume={},
  number={},
  pages={80-87},
  doi={10.1109/LLM4Code66737.2025.00015}}

@misc{ICPC,
  title        = {The ICPC International Collegiate Programming Contest},
  note         = {[Online; accessed 29 October 2025]},
  howpublished = {\url{https://icpc.global/worldfinals/past-problems}}
}

@misc{codet5,
  author       = {Y. Wang and W. Wang and S. Joty and S. C. H. Hoi},
  title        = {CodeT5: Identifier-Aware Unified Pre-Trained Encoder-Decoder Models for Code Understanding and Generation},
  note         = {[Online; accessed 29 October 2025]},
  year         = {2021},
  howpublished = {https://arxiv.org/abs/2109.00859}
}

@misc{humaneval,
  author       = {D. Li and L. Murr},
  title        = {HumanEval on Latest GPT Models -- 2024},
  note         = {[Online; accessed 29 October 2025]},
  year         = {2024},
  howpublished = {\url{https://arxiv.org/abs/2402.14852}}
}

@misc{leetcode,
  title        = {LeetCode -- The World's Leading Online Programming Learning Platform},
  note         = {[Online; accessed 29 October 2025]},
  howpublished = {\url{https://leetcode.com/problemset/?difficulty=HARD}}
}

@misc{haller2024pecc,
  author       = {P. Haller and J. Golde and A. Akbik},
  title        = {PECC: Problem Extraction and Coding Challenges},
  note         = {[Online; accessed 29 October 2025]},
  year         = {2024},
  howpublished = {\url{https://arxiv.org/abs/2404.18766}}
}

@misc{perfcodegen2024,
  author       = {Y. Peng and A. D. Gotmare and M. Lyu and C. Xiong and S. Savarese and D. Sahoo},
  title        = {PerfCodeGen: Improving Performance of LLM-Generated Code with Execution Feedback},
  note         = {[Online; accessed 29 October 2025]},
  year         = {2024},
  howpublished = {\url{https://arxiv.org/abs/2412.03578}}
}

@misc{Wang2025Agentic,
	note = {[Online; accessed 2025-10-31]},
	author = {Wang, Huanting and Gong, Jingzhi and Zhang, Huawei and Xu, Jie and Wang, Zheng},
	year = {2025},
	month = {aug 15},
	title = {AI agentic programming: A survey of techniques, challenges, and opportunities},
	howpublished = {https://arxiv.org/abs/2508.11126},
}

@misc{Xu2025ICPCeval,
	note = {[Online; accessed 2025-11-01]},
	author = {Xu, Shiyi and Hu, Yiwen and Min, Yingqian and Chen, Zhipeng and Zhao, Wayne Xin and Wen, Ji-Rong},
	year = {2025},
	month = {jun 5},
	title = {ICPC-{Eval}: Probing the frontiers of {LLM} reasoning with competitive programming contests},
	howpublished = {https://arxiv.org/abs/2506.04894},
}

@misc{hendrycks2021apps,
  author       = {Dan Hendrycks and Steven Basart and Mantas Mazeika and Andy Zou and Dawn Song},
  title        = {Measuring Coding Challenge Competence With APPS},
  year         = {2021},
  howpublished = {\url{https://arxiv.org/abs/2105.09938}}
}

@misc{yu2024humanevalpro,
  author       = {Zijian Yu and Yuxiang Zhao and Arman Cohan and Xue-Ping Zhang},
  title        = {HumanEval Pro and MBPP Pro: Evaluating Large Language Models on Self-Invoking Code Generation},
  year         = {2024},
  howpublished = {\url{https://arxiv.org/abs/2412.21199}}
}

@misc{Luo2023WizardCoder,
	note = {[Online; accessed 2025-11-01]},
	author = {Luo, Ziyang and Xu, Can and Zhao, Pu and Sun, Qingfeng and Geng, Xiubo and Hu, Wenxiang and Tao, Chongyang and Ma, Jing and Lin, Qingwei and Jiang, Daxin},
	year = {2023},
	month = {jun 14},
	title = {WizardCoder: Empowering code large language models with evol-instruct},
	howpublished = {https://arxiv.org/abs/2306.08568},
}

@misc{Lozhkov2024StarCoder,
	note = {[Online; accessed 2025-11-01]},
	author = {Lozhkov, Anton and Li, Raymond and Allal, Loubna Ben and Cassano, Federico and Lamy-Poirier, Joel and Tazi, Nouamane and Tang, Ao and Pykhtar, Dmytro and Liu, Jiawei and Wei, Yuxiang and Liu, Tianyang and Tian, Max and Kocetkov, Denis and Zucker},
	year = {2024},
	month = {feb 29},
	title = {StarCoder 2 and {The} {Stack} v2: The {Next} {Generation}},
	howpublished = {https://arxiv.org/abs/2402.19173},
}

@misc{Nijkamp2022CodeGen,
	note = {[Online; accessed 2025-11-01]},
	author = {Nijkamp, Erik and Pang, Bo and Hayashi, Hiroaki and Tu, Lifu and Wang, Huan and Zhou, Yingbo and Savarese, Silvio and Xiong, Caiming},
	year = {2022},
	month = {mar 25},
	title = {CodeGen: An open large language model for code with multi-turn program synthesis},
	howpublished = {https://arxiv.org/abs/2203.13474},
}

@misc{zheng2023codegeex,
  author       = {Yifei Zheng and Jiale Xue and Chenghao Xia and Zhipeng Zhang and Zhiyuan Liu and Maosong Sun},
  title        = {CodeGeeX: A Pre-Trained Model for Code Generation with Multilingual Evaluation on HumanEval-X},
  year         = {2023},
  howpublished = {\url{https://arxiv.org/abs/2303.17568}}
}

@misc{Yang2023InterCode,
	note = {[Online; accessed 2025-11-01]},
	author = {Yang, John and Prabhakar, Akshara and Narasimhan, Karthik and Yao, Shunyu},
	year = {2023},
	month = {jun 26},
	title = {InterCode: Standardizing and benchmarking interactive coding with execution feedback},
	howpublished = {https://arxiv.org/abs/2306.14898},
}

@misc{Huang2023AgentCoder,
	note = {[Online; accessed 2025-11-01]},
	author = {Huang, Dong and Zhang, Jie M. and Luck, Michael and Bu, Qingwen and Qing, Yuhao and Cui, Heming},
	year = {2023},
	month = {dec 20},
	title = {AgentCoder: Multi-{Agent}-based code generation with iterative testing and optimisation},
	howpublished = {https://arxiv.org/abs/2312.13010},
}

@misc{Nguyen2024AgileCoder,
	note = {[Online; accessed 2025-11-01]},
	author = {Nguyen, Minh Huynh and Chau, Thang Phan and Nguyen, Phong X. and Bui, Nghi D. Q.},
	year = {2024},
	month = {jun 16},
	title = {AgileCoder: Dynamic {Collaborative} {Agents} for {Software} {Development} based on {Agile} {Methodology}},
	howpublished = {https://arxiv.org/abs/2406.11912},
}

@misc{Gu2024CRUXEval,
	note = {[Online; accessed 2025-11-01]},
	author = {Gu, Alex and Rozi{\` e}re, Baptiste and Leather, Hugh and Solar-Lezama, Armando and Synnaeve, Gabriel and Wang, Sida I.},
	year = {2024},
	month = {jan 5},
	title = {CRUXEval: A benchmark for code reasoning, understanding and execution},
	howpublished = {https://arxiv.org/abs/2401.03065},
}

@misc{Jain2024LiveCodeBench,
	note = {[Online; accessed 2025-11-01]},
	author = {Jain, Naman and Han, King and Gu, Alex and Li, Wen-Ding and Yan, Fanjia and Zhang, Tianjun and Wang, Sida and Solar-Lezama, Armando and Sen, Koushik and Stoica, Ion},
	year = {2024},
	month = {mar 12},
	title = {LiveCodeBench: Holistic and contamination free evaluation of large language models for code},
	howpublished = {https://arxiv.org/abs/2403.07974},
}

@misc{Wang2025CodeFlowBench,
	note = {[Online; accessed 2025-11-01]},
	author = {Wang, Sizhe and Wang, Zhengren and Ma, Dongsheng and Yu, Yongan and Ling, Rui and Li, Zhiyu and Xiong, Feiyu and Zhang, Wentao},
	year = {2025},
	month = {apr 30},
	title = {CodeFlowBench: A multi-turn, iterative benchmark for complex code generation},
	howpublished = {https://arxiv.org/abs/2504.21751},
}

@misc{Islam2024MapCoder,
	note = {[Online; accessed 2025-11-01]},
	author = {Islam, Md. Ashraful and Ali, Mohammed Eunus and Parvez, Md Rizwan},
	year = {2024},
	month = {may 18},
	title = {MapCoder: Multi-{Agent} code generation for competitive problem solving},
	howpublished = {https://arxiv.org/abs/2405.11403},
}

@misc{Qi2025AGENTIF,
	note = {[Online; accessed 2025-11-01]},
	author = {Qi, Yunjia and Peng, Hao and Wang, Xiaozhi and Xin, Amy and Liu, Youfeng and Xu, Bin and Hou, Lei and Li, Juanzi},
	year = {2025},
	month = {may 22},
	title = {AGENTIF: Benchmarking instruction following of large language models in agentic scenarios},
	howpublished = {https://arxiv.org/abs/2505.16944},
}

@misc{zhou2023ifeval,
  author       = {Jeffrey Zhou and Tianjian Lu and Swaroop Mishra and Siddhartha Brahma and Sujoy Basu},
  title        = {Instruction-Following Evaluation for Large Language Models (IFEval)},
  year         = {2023},
  howpublished = {\url{https://arxiv.org/abs/2311.07911}}
}

@inproceedings{jiang2024followbench,
  author       = {Yuxin Jiang and Yufei Wang and Xingshan Zeng and Wanjun Zhong and Liangyou Li and Fei Mi and Lifeng Shang and Xin Jiang and Qun Liu and Wei Wang},
  title        = {FollowBench: A Multi-level Fine-grained Constraints Following Benchmark for Large Language Models},
  booktitle    = {Proceedings of the 62nd Annual Meeting of the Association for Computational Linguistics (Volume 1: Long Papers)},
  pages        = {4667--4688},
  year         = {2024},
  address      = {Bangkok, Thailand},
  publisher    = {Association for Computational Linguistics},
  note         = {[Online; accessed 29 October 2025]}
}

@misc{pei2024complexbench,
  author       = {Pei Ke and others},
  title        = {ComplexBench: Benchmarking Complex Instruction-Following with Multiple Constraints Composition},
  year         = {2024},
  howpublished = {\url{https://arxiv.org/abs/2407.03978}}
}

@misc{Liu2024Large,
	note = {[Online; accessed 2025-11-01]},
	author = {Liu, Junwei and Wang, Kaixin and Chen, Yixuan and Peng, Xin and Chen, Zhenpeng and Zhang, Lingming and Lou, Yiling},
	year = {2024},
	month = {sep 4},
	title = {Large language model-based agents for software engineering: A survey},
	howpublished = {https://arxiv.org/abs/2409.02977},
}

@misc{Wu2024HumanEvalComm,
	note = {[Online; accessed 2025-11-01]},
	author = {Wu, Jie JW and Fard, Fatemeh H},
	year = {2024},
	month = {may 31},
	title = {HumanEvalComm: Benchmarking the communication competence of code generation for llms and {LLM} agent},
	howpublished = {https://arxiv.org/abs/2406.00215},
}

@misc{Dong2023Self,
	note = {[Online; accessed 2025-11-03]},
	author = {Dong, Yihong and Jiang, Xue and Jin, Zhi and Li, Ge},
	year = {2023},
	month = {apr 15},
	title = {Self-collaboration code generation via chatgpt},
	howpublished = {https://arxiv.org/abs/2304.07590},
}

@misc{Luo2024RepoAgent,
	note = {[Online; accessed 2025-11-03]},
	author = {Luo, Qinyu and Ye, Yining and Liang, Shihao and Zhang, Zhong and Qin, Yujia and Lu, Yaxi and Wu, Yesai and Cong, Xin and Lin, Yankai and Zhang, Yingli and Che, Xiaoyin and Liu, Zhiyuan and Sun, Maosong},
	year = {2024},
	month = {feb 26},
	title = {RepoAgent: An llm-powered open-source framework for repository-level code documentation generation},
	howpublished = {https://arxiv.org/abs/2402.16667},
}

@misc{BugSpotter,
	note = {[Online; accessed 2025-11-03]},
	title = {BugSpotter: Automated generation of code debugging exercises},
	howpublished = {https://dl.acm.org/doi/abs/10.1145/3641554.3701974},
}

@article{ZhuAre,
	note = {[Online; accessed 2025-11-03]},
	author = {Zhu, Yizhang and Du, Shiyin and Li, Boyan and Luo, Yuyu and Tang, Nan},
	journal = {Advances in Neural Information Processing Systems},
	pages = {62697--62731},
	title = {Are large language models good statisticians?},
	howpublished = {https://proceedings.neurips.cc/paper\textunderscore{}files/paper/2024/hash/729786203d330da046dd8091c2d92a66-Abstract-Datasets\textunderscore{}and\textunderscore{}Benchmarks\textunderscore{}Track.html},
	volume = {37},
    year={2024}
}

@misc{Sam2025Predicting,
	note = {[Online; accessed 2025-11-03]},
	author = {Sam, Dylan and Finzi, Marc and Kolter, J. Zico},
	year = {2025},
	month = {jan 2},
	title = {Predicting the performance of black-box {LLMs} through self-queries},
	howpublished = {https://arxiv.org/abs/2501.01558},
}

@article{Hughes2025AI,
	note = {[Online; accessed 2025-11-03]},
	author = {Hughes, Laurie and Dwivedi, Yogesh K. and Malik, Tegwen and Shawosh, Mazen and Albashrawi, Mousa Ahmed and Jeon, Il and Dutot, Vincent and Appanderanda, Mandanna and Crick, Tom and De\textquoteright{}, Rahul and Fenwick, Mark and Gunaratnege, Senali Madugoda and Jurcys, Paulius and Kar, Arpan Kumar and Kshetri, Nir and Li, Keyao and Mutasa, Sashah and Samothrakis, Spyridon and Wade, Michael and Walton, Paul},
	journal = {Journal of Computer Information Systems},
	number = {4},
	year = {2025},
	month = {apr 24},
	pages = {489--517},
	publisher = {Informa UK Limited},
	title = {AI agents and agentic systems: a multi-expert analysis},
	volume = {65},
}

@misc{Putta2024Agent,
	note = {[Online; accessed 2025-11-03]},
	author = {Putta, Pranav and Mills, Edmund and Garg, Naman and Motwani, Sumeet and Finn, Chelsea and Garg, Divyansh and Rafailov, Rafael},
	year = {2024},
	month = {aug 13},
	title = {Agent {Q}: Advanced reasoning and learning for autonomous {AI} agents},
	howpublished = {https://arxiv.org/abs/2408.07199},
}

@misc{Xue2025IMPROVE,
	note = {[Online; accessed 2025-11-03]},
	author = {Xue, Eric and Chen, Ke and Huang, Zeyi and Ji, Yuyang and Wang, Haohan},
	year = {2025},
	month = {feb 25},
	title = {IMPROVE: Iterative model pipeline refinement and optimization leveraging {LLM} experts},
	howpublished = {https://arxiv.org/abs/2502.18530},
}

@misc{Li2025From,
	note = {[Online; accessed 2025-11-03]},
	author = {Li, Zhuoyan and Zhu, Hangxiao and Lu, Zhuoran and Xiao, Ziang and Yin, Ming},
	year = {2025},
	month = {feb 17},
	title = {From text to trust: Empowering ai-assisted decision making with adaptive llm-powered analysis},
	howpublished = {https://arxiv.org/abs/2502.11919},
}

@misc{Yang2025survey,
	note = {[Online; accessed 2025-11-04]},
	author = {Yang, Boyang and Cai, Zijian and Liu, Fengling and Le, Bach and Zhang, Lingming and Bissyand{\' e}, Tegawend{\' e} F. and Liu, Yang and Tian, Haoye},
	year = {2025},
	month = {jun 30},
	title = {A survey of llm-based automated program repair: Taxonomies, design paradigms, and applications},
	howpublished = {https://arxiv.org/abs/2506.23749},
}

@misc{Meng2024empirical,
	note = {[Online; accessed 2025-11-04]},
	author = {Meng, Xiangxin and Ma, Zexiong and Gao, Pengfei and Peng, Chao},
	year = {2024},
	month = {nov 15},
	title = {An empirical study on llm-based agents for automated bug fixing},
	howpublished = {https://arxiv.org/abs/2411.10213},
}

@article{Buehler2025PRefLexOR,
	note = {[Online; accessed 2025-11-04]},
	author = {Buehler, Markus J.},
	journal = {npj Artificial Intelligence},
	number = {1},
	year = {2025},
	month = {may 14},
	pages = {1--38},
	title = {PRefLexOR: Preference-based recursive language modeling for exploratory optimization of reasoning and agentic thinking},
	howpublished = {https://www.nature.com/articles/s44387-025-00003-z},
	volume = {1},
}

@article{Lakshmi2025Enhancing,
	note = {[Online; accessed 2025-11-04]},
	author = {Lakshmi, A. Sri and Sigamany, E. S. Sharmila and Traisa, Roopa and Kumar, Raman and Reddy, Karaka Ramakrishna and Chohan, Jasgurpreet Singh and Smerat, Aseel},
	journal = {International Journal of Advanced Computer Science and Applications (IJACSA)},
	number = {9},
	year = {2025},
	month = {sep 30},
	title = {Enhancing code quality through automated refactoring using transformer-based language models},
	howpublished = {https://thesai.org/Publications/ViewPaper?Volume=16&Issue=9&Code=IJACSA&SerialNo=51},
	volume = {16},
}

@misc{Roychoudhury2025Agentic,
	note = {[Online; accessed 2025-11-04]},
	author = {Roychoudhury, Abhik},
	year = {2025},
	month = {aug 24},
	title = {Agentic {AI} for {Software}: Thoughts from {Software} {Engineering} community},
	howpublished = {https://arxiv.org/abs/2508.17343},
}

@misc{Chen2021Evaluating,
	note = {[Online; accessed 2025-11-04]},
	author = {Chen, Mark and Tworek, Jerry and Jun, Heewoo and Yuan, Qiming and Pinto, Henrique Ponde de Oliveira and Kaplan},
	year = {2021},
	month = {jul 7},
	title = {Evaluating large language models trained on code},
	howpublished = {https://arxiv.org/abs/2107.03374},
}

@misc{Wang2022Self,
	note = {[Online; accessed 2025-11-04]},
	author = {Wang, Xuezhi and Wei, Jason and Schuurmans, Dale and Le, Quoc and Chi, Ed and Narang, Sharan and Chowdhery, Aakanksha and Zhou, Denny},
	year = {2022},
	month = {mar 21},
	title = {Self-{Consistency} improves chain of thought reasoning in language models},
	howpublished = {https://arxiv.org/abs/2203.11171},
}

@article{WeiChain,
	note = {[Online; accessed 2025-11-04]},
	author = {Wei, Jason and Wang, Xuezhi and Schuurmans, Dale and Bosma, Maarten and ichter, brian and Xia, Fei and Chi, Ed and Le, Quoc V and Zhou, Denny},
	journal = {Advances in Neural Information Processing Systems},
	pages = {24824--24837},
	title = {Chain-of-{Thought} prompting elicits reasoning in large language models},
    year={2022},
	howpublished = {https://proceedings.neurips.cc/paper/2022/hash/9d5609613524ecf4f15af0f7b31abca4-Abstract-Conference.html?fbclid=IwY2xjawN2bDJleHRuA2FlbQIxMABicmlkETFra3hxU0pLMG1RV3JkMnVxAR4zH\textunderscore{}Wzdq30nWWqbenJD1m1j3Rxc-OxhsKzN8NRomAEnR\textunderscore{}bKQGJFn6\textunderscore{}ugzYvQ\textunderscore{}aem\textunderscore{}g7zBDwPBeKwcis619LArgA},
	volume = {35},
}

@misc{OpenAI2023Gpt,
	note = {[Online; accessed 2025-11-04]},
	author = {{OpenAI} and Achiam, Josh and Adler, Steven and Agarwal, Sandhini and Ahmad, Lama and Akkaya, Ilge and Aleman},
	year = {2023},
	month = {mar 15},
	title = {Gpt-4 technical report},
	howpublished = {https://arxiv.org/abs/2303.08774},
}

@article{Leon2026GPT,
	note = {[Online; accessed 19 March 2026]},
	author = {{OpenAI}},
	year = {2025},
	month = {aug 13},
	title = {GPT-5 System Card},
	howpublished = {\url{https://cdn.openai.com/gpt-5-system-card.pdf}},
}

@misc{DeepSeek2024Deepseek,
	note = {[Online; accessed 19 March 2026]},
	author = {Guo, Daya and Yang, Dejian and Zhang, Haowei and Song, Junxiao and Zhang, Ruoyu and Xu, Runxin and Zhu, Qihao and Ma, Shirong and Wang, Peiyi and Bi, Xiao and others},
	year = {2025},
	month = {jan},
	title = {DeepSeek-R1: Incentivizing reasoning capability in llms via reinforcement learning},
	howpublished = {https://arxiv.org/abs/2501.12948},
}

@misc{Choi2024Linq,
	note = {[Online; accessed 19 March 2026]},
	author = {{Mistral AI}},
	year = {2025},
	month = {aug},
	title = {Codestral 25.08},
	howpublished = {\url{https://docs.mistral.ai/models/codestral-25-08}},
}

@misc{Sawant2025Agentic,
	note = {[Online; accessed 2025-11-04]},
	author = {Sawant, Prashant},
	year = {2025},
	month = {feb 20},
	title = {Agentic {AI}: A quantitative analysis of performance and applications},
	howpublished = {https://www.preprints.org/manuscript/202502.1647},
}

@article{Allam2025Agentic,
	note = {[Online; accessed 2025-11-04]},
	author = {Allam, Hesham and Dempere, Juan},
	journal = {The Artificial Intelligence Business Review},
	number = {1},
	year = {2025},
	month = {aug 5},
	title = {Agentic {AI} for {IT} and beyond: A qualitative analysis of capabilities, challenges, and governance},
	howpublished = {https://theaibr.com/index.php/aibr/article/view/3},
	volume = {1},
}

@misc{CodeforcesGym,
  title        = {Codeforces Gym -- Practice and Training Platform for Competitive Programming},
  howpublished = {\url{https://codeforces.com/gyms}},
  note         = {[Online; accessed 4 November 2025]}
}

@misc{Li2022Competition,
	note = {[Online; accessed 2025-11-07]},
	author = {Li, Yujia and Choi, David and Chung, Junyoung and Kushman, Nate and Schrittwieser, Julian and Leblond, R{\' e}mi and Eccles, Tom and Keeling, James and Gimeno, Felix and Lago, Agustin Dal and Hubert, Thomas and Choy, Peter and d'Autume, Cyprien de Masson and Babuschkin, Igor and Chen, Xinyun and Huang, Po-Sen and Welbl, Johannes and Gowal, Sven and Cherepanov, Alexey and Molloy, James and Mankowitz, Daniel J. and Robson, Esme Sutherland and Kohli, Pushmeet and de Freitas, Nando and Kavukcuoglu, Koray and Vinyals, Oriol},
	year = {2022},
	month = {feb 8},
	title = {Competition-{Level} code generation with alphacode},
	howpublished = {https://arxiv.org/abs/2203.07814},
}

@techreport{AlphaCode2_2023,
  author      = {{Google DeepMind}},
  title       = {{AlphaCode 2} Technical Report},
  institution = {Google DeepMind},
  year        = {2023},
  month       = {dec},
  note        = {Available at \url{https://storage.googleapis.com/deepmind-media/AlphaCode2/AlphaCode2_Tech_Report.pdf}; accessed 12 May 2026}
}

@misc{Grattafiori2024llama,
	note = {[Online; accessed 2025-11-06]},
	author = {Grattafiori, Aaron and Dubey, Abhimanyu and Jauhri, Abhinav and Pandey, Abhinav and Kadian, Abhishek and Al-Dahle},
	year = {2024},
	month = {jul 31},
	title = {The llama 3 herd of models},
	howpublished = {https://arxiv.org/abs/2407.21783},
}

@misc{latex2mathml,
  title        = {LaTeX2MathML: A converter for transforming LaTeX equations to MathML},
  note         = {[Online; accessed 6 November 2025]},
  howpublished = {\url{https://pypi.org/project/latex2mathml/}}
}

@misc{MathJax,
  title        = {MathJax: Beautiful math in all browsers},
  note         = {[Online; accessed 6 November 2025]},
  howpublished = {\url{https://www.mathjax.org/}}
}

@misc{BeautifulSoup4,
  title        = {BeautifulSoup4: HTML and XML parsing library for Python},
  note         = {[Online; accessed 6 November 2025]},
  howpublished = {\url{https://www.crummy.com/software/BeautifulSoup/bs4/doc/}}
}

@misc{sifatGitHub,
	note = {[Online; accessed 2025-11-06]},
	author = {{sifat-hossain-niloy}},
	title = {GitHub - {Sifat}-hossain-niloy/{A-Pros}},
	howpublished = {https://github.com/sifat-hossain-niloy/A-Pros},
}

@misc{CodeforcesTestCasesRepo,
  author       = {{sifat-hossain-niloy}},
  title        = {Codeforces Problems Test Cases},
  note         = {[Online; accessed 19 March 2026]},
  howpublished = {\url{https://github.com/sifat-hossain-niloy/Codeforces-Problems-Test-Cases}}
}

@misc{SeleniumHQ,
  title        = {Selenium},
  note         = {[Online; accessed 19 March 2026]},
  howpublished = {\url{https://www.selenium.dev/}}
}

@misc{Playwright,
  title        = {Playwright},
  note         = {[Online; accessed 19 March 2026]},
  howpublished = {\url{https://playwright.dev/}}
}

@misc{CodeforcesAPIHelp,
  title        = {Codeforces API Help},
  note         = {[Online; accessed 19 March 2026]},
  howpublished = {\url{https://codeforces.com/apiHelp}}
}

@misc{CodeforcesVerdict,
  title        = {Codeforces: Verdicts and Judging System},
  note         = {[Online; accessed 6 November 2025]},
  howpublished = {\url{https://codeforces.com/blog/entry/79}}
}

@misc{Shoshany2024C,
	note = {[Online; accessed 2025-11-07]},
	title = {C++17},
	howpublished = {https://en.cppreference.com/w/cpp/17.html?},
}

@misc{SQLite,
	note = {[Online; accessed 2025-11-07]},
	title = {SQLite home page},
	howpublished = {https://www.sqlite.org/index.html},
}

@misc{Hybrid,
	note = {[Online; accessed 2025-11-07]},
	title = {A-Pros},
	howpublished = {https://sifat-hossain-niloy.github.io/A-Pros/},
}

@inproceedings{10.1145/3696630.3728701,
author = {Al Hasan, Alif and Saha, Subarna and Imran, Mia Mohammad and Zaman, Tarannum Shaila},
title = {LLPut: Investigating Large Language Models for Bug Report-Based Input Generation},
year = {2025},
isbn = {9798400712760},
publisher = {Association for Computing Machinery},
address = {New York, NY, USA},
url = {https://doi.org/10.1145/3696630.3728701},
doi = {10.1145/3696630.3728701},
booktitle = {Proceedings of the 33rd ACM International Conference on the Foundations of Software Engineering},
pages = {1652–1659},
numpages = {8},
location = {Clarion Hotel Trondheim, Trondheim, Norway},
series = {FSE Companion '25}
}

@inproceedings{naeini2015obtaining,
  title     = {Obtaining Well Calibrated Probabilities Using Bayesian Binning},
  author    = {Naeini, Mahdi Pakdaman and Cooper, Gregory F. and Hauskrecht, Milos},
  booktitle = {Proceedings of the AAAI Conference on Artificial Intelligence},
  volume    = {29},
  number    = {1},
  year      = {2015}
}

@book{agresti2002categorical,
  title     = {Categorical Data Analysis},
  author    = {Agresti, Alan},
  year      = {2002},
  edition   = {2},
  publisher = {Wiley-Interscience}
}

@book{cohen1988statistical,
  title     = {Statistical Power Analysis for the Behavioral Sciences},
  author    = {Cohen, Jacob},
  year      = {1988},
  edition   = {2},
  publisher = {Lawrence Erlbaum Associates}
}

@MISC{2024nsf....2348277Z,
       author = {{Zaman}, Tarannum Shaila},
        title = "{CRII: SHF: An Automated and User-centered Framework for Reproducing System-level Concurrency Bugs by Analyzing Bug Reports}",
 howpublished = {NSF Award Number 2348277. Directorate for Computer and Information Science and Engineering, Division of Computing and Communication Foundations. 2024.},
         year = 2024,
        month = jun,
        pages = {48277},
       adsurl = {https://ui.adsabs.harvard.edu/abs/2024nsf....2348277Z}
}

@article{parvez2026depro,
  title={DePro: Understanding the Role of LLMs in Debugging Competitive Programming Code},
  author={Parvez, Nabiha and Pallab, Tanvin Sarkar and Imran, Mia Mohammad and Zaman, Tarannum Shaila},
  journal={arXiv preprint arXiv:2603.19399},
  year={2026}
}

@article{wilson1927probable,
  title     = {Probable Inference, the Law of Succession, and Statistical Inference},
  author    = {Wilson, Edwin B.},
  journal   = {Journal of the American Statistical Association},
  volume    = {22},
  number    = {158},
  pages     = {209--212},
  year      = {1927},
  publisher = {Taylor \& Francis},
  doi       = {10.2307/2276774}
}

@article{brown2001interval,
  title     = {Interval Estimation for a Binomial Proportion},
  author    = {Brown, Lawrence D. and Cai, T. Tony and DasGupta, Anirban},
  journal   = {Statistical Science},
  volume    = {16},
  number    = {2},
  pages     = {101--133},
  year      = {2001},
  publisher = {Institute of Mathematical Statistics},
  doi       = {10.1214/ss/1009213286}
}

@article{edwards1948note,
  title   = {Note on the Correction for Continuity in Testing the Significance of the Difference between Correlated Proportions},
  author  = {Edwards, Allen L.},
  journal = {Psychometrika},
  volume  = {13},
  number  = {3},
  pages   = {185--187},
  year    = {1948},
  doi     = {10.1007/BF02289261}
}

@article{ZAMAN2026112785,
title = {SysPro: Reproducing system-level concurrency bugs from bug reports},
journal = {Journal of Systems and Software},
volume = {236},
pages = {112785},
year = {2026},
issn = {0164-1212},
doi = {https://doi.org/10.1016/j.jss.2026.112785},
url = {https://www.sciencedirect.com/science/article/pii/S0164121226000191},
author = {Tarannum Shaila Zaman and Chadni Islam and Jiangfan Shi and Zihan Shi and Fiona Xian and Tingting Yu},
}

\end{document}